\documentclass[12pt]{article}
\pdfoutput=1
\usepackage[cp1251]{inputenc}
\usepackage{amssymb,psfrag,graphicx}
\usepackage{amsmath}
\usepackage{latexsym}
\usepackage[usenames]{color}
\usepackage{fancybox}
\usepackage{simplewick}
\usepackage{amscd}
\usepackage[all,cmtip]{xy}
\usepackage{mathrsfs}
\usepackage{comment}
\usepackage{cite}
\usepackage{framed}
\definecolor{shadecolor}{rgb}{0.9,0.9,0.95}
\usepackage{setspace}
\usepackage[normalem]{ulem}
\definecolor{darkgreen}{rgb}{0,0.5,0}
\definecolor{darkblue}{cmyk}{0.9,0.9,0,0}
\definecolor{darkred}{rgb}{0.6,0,0.3}

\usepackage{graphicx}
\usepackage{tikz}
\usepackage[setpagesize=false,pagebackref=false, linktocpage, bookmarksopen=true, colorlinks=true, linkcolor=darkblue,citecolor=darkblue,urlcolor=black]{hyperref}
\usepackage{hyperref}

\newcommand{\tr}{{\rm tr}}

\renewcommand{\thefootnote}{\arabic{footnote}}

\def\eqref#1{(\ref{#1})}

\def\beq{\begin{equation}}
\def\eeq{\end{equation}}

\newcommand{\rd}{\mathrm{d}}
\newcommand{\re}{\mathrm{e}}

\newcommand{\ri}{\mathrm{i}}

\newcommand{\rw}{\mathrm{w}}

\newcommand{\rz}{\mathrm{z}}

\newcommand{\rC}{\mathrm{C}}

\newcommand{\rE}{\mathrm{E}}

\newcommand{\rH}{\mathrm{H}}

\newcommand{\rJ}{\mathrm{J}}

\newcommand{\rP}{\mathrm{P}}
\newcommand{\rQ}{\mathrm{Q}}

\newcommand{\rT}{\mathrm{T}}

\numberwithin{equation}{section}

\begin{document}

\thispagestyle{empty}

\renewcommand{\thefootnote}{\fnsymbol{footnote}}
\setcounter{page}{1}
\setcounter{footnote}{0}
\setcounter{figure}{0}
\vspace{0.7cm}
\begin{center}
\Large{\textbf{A pedagogical review on solvable irrelevant deformations of 2d quantum field theory}}

\vspace{1.3cm}

\normalsize{Yunfeng Jiang}
\\ \vspace{1cm}
\footnotesize{\textit{
CERN Theory Department, Geneva, Switzerland
}
\vspace{1cm}
}

\par\vspace{1.0cm}

\textbf{Abstract}\vspace{2mm}
\end{center}
\noindent
This is a pedagogical review on $\rT\overline{\rT}$ deformation of two dimensional quantum field theories. It is based on three lectures which the author gave at ITP-CAS in December 2018. This review consists of four parts. The first part is a general introduction to $\rT\overline{\rT}$ deformation. Special emphasises are put on the deformed classical Lagrangian and the exact solvability of the spectrum. The second part focuses on the torus partition sum of the $\rT\overline{{\rT}}$/$\rJ\overline{\rT}$ deformed conformal field theories and modular invariance/covariance. In the third part, different perspectives of $\rT\overline{\rT}$ deformation are presented, including its relation to random geometry, 2d topological gravity and holography. We summarize more recent developments until January 2021 in the last part.
\setcounter{page}{1}
\renewcommand{\thefootnote}{\arabic{footnote}}
\setcounter{footnote}{0}
\setcounter{tocdepth}{2}
\newpage
\tableofcontents

\section{Introduction}
\label{sec:intro}
Quantum field theory (QFT) provides an important theoretical framework for understanding nature. The space of QFTs can be explored with the help of renormalization group (RG). Conformal field theories constitute the fixed points of RG flow. Understanding conformal field theory is an important question and is now under intensive study.\par

At the same time, it is equally interesting to study the flows away from fixed points. Depending on the operators that trigger the flow, the deformations of QFTs can be divided into three broad classes, namely \emph{relevant}, \emph{marginal} and \emph{irrelevant}. While there have been a lot of works on relevant and marginal deformations of QFT, irrelevant deformations of QFTs are largely unexplored, for good reasons. By standard arguments of renormalization theory, irrelevant deformations of QFTs necessarily require an infinite number of counter terms when computing physical quantities. Therefore the Lagrangian involves a tower of infinitely many terms and becomes highly ambiguous. However, it was discovered recently that for certain special classes of irrelevant deformations in 2 dimensional spacetime, this procedure is under much better control and is even solvable \cite{Smirnov:2016lqw,Cavaglia:2016oda}. This is the so-called $\rT\overline{\rT}$ deformation, which is the main subject of this review. Apart from the $\rT\overline{\rT}$ deformation \cite{Smirnov:2016lqw,Cavaglia:2016oda}, more families of solvable deformations of QFTs have been proposed in the past two years. For theories with a conserved holomorphic $U(1)$ current, one can define the $\rJ\overline{\rT}$ deformation \cite{Guica:2017lia}. For integrable quantum field theories which have higher conserved charges, one can define a whole family of deformations using the conserved currents \cite{Smirnov:2016lqw,LeFloch:2019rut,Conti:2019dxg}.\par

\subsection*{Solvability of $\rT\overline{\rT}$ deformation}
The $\rT\overline{\rT}$ deformation is triggered by the irrelevant operator $\rT\overline{\rT}=-\det(T_{\mu\nu})$.
Although the $\rT\overline{\rT}$ deformation flows towards UV, many physically interesting quantities can be computed \emph{exactly} and \emph{explicitly} in terms of the data from the undeformed theory. More precisely, these quantities include the $S$-matrix, the deformed classical Lagrangian, the finite volume spectrum on an infinite cylinder and the torus partition function. We give a brief discussion on these quantities in what follows.\par

\paragraph{The S-matrix} The $S$-matrix is deformed in a simple way under $\rT\overline{\rT}$ deformation. The deformed $S$-matrix is obtained from the undeformed $S$-matrix by multiplying a phase factor of Castillejo-Dalitz-Dyson type (CDD factor) \cite{Castillejo:1955ed}. This deformation of the $S$-matrix was first conjectured in \cite{Dubovsky:2012wk,Dubovsky:2013ira} to describe a toy model of quantum gravity. The authors of \cite{Dubovsky:2012wk,Dubovsky:2013ira} called the multiplication of the CDD factor ``gravitationl dressing'' because the deformed theory exhibit certain features of gravity. One such feature is non-locality which we will discuss in more detail in this review. Later, the same authors showed that the gravitational dressing of the $S$-matrix corresponds to the $\rT\overline{\rT}$ deformation in the Lagrangian description \cite{Dubovsky:2017cnj}.\par

\paragraph{Deformed Lagrangian} The definition of $\rT\overline{\rT}$ deformation is given in the Lagrangian description of QFT. We will present this definition in section~\ref{sec:def}. From the definition, one can derive the deformed classical Lagrangians. It is found that the deformed Lagrangians take very interesting forms \cite{Cavaglia:2016oda,Bonelli:2018kik,Conti:2018jho}. For example, the deformed Lagrangian of a free massless boson takes the form of the Nambu-Goto action in the static gauge. The derivation of this result will be discussed in detail in section~\ref{sec:def}. More complicated Lagrangians lead to more involved deformed Lagrangians. Due to the complexity of the deformed Lagrangian, it is hard to use them to compute physical quantities at the quantum level. Even so, classical analysis of the deformed Lagrangians give useful hints for some of the features of the deformed quantum theory.\par

\paragraph{Finite volume spectrum} The surprising feature of $\rT\overline{\rT}$ deformation is that despite the complicated form of the deformed Lagrangian, one can find the finite volume spectrum in a universal way. This is based crucially on Zamolodchikov's factorization formula for the expectation value of $\rT\overline{\rT}$ operator. This interesting formula was first proved by Zamolodchikov in 2004 \cite{Zamolodchikov:2004ce}. What makes the $\rT\overline{\rT}$ great again after more than a decade is that the authors of \cite{Smirnov:2016lqw} and \cite{Cavaglia:2016oda} realized that one can perform an irrelevant deformation using this interesting operator. Then the factorization formula results in the solvability of the finite volume spectrum. The deformed spectrum obeys a partial differential equation which is the inviscid Burgers' equation in 1d. This is a well-known equation in fluid dynamics which can be solved explicitly with given initial conditions. We will discuss in detail how to obtain the deformed spectrum in section~\ref{sec:def-spectrum}. In integrable quantum field theories, one can obtain the finite volume spectrum by integrability methods such as thermodynamic Bethe ansatz. The only input for these methods are the factorized $S$-matrix, which is given by the gravitational dressing as we discussed before. Indeed one finds that the two methods give the same deformed spectrum.\par

\paragraph{Torus partition function} Given the deformed spectrum in the finite volume, it is natural to study the torus partition function of the $\rT\overline{\rT}$ deformed theory. The first interesting result concerning the torus partition function is given by Cardy \cite{Cardy:2018sdv} who derived a diffusion type of differential equation for the partition function. It is proven in \cite{Datta:2018thy} that the torus partition function of the $\rT\overline{\rT}$ deformed conformal field theory is still modular invariant although the theory is not conformal any more. More interestingly, it is shown in \cite{Aharony:2018bad} that by \emph{requiring} modular invariance and that the deformed spectrum depends on the undeformed spectrum in a universal way, one can fix the deformed theory uniquely to be that of the $\rT\overline{\rT}$ deformed CFT. A similar analysis can be applied to the $\rJ\overline{\rT}$ deformed CFT where one replaces modular invariance by modular covariance\cite{Aharony:2018ics}. Section~\ref{lecture2} is devoted to a detailed discussion on the torus partition function of the $\rT\overline{\rT}$ and $\rJ\overline{\rT}$ deformed CFTs.\par

\subsection*{Why is $\rT\overline{\rT}$ deformation solvable ?}
As we mentioned above, usually irrelevant deformation is ambiguous, but the $\rT\overline{\rT}$ deformation is solvable. A natural question is what is the reason for this solvability ? There are different point of views on this.

\paragraph{Integrable deformation} One way to see this is to consider integrable QFT. These theories have infinitely many conserved charges. It is shown that these conserved charges remain conserved along the $\rT\overline{\rT}$ flow \cite{Smirnov:2016lqw}. In this sense, it is an \emph{integrable} deformation. This implies that the deformation preserves an infinite amount of symmetry, which strongly constraints the flow and makes it integrable.\par

\paragraph{2d topological gravity} Of course, most QFTs are not integrable. However, the $\rT\overline{\rT}$ deformation is defined for any QFT. There must be other ways to understand the solvability of $\rT\overline{\rT}$ deformation. One such point of view is provided by Cardy's random geometry picture \cite{Cardy:2018sdv}. It is shown that the infinitesimal deformation of the partition function by $\rT\overline{\rT}$ is equivalent to integrating over the variations of the underlying spacetime geometry. The `gravity sector' which governs the possible variations turns out to be a total derivative. This implies that the `gravity sector' related to the infinitesimal $\rT\overline{\rT}$ deformation is solvable. Related to this more geometrical picture, it is proposed in \cite{Dubovsky:2017cnj} that turning on $\rT\overline{\rT}$ deformation for a QFT is equivalent to coupling the theory to a specific 2d topological gravity, which is very similar to the Jackiw-Teitelboim gravity. We will discuss these two points of views in section~\ref{lecture3}.\par

\subsection*{Other motivations and developments}
The $\rT\overline{\rT}$ coupling can have two possible signs. Depending on the convention, for one of the signs, the deformed spectrum become complex if the undeformed energy is large enough. This leads to possible break down of unitarity. We will call this sign of the coupling the `bad sign'\footnote{Sometimes it is also called the `wrong sign'. However, we want to stress that these names should not be taken as moral judgements. We love both signs equally.}. The other sign which leads to well behaved spectrum in the high energy limit is called the `good sign'. These two signs of the deformation parameter lead to very different physics in the UV.

\paragraph{Holography} For the bad sign, we cannot take the original energy spectrum to be too large, otherwise the deformed spectrum becomes complex. There is an interesting holographic dual for this sign proposed in \cite{McGough:2016lol}. The proposal is that for pure gravity (that is, not including matter fields) the $\rT\overline{\rT}$ deformation corresponds to putting a Dirichlet boundary condition in the bulk at finite radius. The ability to `move into the bulk' using $\rT\overline{\rT}$ deformation is very interesting from AdS/CFT point of view and might shed some lights on some fundamental questions such as bulk locality. The relation between $\rT\overline{\rT}$ deformation and the cut-off geometry is further explored in \cite{Kraus:2018xrn,Taylor:2018xcy,Hartman:2018tkw,Caputa:2019pam}.\par

\paragraph{Little string theory} For the good sign, we can take the original energy to be infinitely high. One can investigate the density of states in the deep UV regime. It turns out the asymptotic growth of the density of states in the $\rT\overline{\rT}$ deformed CFT interpolates between the Cardy behavior $\log\rho\sim\sqrt{E}$ and the Hagedorn behavior $\log\rho\sim E$. The Hagedorn behavior indicates that the $\rT\overline{\rT}$ deformed theory is not a local QFT. In fact, this is the expected behavior for a class of non-local theories called little string theory which is dual to gravity theories on linear dilaton background.
Motivated by this fact, a single trace deformation on the worldsheet of string theories in $AdS_3$ is proposed in \cite{Giveon:2017nie}. This deformation share many features of the $\rT\overline{\rT}$ deformation. In particular, it gives the same deformed spectrum. This work is further developed in \cite{Giveon:2017myj,Asrat:2017tzd,Chakraborty:2018kpr,Chakraborty:2018aji}. In this review, we focus on the field theory side and will not discuss this interesting direction. For a complementary review, we refer to the lecture note of Giveon \cite{Giveon:2019fgr}.

\subsection*{Structure of the review}
This review is aimed for non-experts who want to learn about the subject for the first time, so we try to be very pedagogical, especially in section~\ref{sec:def},\ref{sec:def-spectrum} and \ref{lecture2}. Most of the main results are derived in detail. In section~\ref{lecture3}, since we try to cover more topics, the derivations are more sketchy but we try to make the ideas and main steps clear. Finally in section~\ref{newDevp}, we briefly summarize new developments up to January 2021. This section is even more concise and it mainly serves as a literature guide for the interested reader. The structure of this review is as follows.\par

In the first part, which consists section~\ref{sec:def} and \ref{sec:def-spectrum}, we first give the definition of $\rT\overline{\rT}$ deformation and derive the deformed Lagrangian for free massless boson. Then we consider the quantum theory and derive Zamolodchikov's factorization formula which is used to find the finite volume spectrum of the deformed theory on an infinite cylinder. In the last part, we present a different derivation of Zamolodchikov's factorization formula which does not rely on Cartesian coordinate systems. The formalism can be generalized to spacetimes with constant curvature and show that factorization formula does not apply for non-zero curvatures.\par

In section~\ref{lecture2}, we focus on the modular properties of the torus partition function of $\rT\overline{\rT}$ and $\rJ\overline{\rT}$ deformed CFTs. We first prove that the torus partition function of the $\rT\overline{\rT}$ deformed conformal field theory is still modular invariant. Then we show that by requiring modular invariance and that the deformed spectrum only depends on the undeformed energy and momentum of the same state, we can fix the deformation uniquely to be that of the $\rT\overline{\rT}$ deformed CFTs. Finally we discuss some non-perturbative features of the deformed theory and new features of the $\rJ\overline{\rT}$ deformed CFTs.\par

In section~\ref{lecture3}, we first discuss Cardy's random geometry picture for $\rT\overline{\rT}$ deformation. Then we discuss the 2d topological gravity point of view for the $\rT\overline{\rT}$ deformation. Finally we discuss the holographic dual for the bad sign of the deformation parameter.

The topics discussed in the main part of this review are obviously limited. Moreover, $\rT\overline{\rT}$ is a quickly developing field, even for the topics we covered, new insights have been obtained in the past few years. To balance these points, we review briefly other topics and developments of $\rT\overline{\rT}$ deformation until January 2021 in section~\ref{newDevp}.


\section{Definition and deformed lagrangian}
\label{sec:def}
In this and the next section, we give the definition of the $\rT\overline{\rT}$ deformation of 2d quantum field theories in the Lagrangian formulation. In the current section, we discuss how to obtain the deformed classical Lagrangian for the free massless boson and show that the deformed Lagrangian is basically the Naumbu-Goto action in the static gauge. We then discuss the deformed spectrum on an infinite cylinder in section~\ref{sec:def-spectrum}. It turns out that the deformed spectrum can be obtained exactly and non-perturbatively. Our discussions mainly follow the original works \cite{Zamolodchikov:2004ce,Smirnov:2016lqw,Cavaglia:2016oda}. In section~\ref{sec:fac}, we give an alternative proof of the factorization formula without using any specific coordinate system. This method is generalizable to homogeneous symmetric spacetime with non-zero curvatures. We see from this analysis that flat spacetime is special and have particularly nice properties.
In this section, we define the $\rT\bar{\rT}$ deformation of 2d quantum field theory in the Lagrangian formulation and consider a simple example which is the free massless boson in detail.

\subsection{Definition}
\begin{figure}[h!]
\begin{center}
\includegraphics[scale=0.3]{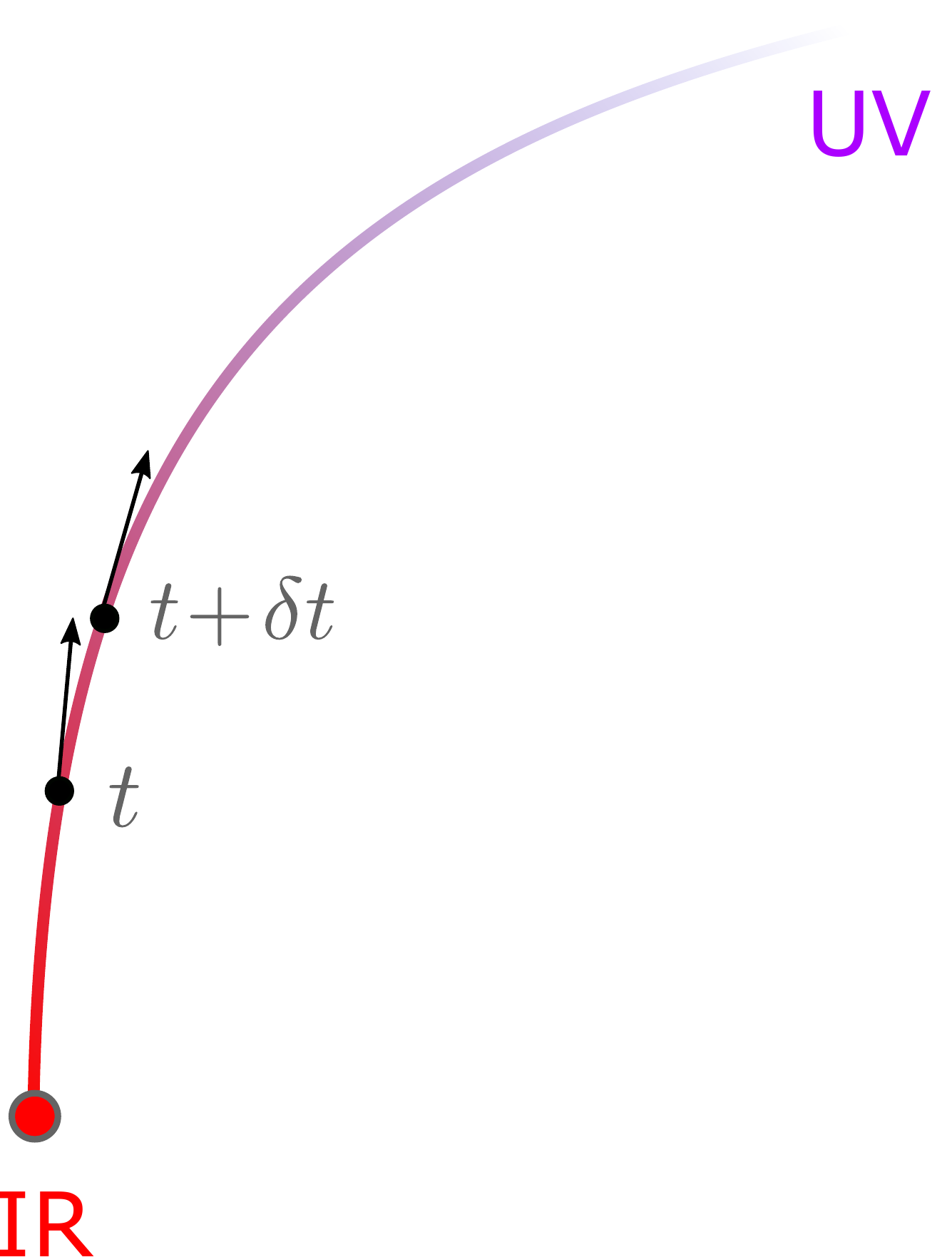}
\caption{The $\rT\overline{\rT}$ flow in the space of 2d quantum field theories. The point $t=0$ corresponds to the undeformed IR theory. As we increase $t$, we go from IR to UV.}
\label{fig:flow}
\end{center}
\end{figure}

Abstractly a 2d quantum field theory can be defined by a Lagrangian $\mathcal{L}$\footnote{For theories that cannot be described by a Lagrangian, there can be other ways to define this deformation. For example, for any CFT, the $\rT\overline{\rT}$ deformation can be described by the trace flow equation.}. We consider a trajectory in the field theory space parameterized by $t$ and denote the Lagrangian at each point of the trajectory by $\mathcal{L}^{(t)}$. The flow for theories on the trajectory is triggered by the operator ``$\det T_{\mu\nu}^{(t)}$''\footnote{We follow the sign convention of \cite{Cavaglia:2016oda}, which is different from \cite{McGough:2016lol}.}
\begin{align}
\label{eq:defL}
\mathcal{L}^{(t+\delta t)}=\mathcal{L}^{(t)}+\delta t\det T_{\mu\nu}^{(t)}=\mathcal{L}^{(t)}-\frac{\delta t}{\pi^2}\rT\overline{\rT}^{(t)}.
\end{align}
This is depicted in figure~\ref{fig:flow}.
To write down the results explicitly, let us fix some conventions. We work in two dimensional Cartesian coordinate $(x,y)$. As usual we can define the holomorphic and anti-holomorphic coordinates $z=(\rz,\bar{\rz})$ by
\begin{align}
\rz=x+\ri y,\qquad \bar{\rz}=x-\ri y.
\end{align}
The components of stress energy tensor in these two coordinates are related by
\begin{align}
\label{eq:TxyTzz}
T_{\rz\rz}=&\,\frac{1}{4}(T_{xx}-T_{yy}-2\ri T_{xy}),\\\nonumber
T_{\bar{\rz}\bar{\rz}}=&\,\frac{1}{4}(T_{xx}-T_{yy}+2\ri T_{xy}),\\\nonumber
T_{\rz\bar{\rz}}=&\,\frac{1}{4}(T_{xx}+T_{yy}).
\end{align}
Following the conventions in \cite{Zamolodchikov:2004ce,Smirnov:2016lqw,Cavaglia:2016oda}, we define
\begin{align}
\label{eq:T}
T=-2\pi T_{\rz\rz},\qquad \bar{T}=-2\pi T_{\bar{\rz}\bar{\rz}},\qquad \Theta=2\pi T_{\rz\bar{\rz}}.
\end{align}
Then the $\rT\overline{\rT}$ operator is given by
\begin{align}
\rT\overline{\rT}=-\pi^2\det T_{\mu\nu}=-\pi^2(T_{xx}T_{yy}-T_{xy}^2)=4\pi^2(T_{\rz\rz}T_{\bar{\rz}\bar{\rz}}-T_{\rz\bar{\rz}}^2)=T\bar{T}-\Theta^2.
\end{align}
where we have used the relations in (\ref{eq:TxyTzz}) and (\ref{eq:T}). We use the notation $\rT\overline{\rT}$ to denote the \emph{composite operator}. This is different from the quantity $T\bar{T}$ in general. The operator $\rT\overline{\rT}$ is of dimension [mass]$^4$. Therefore the coupling $t$ is of dimension [length]$^2$. From RG point of view, this is an irrelevant deformation.\par

Since the operator $\rT\overline{\rT}$ is a composite operator, we should ask how it is defined precisely. We will define it by point splitting and show that it is well-defined up to derivatives of other local operators. This will be discussed in more detail in section~\ref{sec:def-spectrum}.

\subsection{Deformed Lagrangian}
The definition in (\ref{eq:defL}) might seem a bit abstract, so let us work out an explicit example to get some feeling about it. We consider the simplest possible quantum field theory, namely the free massless boson and see what the deformed Lagrangian looks like. This derivation is first presented in \cite{Cavaglia:2016oda}, which we also follow here. The Lagrangian is given by
\begin{align}
\mathcal{L}_{\text{FB}}=\partial\phi\bar{\partial}\phi,
\end{align}
where we have introduced the short-hand notation for the holomorphic and anti-holomorphic derivatives
\begin{align}
\partial\equiv\partial_{\rz}=\frac{1}{2}(\partial_x-\ri\partial_y),\qquad \bar{\partial}\equiv\partial_{\bar{\rz}}=\frac{1}{2}(\partial_x+\ri\partial_y).
\end{align}
We will show that the deformed Lagrangian takes the following form
\begin{align}
\label{eq:deformedL}
\mathcal{L}_{\text{FB}}\mapsto \mathcal{L}_{\text{FB}}^{(t)}=\frac{1}{2t}\left(\sqrt{4t\,\partial\phi\bar{\partial}\phi+1}-1\right)
=-\frac{1}{2t}+\mathcal{L}_{\text{Nambu-Goto}}.
\end{align}
The square root part is exactly the Nambu-Goto action\footnote{This is part of the reason we follow the sign convention of \cite{Cavaglia:2016oda}, otherwise the NG action looks a bit strange with minus signs.} in a 3 dimensional target space
\begin{align}
\mathcal{L}_{\text{Nambu-Goto}}=\frac{1}{2t}\sqrt{\det(\partial_\alpha X\cdot \partial_\beta X)},
\end{align}
in a specific gauge called the \emph{static gauge}\footnote{The static gauge is the gauge choice that one identifies two of the target space coordinates with the worldsheet coordinates.} $X^1=x$, $X^2=y$, $X^3=\sqrt{t}\phi/2$.\par

Now we start the derivation. To simplify the notation, let us introduce
\begin{align}
\tau(z)=T(z)/\pi,\qquad \bar{\tau}(z)=\bar{T}(z)/\pi,\qquad \theta(z)=\Theta(z)/\pi.
\end{align}
Given the Lagrangian, the canonical stress-energy tensor is given by
\begin{align}
T^{\mu\nu}=\frac{\partial\mathcal{L}}{\partial(\partial_\mu\phi)}\partial^\nu\phi-\eta^{\mu\nu}\mathcal{L}.
\end{align}
Written in terms of components,
\begin{align}
\label{eq:Ltotau}
\tau=-\frac{\partial\mathcal{L}}{\partial(\bar{\partial}\phi)}\partial\phi,\quad
\bar{\tau}=-\frac{\partial\mathcal{L}}{\partial(\partial\phi)}\bar{\partial}\phi,\quad
\theta=\frac{1}{2}\left(\frac{\partial\mathcal{L}}{\partial(\partial\phi)}\partial\phi
+\frac{\partial\mathcal{L}}{\partial(\bar{\partial}\phi)}\bar{\partial}\phi
-2\mathcal{L} \right)
\end{align}
The key equation which allows us to find the deformed Lagrangian is the following
\begin{align}
\label{eq:Lttb}
\boxed{\partial_t\mathcal{L}^{(t)}=-\rT\overline{\rT}^{(t)}/\pi^2=-\tau\bar{\tau}+\theta^2.}
\end{align}
This equation comes directly from the definition of $\rT\overline{\rT}$-deformation. Since the left hand side of (\ref{eq:Lttb}) involves a derivative in $t$ while the right hand side does not, we can perform a perturbative expansion in $t$ and set up a recursion procedure to find the coefficients of the $t$-expansion order by order. To this end, let us write
\begin{align}
\label{eq:sumLj}
\mathcal{L}^{(t)}=\sum_{j=0}^{\infty} t^j\,L_j.
\end{align}
Using this ansatz, at the $(j+1)$-th order of $t$ (\ref{eq:Lttb}) takes the form
\begin{align}
\label{eq:Ljtautaub}
L_{j+1}=-\frac{1}{j+1}\sum_{k=0}^j\left(\tau^{(k)}\bar{\tau}^{(j-k)}-\theta^{(k)}\theta^{(j-k)}\right).
\end{align}
The main point here is that the right hand side only involves $\tau^{(k)},\bar{\tau}^{(k)}$ and $\theta^{(k)}$ with $k\le j$. These are determined by $L_k$ as follows
\begin{align}
\label{eq:tautauL}
\tau^{(k)}=-\frac{\partial {L}_k}{\partial(\bar{\partial}\phi)}\partial\phi,\quad
\bar{\tau}^{(k)}=-\frac{\partial {L}_k}{\partial(\partial\phi)}\bar{\partial}\phi,\quad
\theta^{(k)}=\frac{1}{2}\left(\frac{\partial {L}_k}{\partial(\partial\phi)}\partial\phi
+\frac{\partial {L}_k}{\partial(\bar{\partial}\phi)}\bar{\partial}\phi
-2{L}_k \right).
\end{align}
The initial conditions for the recursion relation is
\begin{align}
\tau^{(0)}=-(\partial\phi)^2,\qquad \bar{\tau}^{(0)}=-(\bar{\partial}\phi)^2,\qquad \theta^{(0)}=0.
\end{align}
From these, we find the first coefficient $L_1$ from (\ref{eq:Ljtautaub})
\begin{align}
L_1=-\left(\tau^{(0)}\bar{\tau}^{(0)}-\theta^{(0)2}\right)=-(\partial\phi)^2(\bar{\partial}\phi)^2
\end{align}
From $L_1$, we can work out $\tau^{(1)}$, $\bar{\tau}^{(1)}$ and $\theta^{(1)}$ using (\ref{eq:tautauL})
\begin{align}
\tau^{(1)}=2(\partial\phi)^3(\bar{\partial}\phi),\qquad\bar{\tau}^{(1)}=2(\partial\phi)(\bar{\partial}\phi)^3,\qquad
\theta^{(1)}=-(\partial\phi)^2(\bar{\partial}\phi)^2.
\end{align}
Then we can use again (\ref{eq:Ljtautaub}) to construct $L_2$
\begin{align}
L_2=&\,-\frac{1}{2}\left[\tau^{(0)}\bar{\tau}^{(1)}-\theta^{(0)}\theta^{(1)}+\tau^{(1)}\tau^{(0)}-\theta^{(1)}\theta^{(0)} \right]\\\nonumber
=&\,-\frac{1}{2}\left[-2(\partial\phi)^3(\bar{\partial}\phi)^3-2(\partial\phi)^3(\bar{\partial}\phi)^3\right]\\\nonumber
=&\,2(\partial\phi)^3(\bar{\partial}\phi)^3.
\end{align}
Having $L_2$ at hand, we can derive $\tau^{(2)}$, $\bar{\tau}^{(2)}$ and $\theta^{(2)}$ using (\ref{eq:tautauL}) again. Then we can obtain $L_3$, and so on. Working out the first few orders, one can find a general pattern of $L_j$\footnote{This can be done for example using the `\texttt{FindSequenceFunction}' of \texttt{Mathematica}.}
\begin{align}
\label{eq:explicitLj}
L_j=(-1)^j4^{j}\frac{(1/2)_{j}}{(2)_{j}}\left(L_0\right)^{j+1},\qquad L_0=\mathcal{L}_{\text{FB}}.
\end{align}
where $(a)_n$ is the Pochhammer symbol
\begin{align}
(a)_n=\prod_{k=0}^{n-1}(a-k).
\end{align}
Plugging (\ref{eq:explicitLj}) to (\ref{eq:sumLj}), one finds the deformed Lagrangian (\ref{eq:deformedL}).\par

Alternatively, if it is hard to guess the closed form formula, one can proceed as follows. From the first few orders, we already gain some idea about what the Lagrangian looks like. Therefore we can make the following ansatz
\begin{align}
\mathcal{L}^{(t)}=\frac{1}{t}F(t\,\partial\phi\bar{\partial}\phi).
\end{align}
Using this ansatz, we can write the defining equation (\ref{eq:Lttb}) as a differential equation of $F(x)$ where $x=t\,\partial\phi\bar{\partial}\phi$. In more detail, we have
\begin{align}
\tau=&\,-\frac{\partial\mathcal{L}^{(t)}}{\partial(\bar{\partial}\phi)}\partial\phi=-F'(x)\,(\partial\phi)^2,\\\nonumber
\bar{\tau}=&\,-\frac{\partial\mathcal{L}^{(t)}}{\partial(\partial\phi)}\bar{\partial}\phi=-F'(x)\,(\bar{\partial}\phi)^2,\\\nonumber
\theta=&\,\frac{1}{2}\left[2F'(x)\,(\partial\phi\bar{\partial}\phi)-2\frac{F(x)}{t} \right]=\frac{xF'(x)}{t}-\frac{F(x)}{t},\\\nonumber
\partial_t\mathcal{L}^{(t)}=&\,-\frac{F(x)}{t^2}+\frac{xF'(x)}{t^2}.
\end{align}
Plugging into (\ref{eq:Lttb}), we obtain the following differential equation for $F(x)$
\begin{align}
F^2-2xF'F-xF'+F=0
\end{align}
with the initial condition $\mathcal{L}^{(t=0)}=\partial\phi\bar{\partial}\phi$, we can easily solve this equation and find
\begin{align}
F(x)=\frac{1}{2}(\sqrt{1+4x}-1).
\end{align}
This is exactly the Lagrangian we promised in (\ref{eq:deformedL}).

Several comments are in order for the deformed Lagrangian. First of all, it is clear that similar strategy can be applied for more general Lagrangians, such as more fields and non-trivial potentials. This kind of computation has been performed in \cite{Cavaglia:2016oda,Bonelli:2018kik,Kraus:2018xrn} and in general they lead to \emph{non-local} Lagrangians.\par

It is well-known that the Nambu-Goto Lagrangian describes the propagation of a free bosonic string. We also see that $1/t$ plays the role of string tension in the Lagrangian (\ref{eq:deformedL}). Intuitively, when $t=0$ we have a string with infinite tension which reduces to a point particle whose motion is described by a local Lagrangian. At finite $t$, we have some finite string tension and the Lagrangian describes some extended object. This is another hint of the non-local feature of the $\rT\overline{\rT}$-deformed QFTs.\par

The classical Lagrangian gives some first taste of the deformed theory, but the quantization of such kind of Lagrangians are obviously hard. Even for the Nambu-Goto action, this is a non-trivial problem in the context of effective string theory (see for example \cite{Aharony:2013ipa} and references therein). In fact, it is not clear at the moment how can we make use of these Lagrangian in practice to compute some observable such as the spectrum or correlation functions. Conformal perturbation theory is not well-developed so far (see \cite{Kraus:2018xrn,Chen:2018eqk} for the first order calculation in conformal perturbation theory for certain correlation functions). The main problem is that as usual conformal perturbation theory leads to divergent integrals and a good prescription to regularize such integrals has not been fully worked out yet.\par

One final important remark is that despite of the difficulties we mentioned above, the deformed theory is solvable in the sense that if we know the undeformed spectrum, we can find the deformed spectrum explicitly. This will be the focus of section~\ref{sec:def-spectrum}.

\section{Deformed spectrum}
\label{sec:def-spectrum}
In this section, we derive the deformed spectrum of the $\rT\overline{\rT}$ deformed theories. Consider a QFT on an infinite cylinder with radius $R$. We denote the eigenstates of the Hamiltonian by $|n\rangle$. The starting point for deriving the deformed spectrum is Zamolodchikov's factorization formula
\begin{align}
\label{eq:Zamo}
\langle n|\rT\overline{\rT}|n\rangle=\langle n|T|n\rangle\langle n|\bar{T}|n\rangle-\langle n|\Theta|n\rangle\langle n|\Theta|n\rangle.
\end{align}
We want to address two issues in this section. Firstly, we need to explain more carefully how the $\rT\overline{\rT}$ operator is defined by point splitting as we promised in the previous section. Secondly, we need to prove the key equation (\ref{eq:Zamo}).

\subsection{The $\rT\overline{\rT}$ operator}
Let us start by defining the $\rT\overline{\rT}$ operator. In the $(\rz,\bar{\rz})$ coordinate, the conservation of the stress-energy tensor $\partial_{\mu}T^{\mu\nu}=0$ can be written as
\begin{align}
\label{eq:conserve}
\bar{\partial}T(z)=\partial\Theta(z),\qquad \partial\bar{T}(z)=\bar{\partial}\Theta(z).
\end{align}
We define the operator by point splitting. Consider the following limit
\begin{align}
\label{eq:limitTTb}
\lim_{z\to w}\left(T(z)\bar{T}(w)-\Theta(z)\Theta(w)\right).
\end{align}
In general, taking this limit leads to divergences. We will prove that this operator is in fact well-defined up to derivatives. Why should one expect something special will happen for the specific combination (\ref{eq:limitTTb})? To this end, we can consider the undeformed CFT. The trace of the stress-energy tensor vanishes for CFTs. The two-point function of the stress-energy tensor in $d$ dimensions is given by
\begin{align}
\langle T_{ij}(x)T_{kl}(0)\rangle=\frac{1}{x^{2d}}\left(\frac{1}{2}(I_{ik}I_{jl}+I_{il}I_{jk})-\frac{1}{d}\delta_{ij}\delta_{kl} \right),
\end{align}
where
\begin{align}
I_{ij}=\delta_{ij}-\frac{2x_ix_j}{x^2}.
\end{align}
The specific combination in (\ref{eq:limitTTb}) is given by
\begin{align}
\langle T_{ij}(x)T^{ij}(0)\rangle=\frac{d-2}{x^{2d}}.
\end{align}
We see that when $d=2$, the limit $x\to 0$ is well-defined. This is of course something we know already. In 2d CFT, the holomorphic and anti-holomorphic part of stress-energy tensor do not talk to each other and hence their OPE is regular. On the other hand, it also hints that at higher dimensions, things become more complicated, even for CFTs.

We will show that the limit in (\ref{eq:limitTTb}) is well-defined beyond CFT and along the whole $\rT\overline{\rT}$ deformed trajectory where the trace $\Theta$ is no longer zero. The derivation below follows closely Zamolodchikov's original paper \cite{Zamolodchikov:2004ce}. Let us take the derivative $\partial_{\bar{\rz}}$ of (\ref{eq:limitTTb})
\begin{align}
&\,\partial_{\bar{\rz}}\left(T(z)\bar{T}(w)-\Theta(z)\Theta(w)\right)
=\partial_{\bar{\rz}}T(z)\bar{T}(w)-\partial_{\bar{\rz}}\Theta(z)\Theta(w)\\\nonumber
=&\,\partial_{\rz}\Theta(z)\bar{T}(w)-\partial_{\bar{\rz}}\Theta(z)\Theta(w)\\\nonumber
=&\,\partial_{\rz}\Theta(z)\bar{T}(w)-\partial_{\bar{\rz}}\Theta(z)\Theta(w)
+\left(\Theta(z)\partial_{\rw}\bar{T}(w)-\Theta(z)\partial_{\bar{\rw}}\Theta(w)\right),
\end{align}
where in the second line we have used the conservation law (\ref{eq:conserve}). Also by the conservation law, the term we add in the bracket in the third line is in fact zero. The last line can be rewritten as
\begin{align}
\label{eq:holopartial}
\partial_{\bar{\rz}}\left(T(z)\bar{T}(w)-\Theta(z)\Theta(w)\right)
=(\partial_{\rz}+\partial_{\rw})\left[\Theta(z)\bar{T}(w)\right]
-(\partial_{\bar{\rz}}+\partial_{\bar{\rw}})\left[\Theta(z)\Theta(w) \right].
\end{align}
Similarly, we can find that
\begin{align}
\label{eq:aholopartial}
\partial_{{\rz}}\left(T(z)\bar{T}(w)-\Theta(z)\Theta(w)\right)
=(\partial_{\rz}+\partial_{\rw})\left[T(z)\bar{T}(w)\right]
-(\partial_{\bar{\rz}}+\partial_{\bar{\rw}})\left[T(z)\Theta(w) \right].
\end{align}
The reason of writing the results in the right hand side of (\ref{eq:holopartial}) and (\ref{eq:aholopartial}) can be seen from OPE. For example, consider the OPE of $\Theta(z)\overline{T}(w)$
\begin{align}
\Theta(z)\overline{T}(w)=\sum_i c^i(z-w)\mathcal{O}_i(w),
\end{align}
where the sum is over all operators in the spectrum. The form $c^i(z-w)$ is due to translational invariance. It is clear that $(\partial_{\rz}+\partial_{\rw})$ and $(\partial_{\bar{\rz}}+\partial_{\bar{\rw}})$ annihilate the coefficients $c^i(z-w)$, so it only acts on the operators. The conclusion is that
\begin{align}
\label{eq:ppTT}
\partial_{\rz}\left(T(z)\bar{T}(w)-\Theta(z)\Theta(w)\right)
=&\sum_i A^i(z-w)\partial_{\rw}\mathcal{O}_i(w)+\sum_i B^i(z-w)\partial_{\bar{\rw}}\mathcal{O}_i(w),\\\nonumber
\partial_{\bar{\rz}}\left(T(z)\overline{T}(w)-\Theta(z)\Theta(w)\right)
=&\sum_i C^i(z-w)\partial_{\rw}\mathcal{O}_i(w)+\sum_i D^i(z-w)\partial_{\bar{\rw}}\mathcal{O}_i(w).
\end{align}
The exact form of $A^i,\cdots,D^i$ are not important. The important point is that the operators that appear on the right hand side of (\ref{eq:ppTT}) \emph{all} takes the form of \emph{derivative of some operators}. This gives a constraint for the form of OPE. Let us write the OPE as
\begin{align}
T(z)\bar{T}(w)-\Theta(z)\Theta(w)=\sum_i\mathcal{C}_i(z-w)\mathcal{O}_i(w).
\end{align}
From the previous analysis, we can see that if an operator $\mathcal{O}_k$ is not the derivative of another operator, then the corresponding $\mathcal{C}_i(z-w)$ must be a constant, \emph{i.e.} does not depend on $z$ or $w$. We prove this by contradiction. Suppose we have the term $\mathcal{C}_n(z-w)\mathcal{O}_n(w)$ in the OPE where $\mathcal{O}_n$ is not the derivative of another operator
\begin{align}
T(z)\bar{T}(w)-\Theta(z)\Theta(w)=\mathcal{C}_n(z-w)\mathcal{O}_n(w)+\cdots
\end{align}
Taking derivative on both sides
\begin{align}
\partial_{\rz}\left(T(z)\bar{T}(w)-\Theta(z)\Theta(w) \right)=\partial_{\rz}\mathcal{C}_n(z-w)\mathcal{O}_n(w)
+\mathcal{C}_n(z-w)\partial_{\rz}\mathcal{O}_n(w)+\cdots
\end{align}
Note that the first term is not compatible with the result (\ref{eq:ppTT}) unless $\partial_{\rz}\mathcal{C}_n(z-w)$ is zero, namely $\mathcal{C}_n(z-w)$ is a constant. If the operator $\mathcal{O}_n$ is already a derivative of some other operators, then the above form is compatible with the result (\ref{eq:ppTT}) so we do not get additional constraints. Absorbing the constant coefficient into the definition of the local operator, we conclude that
\begin{align}
\label{eq:defTT}
\lim_{z\to w}\left(T(z)\bar{T}(w)-\Theta(z)\Theta(w)\right)=\rT\overline{\rT}(w)+\texttt{derivatives}.
\end{align}
We see that the operator itself is defined up to total derivatives. However, when putting into the expectation value, the derivative terms vanish. This is because the expectation value is constant in a translational invariant theory. Taking derivatives leads to zero.\par

\subsection{Factorization formula}
Now we are ready to prove the main formula (\ref{eq:Zamo}). As a first step, we want to prove that the following quantity
\begin{align}
\label{eq:CC1}
\mathsf{C}(z,w)=\langle T(z)\bar{T}(w)\rangle-\langle\Theta(z)\Theta(w)\rangle,
\end{align}
which looks like some two-point function is actually a constant. This can be proven as follows. Taking the derivative $\partial_{\bar{\rz}}$ of $\mathsf{C}(z,w)$,
\begin{align}
\partial_{\bar{\rz}}\mathsf{C}(z,w)=&\,\langle\partial_{\bar{\rz}}T(z)\bar{T}(w)\rangle-\langle\partial_{\bar{\rz}}\Theta(z)\Theta(w)\rangle\\\nonumber
=&\,\langle\partial_{\rz}\Theta(z)\bar{T}(w)\rangle+\langle\Theta(z)\partial_{\bar{\rw}}\Theta(w)\rangle\\\nonumber
=&\,-\langle\Theta(z)\partial_{\rw}\bar{T}(w)\rangle+\langle\Theta(z)\partial_{\bar{\rw}}\Theta(w)\rangle=0,
\end{align}
where we have used the conservation laws and the translational invariance to move the derivative from one operator to the other at the price of a minus sign. Similarly we can prove that $\partial_{\rz}\mathsf{C}(z,w)=0$. Hence $\mathsf{C}$ is a constant.\par

Now that $\mathsf{C}$ is a constant, we can write it in two different ways. Firstly, since we are on an infinite cylinder, we can take the two points infinitely separated from each other. In this limit, by clustering decomposition theorem, the two-point functions can be factorized and hence we find
\begin{align}
\label{eq:Clhs}
\mathsf{C}=\lim_{|z-w|\to\infty}\mathsf{C}(z,w)=\langle T\rangle\langle\overline{T}\rangle-\langle\Theta\rangle\langle\Theta\rangle.
\end{align}
On the other hand, we can take the coinciding limit and keeping in mind that the derivative ambiguities vanish within the expectation value, we find that
\begin{align}
\label{eq:Crhs}
\mathsf{C}=\lim_{z\to w}\mathsf{C}(z,w)=\langle\rT\overline{\rT}\rangle.
\end{align}
Combining (\ref{eq:Clhs}) and (\ref{eq:Crhs}), we proved the factorization formula (\ref{eq:Zamo}) for the vacuum state $|n\rangle=|0\rangle$. To complete the proof, we need to show that it holds for general excited state. Let us again define
\begin{align}
\label{eq:defCn}
\mathsf{C}_n(z,w)=\langle n|T(z)\bar{T}(w)|n\rangle-\langle n|\Theta(z)\Theta(w)|n\rangle.
\end{align}
We can show that $\mathsf{C}_n$ is a constant as before. The difference between the vacuum state and excited states is that the asymptotic factorization property (\ref{eq:Clhs}) no longer holds. There's a possibility to pick up contributions from intermediate states. This can be seen from the spectral expansion
\begin{align}
\label{eq:excitedExp}
\langle n|T(z)\bar{T}(z')|n\rangle=\sum_{m}\langle n|T(z)|m\rangle\langle m|\bar{T}(z')|n\rangle\times \re^{(E_n-E_m)|y-y'|+\ri(P_n-P_m)|x-x'|}
\end{align}
and similar spectral expansion for $\langle n|\Theta(z)\Theta(z')|n\rangle$. Note that in (\ref{eq:excitedExp}) we have written the exponential factors in Cartesian coordinates $z=(x,y)$. Now it is clear that for $\mathsf{C}_n$ to be \emph{independent of the coordinates}, all the terms with non-diagonal matrix elements should cancel between the two terms in (\ref{eq:defCn}). We are therefore left with diagonal matrix elements. Taking the limit $z\to z'$, we obtain
\begin{align}
\langle n|\rT\overline{\rT}|n\rangle=\langle n|T|n\rangle\langle n|\bar{T}|n\rangle-\langle n|\Theta|n\rangle\langle n|\Theta|n\rangle.
\end{align}
which is the key result we are after.

\subsection{Burgers' equation and deformed spectrum}
The proof of the factorization formula only depends on Lorentz invariance and conservation of the stress energy tensor. Since both are still valid along the $\rT\overline{\rT}$ flow, we expect the factorization formula holds for the deformed theory at generic point $t$. Then we can use the factorization formula to find the spectrum of the deformed theory. This is done by translating factorization formula into a differential equation of the spectrum. For QFT on a cylinder of circumference $R$, expectation values of components of the stress-energy tensor are related to the spectrum as
\begin{align}
\label{eq:ETelement}
\mathcal{E}_n(R,t)=-R\langle n|T_{yy}|n\rangle,\quad \partial_R\mathcal{E}_n(R,t)=-\langle n|T_{xx}|n\rangle,\quad P_n=-\ri R\langle n|T_{xy}|n\rangle.
\end{align}
In this way, we can write the right hand side of the factorization formula (\ref{eq:Zamo}) in terms of $\mathcal{E}_n$ and $P_n$. On the other hand, by the definition of $\rT\overline{\rT}$ deformation, we have
\begin{align}
\partial_t\mathcal{E}_n(R,t)=-{R}\langle n|\det(T_{\mu\nu})|n\rangle
\end{align}
on the left hand side of the factorization formula. Combining these results, we find
\begin{align}
\label{eq:Burgers}
\boxed{\partial_t\mathcal{E}_n(R,t)=\mathcal{E}_n(R,t)\partial_R\mathcal{E}_n(R,t)+\frac{1}{R}P_n(R)^2.}
\end{align}
This is the inviscid Burgers' equation in one dimension. This equation is well-known in fluid mechanics which describes shock waves and can be solved by method of characteristics. To solve the equation, we need the initial condition $\mathcal{E}_n(R,0)=E_n(R)$. This is particularly simple for CFTs since we have
\begin{align}
\label{eq:initialcondE}
E_n(R)=\frac{1}{R}\left(n+\bar{n}-\frac{c}{12}\right),\qquad P_n(R)=\frac{1}{R}(n-\bar{n})
\end{align}
where $n$ and $\bar{n}$ are the eigenvalues of the Virasoro generator $L_0$ and $\bar{L}_0$.
With the initial condition (\ref{eq:initialcondE}), the solution of Burgers' equation is given by
\begin{align}
\mathcal{E}_n(R,t)=\frac{R}{2t}\left(\sqrt{1+\frac{4t\,E_n}{R}+\frac{4t^2\,P_n^2}{R^2}}-1\right).
\end{align}
The momentum $P_n$ are quantized and remain undeformed.
Several comments are in order for the deformed spectrum. Firstly, if we do not impose the initial condition $\mathcal{E}_n(R,0)=E_n$, we can find two solutions for the Burgers' equation corresponding to the two branches of square root
\begin{align}
\mathcal{E}^{(\pm)}_n(R,t)=\frac{R}{2t}\left(\pm\sqrt{1+\frac{4t\,E_n}{R}+\frac{4t^2\,P_n^2}{R^2}}-1\right).
\end{align}
The solution $\mathcal{E}_n^{(-)}$ does not have a well-defined conformal limit since in the limit $t\to 0$ it diverges. In order to satisfy the initial condition, we take $\mathcal{E}_n^{(+)}$ as our solution. However, the solution $\mathcal{E}_n^{(-)}$ is not completely useless. It is actually very helpful to find the non-perturbative solution of the flow equation which will be discussed in section~\ref{lecture2}.\par

The equation (\ref{eq:Burgers}) is a non-linear partial differential equation. In fact it can be brought to a simpler form and written as an ordinary differential equation. Let us introduce the \emph{dimensionless} quantities
\begin{align}
\mathbb{E}_n(\lambda)=R\mathcal{E}_n(R,t),\qquad \mathbb{P}_n=R P_n\,.
\end{align}
We define the dimensionless coupling constant
\begin{align}
\lambda=\frac{\alpha t}{R^2}
\end{align}
where $\alpha$ is some multiplicative constant such as $1/2\pi$ which we keep arbitrary for later convenience. Using this change of variable, the Burgers' equation can be rewritten as an ordinary differential equation (ODE)
\begin{align}
\label{eq:ODESpec}
\alpha\mathbb{E}_n'+2\lambda\mathbb{E}_n\mathbb{E}_n'+(\mathbb{E}_n^2-\mathbb{P}_n^2)=0,
\end{align}
where $\mathbb{E}_n'=\partial_{\lambda}\mathbb{E}_n(\lambda)$. This ODE is useful for an alternative derivation of the flow equation for the torus partition function.

The solution of the Burgers' equation or ODE takes the form of a square root, therefore $\mathbb{E}_n(\lambda)$ can be taken as the solution of the following quadratic algebraic equation
\begin{align}
\lambda \mathbb{E}_n(\lambda)^2+{\alpha}\mathbb{E}_n(\lambda)-\left(\lambda\mathbb{P}_n^2+{\alpha}\mathbb{E}_n(0)\right)=0.
\end{align}
Or equivalently,
\begin{align}
\alpha\mathbb{E}_n(\lambda)+\lambda(\mathbb{E}_n(\lambda)^2-\mathbb{P}_n^2)=\alpha\mathbb{E}_n(0),
\end{align}
which states that the combination $\alpha\mathbb{E}_n(\lambda)+\lambda(\mathbb{E}_n(\lambda)^2-\mathbb{P}_n^2)$ is a constant along the flow. The right hand side is simply the value of this quantity at $\lambda=0$. This quadratic equation actually has physical meaning in the string theory side. It corresponds to the mass-shell relation. The deformed spectrum for single trace deformation of string theory \cite{Giveon:2017nie,Giveon:2019fgr} is obtained from this algebraic equation instead of a differential equation.

Taking derivative with respect to $\lambda$ on both sides of the quadratic equation, we find the ODE (\ref{eq:ODESpec}) again. Writing the ODE in terms of dimensionful quantities and translate the derivative of $\mathbb{E}_n'$ and $\mathbb{E}_n'\mathbb{E}_n$ as derivatives with respect to $t$ and $R$ respectively, we recover the Burgers' equation.

\subsection{A covariant proof}
\label{sec:fac}
The proof of factorization formula in the previous section uses the Cartesian coordinate system. Physical results should not depend on the choice of coordinate systems. We thus look for an alternative derivation of the factorization which does not depend on any specific choice of coordinate systems. To this end, it turns out to be helpful to be slightly more general and consider spacetimes with maximal symmetry, for example sphere and hyperbolic space. The idea is to exploit the symmetry of spacetime and write the two-point function of the stress energy tensor in a convenient form. We follow the discussions in \cite{Jiang:2019tcq}.

\subsubsection*{An invariant biscalar}
Recall that the central step towards the factorization formula is to prove that the quantity $\mathsf{C}(x,y)$ defined in (\ref{eq:CC1}) is a constant. This quantity can be written in the Cartesian coordinate as
\begin{align}
\label{eq:coord}
\mathsf{C}(x,y)\propto\epsilon^{ik}\epsilon^{jl}\langle T_{ij}(x)T_{kl}(y)\rangle=\langle T^{ij}(x)T_{ij}(y)\rangle-\langle T^i_i(x)T^j_j(y)\rangle,
\end{align}
where we have used the identity $\epsilon^{ik}\epsilon^{jl}=\delta^{il}\delta^{kj}-\delta^{ij}\delta^{kl}$. An alternative proof that $\mathsf{C}(x,y)$ is a constant is given in \cite{Cardy:2018sdv}. Both proofs are done in the Cartesian coordinate. This fact should not depend on which coordinate system we are using. So we should first define $\mathsf{C}(x,y)$ in a way which is invariant under change of coordinates.\par

Below we use Latin letters $i,j,...$ to denote indices of Cartesian system and Greek letters $\mu,\nu,...$ for those of general coordinate system. It is tempting to simply replace $i,j$ by $\mu,\nu$ in (\ref{eq:coord}) and take it as a definition of $\mathsf{C}(x,y)$ in general coordinate. However, this naive replacement does not work because we are contracting indices at different spacetime points. The resulting quantity is not invariant under coordinate transformation. Since the factors at different points do not cancel each other. We need some kind of ``connection'' to make the two points talk to each other. In order to gain some idea, let us consider a similar situation in gauge theory. Suppose we want to make a gauge invariant bilinear in terms of two fermions $\bar{\psi}(x)$ and $\psi(y)$ which transform as
\begin{align}
\bar{\psi}(x)\mapsto \re^{-\ri\alpha(x)}\bar{\psi}(x),\qquad \psi(y)\mapsto \re^{\ri\alpha(y)}\psi(y)
\end{align}
under gauge transformation. Simply taking $\bar{\psi}(x)\psi(y)$ does not work because the two phase factors do not cancel. To make a gauge invariant quantity, we need a Wilson line to connect the two spacetime points. The following quantity
\begin{align}
\bar{\psi}(x)W(x,y)\psi(y)
\end{align}
is gauge invariant. Here $W(x,y)$ is the Wilson line
\begin{align}
W(x,y)=\mathcal{P}\,\exp\left(\ri\int_{\gamma}A_{\mu}(z)\rd z^{\mu} \right)
\end{align}
where $\mathcal{P}$ denotes path ordering and $\gamma$ is a path between the two spacetime points.

To define an invariant quantity $\mathsf{C}(x,y)$, we need a quantity similar to a Wilson line. In fact, such a quantity also exist in gravity theory and is called the \emph{parallel propagator}.
\paragraph{Parallel propagator}
We give a brief introduction to the parallel propagator following \cite{Carroll:1997ar}. Consider a path $\gamma:x^{\mu}(\lambda)$ parameterized by an affine parameter $\lambda$. The equation of parallel transport along this path is given by
\begin{align}
\label{eq:parallel-transport}
\frac{\rd x^\mu}{\rd\lambda}\nabla_{\mu}V^\nu=\frac{\rd x^\mu}{\rd\lambda}\partial_{\mu}V^\nu+\frac{\rd x^\mu}{\rd\lambda}\Gamma^{\nu}_{\rho\sigma}V^{\sigma}=0.
\end{align}
Solving the parallel transportation equation for the vector $V^{\mu}$ amounts to finding a matrix $P^{\mu}_{\phantom{a}\rho}(\lambda,\lambda_0)$ such that it relates the values of the vector $V^{\mu}$ at two different positions $\lambda_0$ and $\lambda$
\begin{align}
\label{eq:substituteP}
V^{\mu}(\lambda)=P^{\mu}_{\phantom{a}\rho}V^{\rho}(\lambda_0).
\end{align}
This matrix $P^{\mu}_{\phantom{a}\rho}(\lambda,\lambda_0)$ is called the \emph{parallel propagator}. It depends on the choice of the path $\gamma$, like the Wilson loop. It can also be expressed as path ordered exponential of connections, like the Wilson loop. In this sense, the parallel propagator is a gravity analog of the Wilson loop in gauge theories. We now give a formal expression of the parallel propagator. Let us define the quantity
\begin{align}
A^\mu_{\phantom{a}\rho}(\lambda)=-\Gamma^{\mu}_{\sigma\rho}\frac{\rd x^{\sigma}}{\rd\lambda},
\end{align}
then the parallel transport equation (\ref{eq:parallel-transport}) can be written as
\begin{align}
\label{eq:VA}
\frac{\rd}{\rd\lambda}V^{\mu}=A^{\mu}_{\phantom{a}\rho}V^{\rho}.
\end{align}
Substituting (\ref{eq:substituteP}) into (\ref{eq:VA}), we obtain
\begin{align}
\frac{\rd}{\rd\lambda}P^{\mu}_{\phantom{a}\rho}(\lambda,\lambda_0)=A^{\mu}_{\phantom{a}\sigma}(\lambda)P^{\sigma}_{\phantom{a}\rho}(\lambda,\lambda_0).
\end{align}
A formal solution of this equation is given by the following path-ordered exponential
\begin{align}
P^{\mu}_{\phantom{a}\nu}(\lambda,\lambda_0)=\mathcal{P}\exp\left(-\int_{\gamma}\Gamma^{\mu}_{\sigma\nu}\frac{\rd x^{\sigma}}{\rd\eta}\rd\eta \right).
\end{align}
The parallel propagator transforms as a bi-vector which connects two spacetime points. For more detailed introduction, we refer \cite{Carroll:1997ar} and also \cite{Allen:1985wd,Osborn:1999az}.

The definition of the parallel propagator depends on the choice of the path. For our purpose, we choose the path between two points $x$ and $y$ to be the \emph{geodesic}. We denote such parallel propagator as $I^{\mu}_{\phantom{a}\alpha'}(x,y)$. Using parallel propagators, we propose that the invariant biscalar $\mathsf{C}(x,y)$ can be defined as
\begin{align}
\label{eq:defC}
\mathsf{C}(x,y)=\left[I_{\mu\alpha'}(x,y)I_{\nu\beta'}(x,y)-g_{\mu\nu}(x)g_{\alpha'\beta'}(y)\right]\langle T^{\mu\nu}(x)T^{\alpha'\beta'}(y) \rangle.
\end{align}

\subsubsection*{Maximally symmetric bi-tensors}
In a maximally symmetric spacetime, two-point function of two scalar operators $\mathcal{O}_1(x)$ and $\mathcal{O}_2(y)$ is a function of the geodesic distance between the two points $\theta(x,y)$, namely
\begin{align}
\label{eq:Ftheta}
\langle\mathcal{O}_1(x)\mathcal{O}_2(y)\rangle=F(\theta(x,y)).
\end{align}
On flat spacetime this reduces to the familiar result that $\langle\mathcal{O}_1(x)\mathcal{O}_2(y)\rangle=F(x-y)$.
(\ref{eq:Ftheta}) has a non-trivial generalization to two-point functions of symmetric tensors of arbitrary rank. The statement is that the two-point function $\langle T^{\mu\nu\cdots}(x)T^{\alpha'\beta'\cdots}(y)\rangle$ can be decomposed into different tensor structures. All the tensor structures are constructed in terms of the vectors $n_{\mu},m_{\alpha'}$, the metric $g_{\mu\nu},g_{\alpha'\beta'}$ and the parallel propagator $I_{\mu\alpha'}$. Here the vectors $n_{\mu}$ and $m_{\alpha'}$ are derivatives of the geodesic distance
\begin{align}
n_{\mu}(x,x')= \nabla_{\mu}\theta(x,x'),\qquad m_{\alpha'}(x,x')=\nabla_{\alpha'}\theta(x,x'),
\end{align}
where indices with a prime ($\nabla_{\alpha'}$) means we take derivatives with respect to the coordinate at the second position $x'$. These two vectors are normalized as $n_{\mu}n^{\mu}=m_{\alpha'}m^{\alpha'}=1$ and are related by the parallel propagator as
\begin{align}
I_{\mu}^{\phantom{a}\alpha'}(x,y)m_{\alpha'}(y)+n_{\mu}(x)=0.
\end{align}
The proof can be found in the paper \cite{Allen:1985wd,Osborn:1999az}. For the stress-energy tensor, we have the following tensor decomposition
\begin{align}
\label{eq:ansatzTT}
\langle T^{\mu\nu}(x)T^{\alpha'\beta'}(y) \rangle=&\,A_1\,n^{\mu}n^{\nu}m^{\alpha'}m^{\beta'}\\\nonumber
+&\,A_2\left(I^{\mu\alpha'}n^{\nu}m^{\beta'}+I^{\mu\beta'}n^{\nu}m^{\alpha'}+I^{\nu\alpha'}n^{\mu}m^{\beta'}+I^{\nu\beta'}n^{\mu}m^{\alpha'} \right)\\\nonumber
+&\,A_3\left(I^{\mu\alpha'}I^{\nu\beta'}+I^{\mu\beta'}I^{\nu\alpha'}\right)\\\nonumber
+&\,A_4\left(n^{\mu}n^{\nu}g^{\alpha'\beta'}+g^{\mu\nu}m^{\alpha'}m^{\beta'} \right)\\\nonumber
+&\,A_5\,g^{\mu\nu}g^{\alpha'\beta'}.
\end{align}
where $A_i(\theta)$ $(i=1,\cdots,5)$ are scalar functions that only depend on $\theta$. In what follows, we will need the covariant derivatives of the quantities $I_{\mu\alpha'}$, $n_{\mu}$ and $m_{\alpha'}$ which are \cite{Allen:1985wd}
\begin{align}
\label{eq:covD}
\nabla_{\mu}n_{\nu}=&\,\mathcal{A}(\theta)(g_{\mu\nu}-n_{\mu}n_{\nu}),\\\nonumber
\nabla_{\mu}m_{\alpha'}=&\,\mathcal{C}(\theta)(I_{\mu\alpha'}+n_{\mu}m_{\alpha'}),\\\nonumber
\nabla_{\mu}I_{\nu\alpha'}=&\,-(\mathcal{A}(\theta)+\mathcal{C}(\theta))(g_{\mu\nu}m_{\alpha'}+I_{\mu\alpha'}n_{\nu}),
\end{align}
where $\mathcal{A}(\theta)$ and $\mathcal{C}(\theta)$ are scalar functions of the geodesic distance $\theta(x,x')$. For different spacetime, they are given by
\begin{itemize}
\item Flat spacetime
\begin{align}
\label{eq:flatAC}
\mathcal{A}(\theta)=\frac{1}{\theta},\qquad \mathcal{C}(\theta)=-\frac{1}{\theta}.
\end{align}
\item Positive curvature spacetime (scalar curvature $\mathcal{R}=d(d-1)/R^2$)
\begin{align}
\mathcal{A}(\theta)=\frac{1}{R}\cot\left(\frac{\theta}{R}\right),\qquad \mathcal{C}(\theta)=-\frac{1}{R}\csc\left(\frac{\theta}{R}\right).
\end{align}
\item Negative curvature spacetime (scalar curvature $\mathcal{R}=-d(d-1)/R^2$)
\begin{align}
\mathcal{A}(\theta)=\frac{1}{R}\coth\left(\frac{\theta}{R}\right),\qquad \mathcal{C}(\theta)=-\frac{1}{R}\text{csch}\left(\frac{\theta}{R}\right).
\end{align}
\end{itemize}
Using the decomposition (\ref{eq:ansatzTT}) and the definition (\ref{eq:defC}), we can write the invariant biscalar $\mathsf{C}(x,y)$ in terms of $A_i(\theta)$
\begin{align}
\mathsf{C}(x,y)=2(1-d)A_2+d(d-1)A_3+2(1-d)A_4+d(1-d)A_5.
\end{align}
where $d$ is the dimension of the spacetime. For $d=2$, we simply have
\begin{align}
\label{eq:Cxy2}
\mathsf{C}(x,y)=-2(A_2-A_3+A_4+A_5).
\end{align}
To prove the factorization formula, we need to show that $\partial_{\mu}\mathsf{C}(x,y)=0$. To this end, we need to take into account the conservation law of the stress energy tensor.

\subsubsection*{Ward identity and factorization}
The five unfixed coefficients $A_i(\theta)$ are not independent. They are related to each other by the Ward identity
\begin{align}
\label{eq:conserveT}
\nabla_{\mu}\langle T^{\mu\nu}(x)T^{\alpha'\beta'}(y) \rangle=0.
\end{align}
We act $\nabla_{\mu}$ on the right hand side of (\ref{eq:ansatzTT}) and then make use of the relations (\ref{eq:covD}). The result can be written as
\begin{align}
\mathcal{X}\,n^{\nu}m^{\alpha'}m^{\beta'}+\mathcal{Y}(I^{\nu\alpha'}m^{\beta'}+I^{\nu\beta'}m^{\alpha'})+\mathcal{Z}\,n^{\nu}g^{\alpha'\beta'}=0.
\end{align}
Since the three tensor structures are independent, this is equivalent to three equations $\mathcal{X}=0,\mathcal{Y}=0,\mathcal{Z}=0$ where $\mathcal{X}$, $\mathcal{Y}$ and $\mathcal{Z}$ are given by
\begin{align}
\label{eq:conserved}
\mathcal{X}=&\,A'_1-2A'_2+A'_4+(d-1)\left[\mathcal{A}\,A_1-2(\mathcal{A}+\mathcal{C})\,A_2 \right]+2(\mathcal{A}-\mathcal{C})\,A_2+2\mathcal{C}\,A_4,\\\nonumber
\mathcal{Y}=&\,A'_2-A'_3+d\,\mathcal{A}\,A_2-d(\mathcal{A}+\mathcal{C})\,A_3+\mathcal{C}\,A_4,\\\nonumber
\mathcal{Z}=&\,A'_4+A'_5+(d-1)\mathcal{A}\,A_4+2\mathcal{C}\,A_2-2(\mathcal{A}+\mathcal{C})\,A_3.
\end{align}
For $d=2$, these become
\begin{align}
\label{eq:2dconserved}
\mathcal{X}=&\,A'_1-2A'_2+A'_4+\left[\mathcal{A}\,A_1-2(\mathcal{A}+\mathcal{C})\,A_2 \right]+2(\mathcal{A}-\mathcal{C})\,A_2+2\mathcal{C}\,A_4,\\\nonumber
\mathcal{Y}=&\,A'_2-A'_3+2\,\mathcal{A}\,A_2-2(\mathcal{A}+\mathcal{C})\,A_3+\mathcal{C}\,A_4,\\\nonumber
\mathcal{Z}=&\,A'_4+A'_5+\mathcal{A}\,A_4+2\mathcal{C}\,A_2-2(\mathcal{A}+\mathcal{C})\,A_3
\end{align}
Now we take the derivative of $\mathsf{C}(x,y)$ defined in (\ref{eq:Cxy2}). We have
\begin{align}
\label{eq:curveDC}
\partial_{\mu}\mathsf{C}(x,y)=-2(A'_2-A'_3+A'_4+A'_5)n_{\mu}.
\end{align}
It is interesting to notice that the equation $\mathcal{Y}+\mathcal{Z}=0$ leads to
\begin{align}
A'_2-A'_3+A'_4+A'_5+(\mathcal{A}+\mathcal{C})(2A_2-4A_3+A_4)=0,
\end{align}
which allows us to get rid of the derivatives in (\ref{eq:curveDC}) completely and we have
\begin{align}
\partial_{\mu}\mathsf{C}(x,y)=2n_{\mu}(\mathcal{A}+\mathcal{C})(2A_2-4A_3+A_4).
\end{align}
In flat spacetime, we have $\mathcal{A}+\mathcal{C}=0$ from (\ref{eq:flatAC}) so that $\partial_{\mu}\mathsf{C}(x,y)=0$. For curved spacetime, however, we have $\mathcal{A}+\mathcal{C}\ne 0$. For example,
\begin{align}
\mathcal{A}+\mathcal{C}=-\frac{1}{R}\tan\left(\frac{\theta}{2R}\right).
\end{align}
for the sphere. This shows that flat spacetime is special and the fact that $\mathsf{C}(x,y)$ is a constant depends crucially on the flatness of spacetime.

In order to prove $\partial_{\mu}\mathsf{C}(x,y)=0$ in flat spacetime, our machinery seems a bit heavy compare to the previous proofs. However, the merit of this approach is that it can be easily to generalized to curved spacetime. We see that the invariant biscalar $\mathsf{C}(x,y)$ is no longer a constant in curved spacetime. In fact, one can do better and derive an explicit formula for the expectation value of the $\rT\overline{\rT}$ operator in maximally symmetric curved spacetime. The conclusion is that the factorization formula does not apply, unless in the large-$c$ limit. There are corrections to the factorization formula at finite curvature and central charge. For more details, we refer to \cite{Jiang:2019tcq}.

\section{Modular bootstrap and uniqueness}
\label{lecture2}
In this section, we study the torus partition sum of the $\rT\overline{\rT}$ deformed conformal field theory. We are particularly interested in the modular property of the torus partition sum. More specifically, we ask the following two questions.\par

Firstly, as is well known, torus partition sum of 2d conformal field theory is modular invariant. After turning on the $\rT\overline{\rT}$ deformation, we know the deformed spectrum. A natural question is what happens to the modular property of the deformed partition sum. The answer turns out to be that the deformed partition function is still modular invariant. This is a very nice property and again shows the internal simplicity of the $\rT\overline{\rT}$ deformation. In addition, we can use this property to derive a Cardy like formula for the asymptotic density of states for $\rT\overline{\rT}$ deformed CFTs.\par

To introduce the second question, we can take a slightly more general point of view. In fact, constructing solvable deformations in the Hamiltonian formulation is easy and one can construct infinitely many such deformations. Of course, not all of these deformations are physical. To constraint the `physical' ones, we need to impose some consistency conditions. One natural condition is requiring the torus partition sum of the deformed theory to be modular invariant. This resembles the idea of bootstrap. On general grounds, we expect that the consistency conditions will lead to some constraints of the infinite many deformations and select a subset of these theories. Our second question is how constraining is the requirement of modular invariance ? Somewhat surprisingly, the constraint turns out to be so strong that it restricts the infinite family to a single theory. What's more, this theory is exactly the $\rT\overline{\rT}$ deformed conformal field theory ! We call this property \emph{uniqueness}.

\subsection{Modular invariance of torus partition sum}
In this section, we investigate the first question. Namely, what is the modular property of the $\rT\overline{\rT}$ deformed conformal field theory.
\subsubsection*{The set-up}
Consider a CFT on a torus. The partition function of the CFT is given by
\begin{align}
\label{eq:undeformedZ0}
Z_0(\tau,\bar{\tau})=\tr\left[\re^{2\pi \ri\tau(L_0-\frac{c}{24})}\re^{-2\pi \ri\bar{\tau}(\bar{L}_0-\frac{c}{24})}\right]
=\sum_n \re^{2\pi \ri\tau_1 R P_n-2\pi\tau_2 R E_n},
\end{align}
where $\tau=\tau_1+\ri\tau_2$ is the modular parameter that specifies the torus and $\bar{\tau}=\tau_1-\ri\tau_2$. The energy and momentum are given by
\begin{align}
H|n\rangle=E_n|n\rangle,\qquad P|n\rangle=P_n|n\rangle,
\end{align}
where
\begin{align}
H=\frac{1}{R}\left(L_0+\bar{L}_0-\frac{c}{12}\right),\qquad P=\frac{1}{R}\left(L_0-\bar{L}_0\right).
\end{align}
The CFT torus partition sum is modular invariant
\begin{align}
Z_0\left(\frac{a\tau+b}{c\tau+d},\frac{a\bar{\tau}+b}{c\bar{\tau}+d}\right)=Z_0(\tau,\bar{\tau}).
\end{align}
where $a,b,c,d\in\mathbb{Z}$ and $ad-bc=1$. Modular invariance comes from a seemingly trivial observation. A torus can be identified with a lattice of the plane. The lattice can be parameterized by the complex parameter $\tau$. However, this parametrization is not unique. Any $\mathrm{PSL}(2,\mathbb{Z})$ (modular) transformation on $\tau$ leads to the same lattice and hence the same torus. Physical results should not depend on how we parameterize the torus. This is the statement of modular invariance.

\subsubsection*{Deformed partition function}
Now we consider the $\rT\overline{\rT}$ deformation. We consider the spectrum on a cylinder of circumference $R$. The deformed energy has been derived in section~\ref{sec:def-spectrum}
\begin{align}
\mathcal{E}_n(\lambda)=\frac{1}{\lambda \pi R}\left(\sqrt{1+2\pi\lambda R E_n+\lambda^2\pi^2 R^2 P_n^2}-1\right),
\end{align}
where we have introduced the dimensionless parameter $\lambda=2t/(\pi R^2)$. The deformed partition sum is defined as
\begin{align}
\mathcal{Z}_{\rT\overline{\rT}}(\tau,\bar{\tau}|\lambda)=\sum_n \re^{2\pi \ri\tau_1R P_n-2\pi\tau_2 R\mathcal{E}_n(\lambda)}.
\end{align}
In order to study the modular properties of the partition sum, we can perform the perturbative expansion of $\mathcal{Z}_{\rT\overline{\rT}}(\tau,\bar{\tau}|\lambda)$ in $\lambda$
\begin{align}
\mathcal{Z}_{\rT\overline{\rT}}(\tau,\bar{\tau}|\lambda)=\sum_{k=0}^\infty Z_k(\tau,\bar{\tau})\,\lambda^k,
\end{align}
where each $Z_k(\tau,\bar{\tau})$ has the following structure
\begin{align}
\label{eq:defZk}
Z_k(\tau,\bar{\tau})=\sum_n F_n^{(k)}\, \re^{2\pi \ri R\tau_1 P_n-2\pi R\tau_2 E_n}.
\end{align}
Here $F_n^{(k)}=F_n^{(k)}(E_n,P_n)$ are polynomials of $E_n$, $P_n$ and $\tau_2$. The first few $F_n^{(k)}$ are given by
\begin{align}
F_n^{(1)}=&\,(E_n^2-P_n^2)(\pi R)^2\tau_2,\\\nonumber
F_n^{(2)}=&\,\frac{1}{2}(E_n^2-P_n^2)^2(\pi R)^4\tau_2^2-\frac{1}{2}E_n(E_n^2-P_n^2)(\pi R)^3\tau_2,\\\nonumber
F_n^{(3)}=&\,\frac{1}{6}(E_n^2-P_n^2)^3(\pi R)^6\tau_2^3-E_n(E_n^2-P_n^2)^2(\pi R)^5\tau_2^2+\frac{1}{4}(E_n^2-P_n^2)(5E_n^2-P_n^2)(\pi R)^4\tau_2.
\end{align}
It is easy to see that inserting monomials of $E_n$ and $P_n$ in the sum like (\ref{eq:defZk}) can be achieved by taking derivatives of the undeformed partition function (\ref{eq:undeformedZ0}) with respect to $\tau_1$ and $\tau_2$. By this we simply mean, for example,
\begin{align}
\sum_n (E_n)^a(P_n)^b \re^{2\pi \ri\tau_1 R P_n-2\pi\tau_2 R E_n}\propto \partial_{\tau_1}^a\partial_{\tau_2}^b
\sum_n \re^{2\pi \ri\tau_1 R P_n-2\pi\tau_2 R E_n}=\partial_{\tau_1}^a\partial_{\tau_2}^b Z_0(\tau,\bar{\tau}).
\end{align}
Notice that we can do this because the deformation of the energy spectrum is \emph{universal}, namely all the energy levels are deformed in the same way. Using the replacement rule
\begin{align}
E_n\mapsto \frac{1}{2\pi \ri R}(\partial_{\tau}-\partial_{\bar{\tau}}),\qquad P_n\mapsto\frac{1}{2\pi \ri R}(\partial_{\tau}+\partial_{\bar{\tau}}).
\end{align}
We can rewrite
\begin{align}
Z_k(\tau,\bar{\tau})=\hat{\mathcal{D}}^{(k)}(\partial_{\tau},\partial_{\bar{\tau}})Z_0(\tau,\bar{\tau}),
\end{align}
where $\hat{\mathcal{D}}^{(k)}(\partial_{\tau},\partial_{\bar{\tau}})$ is certain differential operator made of $\partial_{\tau}$, $\partial_{\bar{\tau}}$ and $\tau_2$. The first few differential operators are given by
\begin{align}
\label{eq:firstfewD}
\hat{\mathcal{D}}^{(1)}(\partial_{\tau},\partial_{\bar{\tau}})=&\,\tau_2\partial_{\tau}\partial_{\bar{\tau}},\\\nonumber
\hat{\mathcal{D}}^{(2)}(\partial_{\tau},\partial_{\bar{\tau}})=&\,\frac{1}{2}\tau_2^2\partial_{\tau}^2\partial_{\bar{\tau}}^2
+\frac{\ri}{2}\tau_2(\partial_{\tau}-\partial_{\bar{\tau}})\partial_{\tau}\partial_{\bar{\tau}},\\\nonumber
\hat{\mathcal{D}}^{(3)}(\partial_{\tau},\partial_{\bar{\tau}})=&\,\frac{1}{6}\tau_2^3\partial_{\tau}^3\partial_{\bar{\tau}}^3
+\frac{\ri}{2}\tau_2^2(\partial_{\tau}-\partial_{\bar{\tau}})\partial_{\tau}^2\partial_{\bar{\tau}}^2
-\frac{1}{4}\tau_2(\partial_{\tau}^2-3\partial_{\tau}\partial_{\bar{\tau}}+\partial_{\bar{\tau}}^2)\partial_{\tau}\partial_{\bar{\tau}}.
\end{align}
In what follows, we will study these differential operators in more detail. It turns out that these operators can be organized in a nice way which make it manifest that $Z_k(\tau,\bar{\tau})$ transforms as modular form of weight $(k,k)$. To this end, we will need some mathematics about the modular forms which we will introduce in the next subsection.

\subsubsection*{Some mathematics}
A function $f_{k,\bar{k}}(\tau,\bar{\tau})$ is called a \emph{modular form} of weight $(k,\bar{k})$ if it satisfies the following property
\begin{align}
\label{mfkkb}
f_{k,\bar{k}}\left(\tau',\bar{\tau}'\right)=(c\tau+d)^k(c\bar{\tau}+d)^{\bar{k}}f_{k,\bar{k}}(\tau,\bar{\tau})
\end{align}
where
\begin{align}
\tau'=\frac{a\tau+b}{c\tau+d},\qquad \bar{\tau}'=\frac{a\bar{\tau}+b}{c\bar{\tau}+d}.
\end{align}
One simple and useful example is $\tau_2=\text{Im}\tau$. Computing the imaginary part of $(a\tau+b)/(c\tau+d)$, we find that $\tau_2$ transforms as
\begin{align}
\label{eq:tau2modular}
\tau'_2=\frac{1}{(c\tau+d)(c\bar{\tau}+d)}\tau_2.
\end{align}
So we can say that $\tau_2$ is a modular form of weight $(-1,-1)$. Usually the derivative of a modular form is no longer a modular form, but becomes \emph{quasi-modular}. Let us consider $\partial_{\tau}f_{k,\bar{k}}(\tau,\bar{\tau})$. Acting $\partial_{\tau}$ on both sides of (\ref{mfkkb}) and using the fact $\partial_{\tau'}=(c\tau+d)^2\partial_{\tau}$, we obtain
\begin{align}
\frac{1}{(c\tau+d)^2}\partial_{\tau'}f_{k,\bar{k}}\left(\tau',\bar{\tau}'\right)=&\,
(c\tau+d)^k(c\bar{\tau}+d)^{\bar{k}}\partial_{\tau}f_{k,\bar{k}}(\tau,\bar{\tau})\\\nonumber
&\,+c k(c\tau+d)^{k-1}(c\bar{\tau}+d)^{\bar{k}}f_{k,\bar{k}}(\tau,\bar{\tau}).
\end{align}
Or equivalently,
\begin{align}
\label{eq:qausimodulardtau}
\partial_{\tau'}f_{k,\bar{k}}\left(\tau',\bar{\tau}'\right)=&\,
(c\tau+d)^{k+2}(c\bar{\tau}+d)^{\bar{k}}\partial_{\tau}f_{k,\bar{k}}(\tau,\bar{\tau})\\\nonumber
&\,+c k(c\tau+d)^{k+1}(c\bar{\tau}+d)^{\bar{k}}f_{k,\bar{k}}(\tau,\bar{\tau}).
\end{align}
The first term is covariant, but the second term is anomalous. This means $\partial_{\tau}f_{k,\bar{k}}(\tau,\bar{\tau})$ is no longer modular for $k>0$. We can cancel the anomalous term by adding a term to the derivative and make a \emph{covariant derivative}
\begin{align}
\mathsf{D}_{\tau}^{(k)}=\partial_{\tau}-\frac{\ri k}{2\tau_2}.
\end{align}
Multiplying $-\ri k/(2\tau_2)$ to both sides of (\ref{mfkkb}) and using (\ref{eq:tau2modular}), we obtain
\begin{align}
\label{eq:quasimodularik}
-\frac{\ri k}{2\tau'_2}f_{k,\bar{k}}(\tau',\bar{\tau}')=&\,-\frac{\ri k}{2\tau_2}(c\tau+d)^{k+1}(c\bar{\tau}+d)^{\bar{k}+1}f_{k,\bar{k}}(\tau,\bar{\tau})\\\nonumber
=&\,-\frac{\ri k}{2\tau_2}(c\tau+d)^{k+1}(c\bar{\tau}+d)^{\bar{k}}(c\tau+d-2\ri c\tau_2)f_{k,\bar{k}}(\tau,\bar{\tau})\\\nonumber
=&\,-\frac{\ri k}{2\tau_2}(c\tau+d)^{k+2}(c\bar{\tau}+d)^{\bar{k}}f_{k,\bar{k}}(\tau,\bar{\tau})-c k(c\tau+d)^{k+1}(c\bar{\tau}+d)^{\bar{k}}f_{k,\bar{k}}(\tau,\bar{\tau}),
\end{align}
where in the second line we used the simple fact that
\begin{align}
c\bar{\tau}+d=c{\tau}+d-2\ri c\tau_2.
\end{align}
Taking the sum of (\ref{eq:qausimodulardtau}) and (\ref{eq:quasimodularik}), we see that the anomalous terms cancel each other exactly and we obtain
\begin{align}
\mathsf{D}_{\tau'}^{(k)}f_{k,\bar{k}}(\tau',\bar{\tau}')=
(c\tau+d)^{k+2}(c\bar{\tau}+d)^{\bar{k}}\,\mathsf{D}_{\tau}^{(k)}f_{k,\bar{k}}(\tau,\bar{\tau}).
\end{align}
Therefore we have proved that $\mathsf{D}^{(k)}_{\tau}f_{k,\bar{k}}(\tau,\bar{\tau})$ is a modular form of weight $(k+2,\bar{k})$. Similarly, we can define the anti-holomorphic covariant derivative
\begin{align}
\mathsf{D}_{\bar{\tau}}^{(\bar{k})}=\partial_{\bar{\tau}}+\frac{\ri\bar{k}}{2\tau_2}.
\end{align}
Acting this covariant derivative on a modular form of weight $(k,\bar{k})$, we obtain a modular form of $(k,\bar{k}+2)$. Notice that the definition of $\mathsf{D}_{\tau}^{(k)}$ and $\mathsf{D}_{\bar{\tau}}^{(\bar{k})}$ depend on the weights of the modular form that they act on.\par

The covariant derivatives $\mathsf{D}_{\tau}^{(k)}$ and $\mathsf{D}_{\bar{\tau}}^{(k)}$ are called \emph{Maass-Shimura derivatives}. Using these, we can define the following operators which are useful later
\begin{align}
\label{eq:curlyD}
\mathscr{D}_{\tau}^{(k)}=\prod_{j=0}^{k-1}\mathsf{D}_{\tau}^{(2j)},\qquad
\mathscr{D}_{\bar{\tau}}^{(k)}=\prod_{j=0}^{k-1}\mathsf{D}_{\bar{\tau}}^{(2j)}.
\end{align}
From the definition of these operators, it is easy to see that the operator $\mathscr{D}_{\tau}^{(k)}$ takes a modular form of weight $(0,m)$ to a modular form of weight $(2k,m)$ for any $m$. Similarly, $\mathscr{D}_{\bar{\tau}}^{(k)}$ takes a modular form of weight $(m,0)$ to a modular form of $(m,2k)$ for any $m$.\par

Notice that in the definition of the differential operators in (\ref{eq:curlyD}), the operators on the left actually act on the ones to the right since the latter contain $1/\tau_2$. So if we expand explicitly in terms of $\partial_{\tau}$ and $\partial_{\bar{\tau}}$, the result is more complicated. As an example, let us write explicitly
\begin{align}
\mathscr{D}_{\tau}^{(3)}=&\,\mathsf{D}_{\tau}^{(4)}\mathsf{D}_{\tau}^{(2)}\mathsf{D}_{\tau}^{(0)}
=\left(\partial_{\tau}-\frac{2\ri}{\tau_2}\right)\left(\partial_{\tau}-\frac{\ri}{\tau_2}\right)\partial_{\tau}
=\partial_{\tau}^3-\frac{3\ri}{\tau_2}\partial_{\tau}^2-\frac{3}{2\tau_2^2}\partial_{\tau},
\end{align}
where we have used the fact that
\begin{align}
\partial_{\tau}(1/\tau_2)=\frac{\ri}{2\tau_2^2}.
\end{align}

\subsubsection*{Modular invariance of deformed partition function}
Now we are equipped with necessary mathematical tools, we start to investigate the modular properties of $Z_k(\tau,\bar{\tau})$. We find that the first few differential operators given in (\ref{eq:firstfewD}) can be written in terms of the covariant operators defined in (\ref{eq:curlyD}) in a surprisingly simple way
\begin{align}
\hat{\mathcal{D}}^{(1)}(\partial_{\tau},\partial_{\bar{\tau}})=&\,\frac{1}{1!}\tau_2\mathscr{D}_{\tau}^{(1)}\mathscr{D}_{\bar{\tau}}^{(1)},\\\nonumber
\hat{\mathcal{D}}^{(2)}(\partial_{\tau},\partial_{\bar{\tau}})=&\,\frac{1}{2!}\tau_2^2\mathscr{D}_{\tau}^{(2)}\mathscr{D}_{\bar{\tau}}^{(2)},\\\nonumber
\hat{\mathcal{D}}^{(3)}(\partial_{\tau},\partial_{\bar{\tau}})=&\,\frac{1}{3!}\tau_2^3\mathscr{D}_{\tau}^{(3)}\mathscr{D}_{\bar{\tau}}^{(3)}.
\end{align}
We can check further and find that to very high orders of $k$ we have
\begin{align}
\hat{\mathcal{D}}^{(k)}(\partial_{\tau},\partial_{\bar{\tau}})=&\,\frac{1}{k!}\tau_2^k\mathscr{D}_{\tau}^{(k)}\mathscr{D}_{\bar{\tau}}^{(k)}.
\end{align}
Again we emphasis that the differential operators on the left act non-trivially on the ones on the right. From the modular properties of the operators $\mathscr{D}_{\tau}^{(k)}$ and $\mathscr{D}_{\bar{\tau}}^{(k)}$, we see that $\mathscr{D}_{\tau}^{(k)}\mathscr{D}_{\bar{\tau}}^{(k)}$ takes a modular function (which is a modular form of weight $(0,0)$ to a modular form of weight $(2k,2k)$. Multiplying with $\tau_2^k$, which reduces the weights by $(-k,-k)$, we find that $\hat{\mathcal{D}}^{(k)}(\partial_{\tau},\partial_{\bar{\tau}})$ maps a modular function to a modular form of weight $(k,k)$. If we require the dimensionless parameter $\lambda$ transforms as a $(-1,-1)$ form, then $Z_k(\tau,\bar{\tau})\lambda^k$ is invariant under modular transformation. This is of course a strong hint that the deformed partition function is modular invariant, but not yet a rigorous proof. To complete the proof, we can use the flow equation for the deformed partition function. This can be derived from Cardy's random geometry point of view, as we will discuss in section~\ref{lecture3}. In fact it can also be derived directly from the Burgers' equation, as we will show now.

\paragraph{Flow equation} Let us first write the deformed partition function in terms of dimensionless quantities $\mathbb{E}_n,\mathbb{P}_n$ and $\lambda$
\begin{align}
\mathcal{Z}_{\rT\overline{\rT}}(\tau,\bar{\tau}|\lambda)=\sum_n \re^{2\pi \ri\tau_1\mathbb{P}_n-2\pi\tau_2\mathbb{E}_n(\lambda)}.
\end{align}
We quote the ODE (\ref{eq:ODESpec}) which is equivalent to Burgers' equation for CFTs (taking $\alpha=1/\pi$)
\begin{align}
\label{eq:ODEBurg}
\mathbb{E}_n'+2\pi\lambda\mathbb{E}_n\mathbb{E}_n'+\pi(\mathbb{E}_n^2-\mathbb{P}_n^2)=0.
\end{align}
We can take derivative of the deformed partition function with respect to $\tau_1,\tau_2$ and $\lambda$. These derivatives will act on the exponent and bring down some factors. For example, taking derivative with respect to $\lambda$, we have
\begin{align}
\partial_{\lambda}\mathcal{Z}_{\rT\overline{\rT}}=\sum_n\left(-2\pi\tau_2\mathbb{E}_n' \right)\re^{2\pi \ri\tau_1\mathbb{P}_n-2\pi\tau_2\mathbb{E}_n(\lambda)}.
\end{align}
We can denote this relation by the replacement rule
\begin{align}
\label{eq:plambda1}
\partial_{\lambda}\mapsto-2\pi\tau_2\mathbb{E}_n'.
\end{align}
Or equivalently, we write
\begin{align}
\mathbb{E}_n'\mapsto-\frac{1}{2\pi\tau_2}\partial_{\lambda}.
\end{align}
Taking again the derivative with respect to $\tau_2$, we obtain the replacement rule
\begin{align}
\label{eq:plambda2}
\partial_{\tau_2}\partial_{\lambda}\mapsto -2\pi\mathbb{E}_n'+4\pi^2\mathbb{E}_n'\mathbb{E}_n.
\end{align}
Combining (\ref{eq:plambda1}) and (\ref{eq:plambda2}), we have
\begin{align}
\mathbb{E}_n'\mathbb{E}_n\mapsto \frac{1}{4\pi^2}\left(\partial_{\tau_2}-\frac{1}{\tau_2}\right)\partial_{\lambda}.
\end{align}
Finally, it is easy to see that
\begin{align}
\mathbb{E}_n^2-\mathbb{P}_n^2\mapsto\frac{1}{4\pi^2}\left(\partial_{\tau_1}^2+\partial_{\tau_2}^2\right)
=\frac{1}{\pi^2}\partial_{\tau}\partial_{\bar{\tau}},
\end{align}
where we have used the relation
\begin{align}
\partial_{\tau_1}=\partial_{\tau}+\partial_{\bar{\tau}},\qquad\partial_{\tau_2}=\ri(\partial_{\tau}-\partial_{\bar{\tau}}).
\end{align}
Now we can use the replacement rule to rewrite the ODE (\ref{eq:ODEBurg}) as
\begin{align}
\label{eq:flowEq}
\partial_{\lambda}\mathcal{Z}_{\rT\overline{\rT}}=\left[\tau_2\partial_{\tau}\partial_{\bar{\tau}}+\frac{1}{2}\left(\partial_{\tau_2}-\frac{1}{\tau_2} \right)
\lambda\partial_{\lambda}\right]\mathcal{Z}_{\rT\overline{\rT}}.
\end{align}
This is the flow equation we need.\par

\paragraph{A recursion relation} Now we plug the perturbative expansion of the partition function
\begin{align}
\mathcal{Z}_{\rT\overline{\rT}}(\tau,\bar{\tau}|\lambda)=\sum_{k=0}^\infty Z_k(\tau,\bar{\tau})\lambda^k
\end{align}
into the flow equation (\ref{eq:flowEq}), we obtain a recursion relation between $Z_p$ and $Z_{p+1}$
\begin{align}
Z_{p+1}=\frac{\tau_2}{p+1}\left(\mathsf{D}_{\tau}^{(p)}\mathsf{D}_{\bar{\tau}}^{(p)}-\frac{p(p+1)}{4\tau_2^2}\right)Z_p.
\end{align}
From this recursion relation, it is clear that if $Z_p$ is a modular form of weight $(p,p)$, then $Z_{p+1}$ is a modular form of weight $(p+1,p+1)$. By requiring that $\lambda$ transforms as
\begin{align}
\lambda'=\frac{\lambda}{(c\tau+d)(c\bar{\tau}+d)}=\frac{\lambda}{|c\tau+d|^2},
\end{align}
we see that the $\rT\overline{\rT}$ deformed partition function is modular invariant
\begin{align}
\mathcal{Z}_{\rT\overline{\rT}}\left(\left.\frac{a\tau+b}{c\tau+d},\frac{a\bar{\tau}+b}{c\bar{\tau}+d}\right|\frac{\lambda}{|c\tau+d|^2}\right)
=\mathcal{Z}_{\rT\overline{\rT}}(\tau,\bar{\tau}|\lambda).
\end{align}
We notice that the quantization radius $R$ also transforms under modular transformation as
\begin{align}
R\mapsto |c\tau+d|R.
\end{align}
Therefore the transformation rule for $\lambda=2t/(\pi R^2)$ comes solely from the transformations of $R$. If we write the deformed partition function in terms of the \emph{dimensionful parameter} $t$, we have
\begin{align}
\mathcal{Z}_{\rT\overline{\rT}}\left(\left.\frac{a\tau+b}{c\tau+d},\frac{a\bar{\tau}+b}{c\bar{\tau}+d}\right|t\right)
=\mathcal{Z}_{\rT\overline{\rT}}(\tau,\bar{\tau}|t).
\end{align}

\subsection{Modular bootstrap}
Now we move to the second question. We consider the same setup as in the previous section. Let us forget about the $\rT\overline{\rT}$ deformation for the moment and recover it from modular bootstrap.

\subsubsection*{Solvable deformations} For a given CFT on a cylinder of radius $R$, we can consider a \emph{trivially solvable deformation}, depending on a \emph{dimensionless parameter} $\lambda$
\begin{align}
{H}\mapsto\mathcal{H}({H},{P},\lambda),\qquad {P}\mapsto {P}.
\end{align}
It is clear that the state $|n\rangle$ automatically diagonalize the deformed Hamiltonian $\mathcal{H}$
\begin{align}
\mathcal{H}({H},{P},\lambda)|n\rangle=\mathcal{H}(E_n,P_n,\lambda)|n\rangle.
\end{align}
The momentum is not deformed. Let us denote $\mathcal{E}_n(\lambda)=\mathcal{H}(E_n,P_n,\lambda)$ and the initial condition is $\mathcal{E}_n(0)=E_n$. We define the deformed torus partition function as
\begin{align}
\label{eq:defModZZ}
\mathcal{Z}(\tau,\bar{\tau}|\lambda)=\sum_n \re^{2\pi \ri\tau_1 R P_n-2\pi\tau_2 R \mathcal{E}_n(\lambda)}.
\end{align}
There are infinitely many possible choices for the function $\mathcal{H}(E,P,\lambda)$, but not all the deformations are physical. To restrict to a subset family of the solvable deformations, we need to impose some consistency conditions. One natural consistency condition for partition function on the torus is modular invariance. Let us impose that and see which kinds of theories do we get. More explicitly, we require
\begin{align}
\label{eq:modularAss}
\mathcal{Z}\left(\left.\frac{a\tau+b}{c\tau+d},\frac{a\bar{\tau}+b}{c\bar{\tau}+d}\right|\frac{\lambda}{|c\tau+d|^2}\right)
=\mathcal{Z}(\tau,\bar{\tau}|\lambda).
\end{align}
We will see that this condition is strong enough to fix the function $\mathcal{H}(E,P,\lambda)$ completely. Before that, let us make one comment here. One might ask that why should we require that $\lambda$ transforms as in (\ref{eq:modularAss}), can we instead require that $\lambda$ transforms as a $(m,n)$ form
\begin{align}
\lambda'=\frac{\lambda}{(c\tau+d)^m(c\bar{\tau}+d)^n}.
\end{align}
The answer is no. One can show that the only non-trivial deformations that is compatible with modular invariance is $(m,n)=(1,1)$. This will be clear after our proof of uniqueness.

\subsubsection*{Perturbative expansion}
We need to make another assumption for the deformed spectrum. We assume that the deformed energy $\mathcal{E}_n(\lambda)$ has a regular Taylor expansion in $\lambda$
\begin{align}
\mathcal{E}_n(\lambda)=\sum_{k=0}^{\infty}\rE_n^{(k)}\lambda^k,\qquad \rE_n^{(0)}=E_n.
\end{align}
here $\rE_n^{(k)}(E_n,P_n)$ are functions of $E_n$ and $P_n$. Since the deformed energy has regular Taylor expansion, this is also true for the partition function
\begin{align}
\mathcal{Z}(\tau,\bar{\tau}|\lambda)=\sum_{k=0}^\infty Z_k\,\lambda^k,
\end{align}
where $Z_0$ is the undeformed partition function. It is easy to work out the first few orders
\begin{align}
Z_1=&\,\sum_n\left(-2\pi R\tau_2\rE_n^{(1)}\right)\re^{2\pi \ri\tau_1 R P_n-2\pi\tau_2 R E_n},\\\nonumber
Z_2=&\,\sum_n\left(\frac{\tau_2^2}{2}(2\pi R\rE_n^{(1)})^2-2\pi R\tau_2\rE_n^{(2)}  \right)\re^{2\pi \ri\tau_1 R P_n-2\pi\tau_2 R E_n},\\\nonumber
&\,\vdots\\\nonumber
Z_p=&\,\sum_n\left(\frac{\tau_2^p}{p!}(2\pi R\rE_n^{(1)})^p+\cdots-2\pi R\tau_2\rE_n^{(p)} \right)\re^{2\pi \ri\tau_1 R P_n-2\pi\tau_2 R E_n},
\end{align}
where $\rE_n^{(k)}(E_n,P_n)$ are functions of $E_n$ and $P_n$. Recall from the last section that inserting any monomial of $E_n^i P_n^j$ in the summand of the partition sum can be obtained by taking derivatives with respect to the original partition function with the following replacement rule
    \begin{align}
    E_n\mapsto \frac{1}{2\pi \ri R}(\partial_{\tau}-\partial_{\bar{\tau}}),\quad P_n\mapsto\frac{1}{2\pi \ri R}(\partial_{\tau}+\partial_{\bar{\tau}}).
    \end{align}
The Taylor coefficients $Z_p$ can be written in terms of acting certain differential operator on $Z_0$. They have the following structure
\begin{enumerate}
\item The differential operator is a polynomial in terms of $\tau_2$
    \begin{align}
    \label{eq:diffZponZ0}
    Z_p=\left[\tau_2^p\widehat{\mathcal{O}}_1^{(p)}(\partial_{\tau},\partial_{\bar{\tau}})
    +\tau_2^{p-1}\widehat{\mathcal{O}}_2^{(p)}(\partial_{\tau},\partial_{\bar{\tau}})+\cdots
    +\tau_2\widehat{\mathcal{O}}_p^{(p)}(\partial_{\tau},\partial_{\bar{\tau}}) \right]Z_0,
    \end{align}
    where $\widehat{\mathcal{O}}_j^{(p)}(\partial_{\tau},\partial_{\bar{\tau}})$ are differential operators. Notice that these operators do not depend on $\tau_2$ and are only made of $\partial_{\tau}$, $\partial_{\bar{\tau}}$.
\item The highest order term of $\tau_2$ is given by ${\tau_2^p}(2\pi R\rE_n^{(1)})^p/p!$. Therefore from (\ref{eq:diffZponZ0}), we see that $\widehat{\mathcal{O}}_1^{(p)}(\partial_{\tau},\partial_{\bar{\tau}})$ is fixed by $\rE_n^{(1)}$. Notice that $\rE_n^{(1)}$ fixes $Z_1$ completely.
\item The lowest order term of $\tau_2$ is given by $\tau_2(-2\pi R)\rE_n^{(p)}$. Notice that $\rE_n^{(p)}$ is the only quantity that does not appear in lower orders $Z_0,Z_1,\cdots,Z_{p-1}$.
\item There are no terms with $\tau_2^{m}$ such that $m\le 0$.
\end{enumerate}
Let us first notice that due to our assumption of the modular property (\ref{eq:modularAss}), the Taylor coefficient $Z_p$ transforms as a modular form of weight $(p,p)$
\begin{align}
Z_p\left(\frac{a\tau+b}{c\tau+d},\frac{a\bar{\tau}+b}{c\bar{\tau}+d}\right)=(c\tau+d)^p(c\bar{\tau}+d)^p Z_p(\tau,\bar{\tau}).
\end{align}
Our proof involves two steps
\begin{enumerate}
\item \textbf{Uniqueness}. Show that all $Z_p$ \emph{are uniquely determined} by the modular property,
\item \textbf{Existence}. Give a \emph{method to construct} $Z_p$ explicitly.
\end{enumerate}

\subsubsection*{Uniqueness}
We shall prove uniqueness by induction. First consider $Z_1$
\begin{align}
Z_1=\tau_2\widehat{\mathcal{O}}_1^{(1)}(\partial_{\tau},\partial_{\bar{\tau}})Z_0.
\end{align}
From our assumption, $Z_1$ is a $(1,1)$ modular form. Therefore $\widehat{\mathcal{O}}_1^{(1)}$ is an operator which takes a $(0,0)$ form to $(2,2)$ form. It is not hard to see that the only possibility for $\widehat{\mathcal{O}}_1^{(1)}$ is that
\begin{align}
\widehat{\mathcal{O}}_1^{(1)}(\partial_{\tau},\partial_{\bar{\tau}})=\alpha\partial_{\tau}\partial_{\bar{\tau}},
\end{align}
where $\alpha$ is an unimportant multiplicative constant that can be absorbed into the coupling constant $\lambda$. Now we consider the induction. Assuming $Z_0,Z_1,\cdots,Z_p$ are completely fixed, we want to show that $Z_{p+1}$ is also fixed. As we discussed before, fixing $Z_0,\cdots,Z_p$ imply that we have completely fixed the energy shifts $\rE_n^{(j)}$ with $j=1,2,\cdots,p$. Then in the expression of $Z_{p+1}$, the only unknown operator is $\widehat{\mathcal{O}}_{p+1}^{(p+1)}$ that couples to $\tau_2$.\par
Suppose two such operators exist, which we denote by $\widehat{\mathcal{O}}_{p+1}^{(p+1)}$ and $\widehat{\mathcal{O}'}_{p+1}^{(p+1)}$. Then there are two quantities $Z_{p+1}$ and $Z'_{p+1}$ which are $(p+1,p+1)$ forms. Their difference
\begin{align}
Z_{p+1}-Z'_{p+1}=\tau_2\,\left[\widehat{\mathcal{O}}(\partial_{\tau},\partial_{\bar{\tau}})
-\widehat{\mathcal{O}'}(\partial_{\tau},\partial_{\bar{\tau}})\right]Z_0
\end{align}
is also a $(p+1,p+1)$ form. Notice that higher order terms in $\tau_2$ cancel because they are fixed by $\rE_n^{(j)}$ $(j=0,\cdots,p)$. This implies that there must be an operator $\delta\widehat{\mathcal{O}}_{p+1}^{(p+1)}(\partial_{\tau},\partial_{\bar{\tau}})$ such that
\begin{align}
\delta\widehat{\mathcal{O}}_{p+1}^{(p+1)}(\partial_{\tau},\partial_{\bar{\tau}})Z_0\equiv\left[\widehat{\mathcal{O}}(\partial_{\tau},\partial_{\bar{\tau}})
-\widehat{\mathcal{O}'}(\partial_{\tau},\partial_{\bar{\tau}})\right]Z_0
\end{align}
is a $(p+2,p+2)$ form. We will show that such an operator does not exist for $p>0$. Notice that $\widehat{\mathcal{O}}(\partial_{\tau},\partial_{\bar{\tau}})$ is a differential operator involving only the derivatives $\partial_{\tau}$ and $\partial_{\bar{\tau}}$ and some numerical coefficients. It does not involves other modular quantities such as $\tau_2$. Let us analyze the action of $\partial_{\tau}$ and $\partial_{\bar{\tau}}$ on a modular form of $(k,\bar{k})$. Using the covariant derivatives we have the following
\begin{align}
\partial_{\tau}f_{k,\bar{k}}(\tau,\bar{\tau})=&\,\mathsf{D}^{(k)}_{\tau}f_{k,\bar{k}}(\tau,\bar{\tau})
+\frac{\ri k}{2\tau_2}f_{k,\bar{k}}(\tau,\bar{\tau}),\\\nonumber
\partial_{\bar{\tau}}f_{k,\bar{k}}(\tau,\bar{\tau})=&\,\mathsf{D}^{(k)}_{\bar{\tau}}f_{k,\bar{k}}(\tau,\bar{\tau})
-\frac{\ri\bar{k}}{2\tau_2}f_{k,\bar{k}}(\tau,\bar{\tau}).
\end{align}
The two terms on the right hand side are modular forms of weights $(k+2,\bar{k})$ and $(k+1,\bar{k}+1)$. Similarly, acting $\partial_{\bar{\tau}}$ leads to modular forms of weight $(k,\bar{k}+2)$ and $(k+1,\bar{k}+1)$. The total increase of the weights are 2. To obtain the desired form, we need to act a total $p+2$ derivatives of $\partial_{\tau}$ and $\partial_{\bar{\tau}}$. However, apart from the wanted terms, we will also generate many unwanted terms along the way which we cannot get rid of in general.\par

As an example, consider $p=1$. Suppose we want to generate a $(3,3)$ form from $(0,0)$ form by acting some operators made of $\partial_{\tau}$ and $\partial_{\bar{\tau}}$. The total weight is $3+3=6$. There are four operators which generate such total weight: $\partial_{\tau}^3,\partial^2_{\tau}\partial_{\bar{\tau}},\partial_{\tau}\partial_{\bar{\tau}}^2$ and $\partial_{\bar{\tau}}^3$. These operators generate the following modular forms
\begin{align}
\partial_{\tau}^3:&\qquad (6,0)\quad (5,1)\quad (4,2),\\\nonumber
\partial_{\tau}^2\partial_{\bar{\tau}}:&\qquad (5,1)\quad (4,2)\quad (3,3),\\\nonumber
\partial_{\tau}\partial^2_{\bar{\tau}}:&\qquad (3,3)\quad (2,4)\quad (1,5),\\\nonumber
\partial_{\bar{\tau}}^3:&\qquad (2,4)\quad (1,5)\quad (0,6).
\end{align}
To generate the desired $(3,3)$ modular form, we need to act $\partial_{\tau}^2\partial_{\bar{\tau}}$ and $\partial_{\tau}\partial^2_{\bar{\tau}}$ on $Z_0$. However, these operators also generate the unwanted modular forms of weights $(4,2),(5,1)$ and $(2,4),(1,5)$. To cancel these unwanted modular forms, we need to take into account the operators $\partial_{\tau}^3$ and $\partial_{\bar{\tau}}^3$. However, these operators will then generate the unwanted modular forms of weights $(6,0)$ and $(0,6)$, which cannot get canceled. Therefore there is no way to get rid of all the unwanted terms. Similar analysis can be performed for modular forms of of weight $(p,p)$ for $p>3$. To conclude, \emph{there is no differential operator $\mathcal{O}(\partial_{\tau},\partial_{\bar{\tau}})$ made of $\partial_{\tau}$ and $\partial_{\bar{\tau}}$ that takes a $(0,0)$ form to a $(p,p)$ form for $p\ge 3$.} As a result, there cannot be two differential operators $\widehat{\mathcal{O}}_{p+1}^{(p+1)}$ and $\widehat{\mathcal{O}'}_{p+1}^{(p+1)}$. So the operator $\widehat{\mathcal{O}}_{p+1}^{(p+1)}$ is unique. This implies that $Z_{p+1}$ is uniquely fixed.

\subsubsection*{Existence}
We have shown that the differential operator which leads to $Z_p$ is unique. The next step is to give an explicit way to construct such an operator. Of course, since we have proved that $\rT\overline{\rT}$ deformed CFT satisfies the modular invariance. By the uniqueness, the theory is nothing but the $\rT\overline{\rT}$ deformed CFT. Suppose we do not know the deformed spectrum, can we still have a practical way to construct such a differential operator ? This kind of construction will be useful for more general cases.

\paragraph{A covariant ansatz} Let us take a closer look at the following expression
\begin{align}
\label{eq:expandZp}
Z_p=\left[\tau_2^p\widehat{\mathcal{O}}_1^{(p)}(\partial_{\tau},\partial_{\bar{\tau}})
    +\tau_2^{p-1}\widehat{\mathcal{O}}_2^{(p)}(\partial_{\tau},\partial_{\bar{\tau}})+\cdots
    +\tau_2\widehat{\mathcal{O}}_p^{(p)}(\partial_{\tau},\partial_{\bar{\tau}}) \right]Z_0.
\end{align}
Each operator $\widehat{\mathcal{O}}_k^{(p)}(\partial_{\tau},\partial_{\bar{\tau}})$ can be expanded in terms of $\partial_{\tau}^i$ and $\partial_{\bar{\tau}}^j$. It can be proven that the usual derivatives can be written in terms of the covariant derivatives defined in (\ref{eq:curlyD})
\begin{align}
(\partial_{\tau})^n=\sum_{r=0}^n{n\choose r}\frac{(k+r)_{n-r}}{(2\pi i\tau_2)^{n-r}}\mathscr{D}_{\tau}^{(r)},
\end{align}
where $(a)_m$ is the Pochhammer symbol. A similar expression exists for $(\partial_{\bar{\tau}})^n$. This implies that we can equivalently express the differential operator in terms of $\mathscr{D}_{\tau}^{(i)}$ and $\mathscr{D}_{\bar{\tau}}^{(j)}$ and $\tau_2$. Therefore we can write the operator in terms of linear combinations of $\tau_2^k\mathscr{D}_{\tau}^{(i)}\mathscr{D}_{\bar{\tau}}^{(j)}$. The advantage of this rewriting is that each term increases the modular weight by a \emph{definite} amount. For example, the operator $\tau_2^k\mathscr{D}_{\tau}^{(i)}\mathscr{D}_{\bar{\tau}}^{(j)}$ acting on a modular function leads to a modular form of weight $(2i-k,2j-k)$. At each order $p$, we want $Z_p$ to be modular form of weight $(p,p)$. We can make an ansatz so that each term in the ansatz leads to the desired weight of the modular form. We have
\begin{align}
2i-k=p,\qquad 2j-k=p.
\end{align}
We see therefore $i=j$, so the ansatz consists of operators $\tau_2^{2i-p}\mathscr{D}_{\tau}^{(i)}\mathscr{D}_{\bar{\tau}}^{(i)}$ for any integer $i$. In addition, the possible values of $i$ are bounded from the structure of perturbative expansion (\ref{eq:expandZp}).
\begin{itemize}
\item It is clear that $i\ge 0$ since negative numbers of $\mathscr{D}_{\tau}^{(i)}$ do not make sense here;
\item From (\ref{eq:expandZp}), it is clear that the highest power of $\tau_2$ for $Z_p$ is $\tau_2^p$. There are also factors of $\tau_2$ in the covariant derivatives $\mathscr{D}_{\tau}^{(i)}$, but they are all negative powers. Therefore we have $2i-p\le p$, which implies $i\le p$.
\end{itemize}
To sum up, the possible range of values of $i$ is given by $0\le i\le p$ and $i\in\mathbb{Z}$. Therefore at each order $p$, we can make the following ansatz
\begin{align}
\hat{\mathcal{D}}^{(p)}(\partial_{\tau},\partial_{\bar{\tau}})=\sum_{j=0}^p a^{(p)}_j\tau_2^{2j-p}\mathscr{D}_{\tau}^{(j)}\mathscr{D}_{\bar{\tau}}^{(j)}.
\end{align}
This ansatz has used the modular property of $Z_p$. Now we need to fix the coefficients $a^{(p)}_j$. These coefficients are fixed by the structure of perturbative expansion. Firstly, we notice that $a^{(p)}_p$ is completely fixed because $\tau_2^p$ is always coupled to $(-2\pi R\rE_n^{(1)})^p/p!$ and $\rE_n^{(1)}$ is already fixed by $Z_1$ to be
\begin{align}
R\rE_n^{(1)}\mapsto -\alpha\partial_{\tau}\partial_{\bar{\tau}}
\end{align}
where $\alpha$ is some constant that can be absorbed in the dimensionless coupling constant. We have
\begin{align}
Z_p=\frac{\alpha^p}{p!}\left[\tau_2^p(\partial_{\tau}\partial_{\bar{\tau}})^p+\cdots\right],
\end{align}
therefore $a^{(p)}_p=\alpha^p/p!$. The rest of the coefficients can be fixed by the observation that in (\ref{eq:expandZp}) there are no non-positive powers of $\tau_2$. Let us consider one example $Z_2$. The covariant ansatz for $Z_2$ is given by
\begin{align}
Z_2(\tau,\bar{\tau})=\left(\frac{\alpha^2}{2!}\tau_2^2\,\mathscr{D}_{\tau}^{(2)}\mathscr{D}_{\bar{\tau}}^{(2)}
+a^{(2)}_1\,\mathscr{D}_{\tau}^{(1)}\mathscr{D}_{\bar{\tau}}^{(1)}
+\frac{a^{(2)}_0}{\tau_2^2}\,\mathscr{D}_{\tau}^{(0)}\mathscr{D}_{\bar{\tau}}^{(0)}\right)Z_0(\tau,\bar{\tau}).
\end{align}
Expanding the covariant derivatives, we find
\begin{align}
\label{eq:Z2expand}
Z_2(\tau,\bar{\tau})=\left[\frac{\alpha^2}{2}\tau_2^2\partial_{\tau}^2\partial_{\bar{\tau}}^2
+\frac{\ri\alpha^2}{2}\tau_2(\partial_{\tau}-\partial_{\bar{\tau}})\partial_{\tau}\partial_{\bar{\tau}}
+a^{(2)}_1\,\partial_{\tau}\partial_{\bar{\tau}}+\frac{a^{(2)}_0}{\tau_2^2}
\right]Z_0(\tau,\bar{\tau}).
\end{align}
It is clear that the last two terms in (\ref{eq:Z2expand}) couple to non-positive powers of $\tau_2$ and are not compatible with the structure of perturbative expansion. Therefore the coefficients need to be put to zero. We thus find
\begin{align}
a^{(2)}_1=0,\qquad a^{(2)}_0=0,
\end{align}
and fix $Z_2(\tau,\bar{\tau})$ to be
\begin{align}
Z_2(\tau,\bar{\tau})=\frac{\alpha^2}{2}\tau_2^2\,\mathscr{D}_{\tau}^{(2)}\mathscr{D}_{\bar{\tau}}^{(2)}\,Z_0(\tau,\bar{\tau}).
\end{align}
Similarly, one can find a few more terms. This gives a systematic way to construct $Z_p(\tau,\bar{\tau})$. In order to fix all the $Z_p$ once and for all, it is helpful to set up a recursion relation between $Z_p$ and $Z_{p+1}$. From our previous analysis, we see that we need an operator which maps a $(p,p)$ modular form to a $(p+1,p+1)$ modular form. There are only two natural candidates $\tau_2\mathsf{D}_{\tau}^{(p)}\mathsf{D}_{\bar{\tau}}^{(p)}$ and $1/\tau_2$. Therefore we can make the following ansatz for the recursion relation
\begin{align}
\label{eq:recursAns}
Z_{p+1}=d_p\,\tau_2\left(\mathsf{D}_{\tau}^{(p)}\mathsf{D}_{\bar{\tau}}^{(p)}-\frac{b_p}{\tau_2^2}\right)Z_p.
\end{align}
The coefficient $b_p$ can be fixed by requiring that there is no terms at order $\tau_2^0$. More explicitly, we notice that $Z_p(\tau,\bar{\tau})$ has the following expansion
\begin{align}
Z_p(\tau,\bar{\tau})=\sum_{k=0}^{p-1}\tau_2^{p-k}Y_k(\tau,\bar{\tau}).
\end{align}
Acting the operator (\ref{eq:recursAns}) on the above ansatz, it leads to the following terms at order $\tau_2^0$
\begin{align}
\frac{p}{2}Y_{p-1}+\frac{p(p-1)-4b_p}{4}Y_{p-1}.
\end{align}
Requiring this to be zero, we can fix $b_p$ to be
\begin{align}
b_p=\frac{p(p+1)}{4}.
\end{align}
The coefficient $d_p$ is fixed by the global normalization and we have
\begin{align}
d_p=\frac{\alpha}{p+1}.
\end{align}
Therefore we fix the recursion relations to be
\begin{align}
\label{eq:recursionZppp}
Z_{p+1}=\frac{\alpha\tau_2}{p+1}\left[\mathsf{D}_{\tau}^{(p)}\mathsf{D}_{\bar{\tau}}^{(p)}-\frac{p(p+1)}{4\tau_2^2}\right]Z_p.
\end{align}
Fixing $\alpha=1$, we see that this is exactly the same recursion relation for the $\rT\overline{\rT}$ deformed CFT. Then we can run the story in the reverse order. Namely, from the recursion relation we can write down the flow equation. Having the flow equation and assuming torus partition sum takes the form of sum over Boltzman weights (\ref{eq:defModZZ}), we can recover the Burgers' equation for the deformed spectrum. This shows that the result is exactly the $\rT\overline{\rT}$ deformed CFT.

\subsection{Non-perturbative features}
In the previous sections, we have shown that the $\rT\overline{\rT}$ deformed CFT is modular invariant and that by requiring modular invariance, one can fix the solvable deformation to be that of the $\rT\overline{\rT}$ deformed CFT. In this section, we will explore some interesting non-perturbative features of the deformed partition sum.

\subsubsection*{Different signs and non-perturbative ambiguity}
We have seen in section~\ref{sec:def-spectrum} that the deformed spectrum for the two signs of the coupling constant $t$ (or equivalently in terms of the dimensionless coupling constant $\lambda$) have different properties. For the good sign, the spectrum is well-defined for arbitrary high energy states. On the other hand, for the `bad' sign, the deformed spectrum become complex for high energy states in the original theory and signifies a possible break down of unitarity. We will see that similar phenomena appears at the level of partition function. The solutions of the flow equation are different for the two signs of the coupling constant. For the good sign (corresponds to $\lambda>0$ in our convention), the solution seems to be unique. For the bad sign, there are non-perturbative ambiguities. Let us first recall the recursion relation
\begin{align}
\partial_{\lambda}\mathcal{Z}(\tau,\bar{\tau}|\lambda)=\left[\tau_2\partial_{\tau}\partial_{\bar{\tau}}+\frac{1}{2}\left(\partial_{\tau_2}-\frac{1}{\tau_2} \right)
\lambda\partial_{\lambda}\right]\mathcal{Z}(\tau,\bar{\tau}|\lambda).
\end{align}
We want to study the solution of this equation with the initial condition
\begin{align}
\mathcal{Z}(\tau,\bar{\tau}|0)=Z_0(\tau,\bar{\tau}),
\end{align}
where $Z_0(\tau,\bar{\tau})$ is the partition function of a CFT. We have seen that perturbatively we can construct $\mathcal{Z}(\tau,\bar{\tau}|\lambda)$ by the recursion relation (\ref{eq:recursionZppp}). However, the flow equation also allows non-perturbative solutions in $\lambda$.

\paragraph{A non-perturbative solution} The simplest possible non-perturbative solution we can find is
\begin{align}
f_{\text{NP}}(\tau,\bar{\tau}|\lambda)=\exp\left(\frac{4\tau_2}{\lambda}\right).
\end{align}
This solution is modular invariant and is indeed a solution of the flow equation. One interesting observation, that is also generalizable to the $\rJ\overline{\rT}$ case is that the exponent can be obtained from the Burgers' equation. In fact the Burgers' equation allows two solutions if we do not impose the initial condition $\lim_{\lambda\to 0}\mathcal{E}_n(\lambda)=E_n$ which correspond to the two branches of the square root. Let us denote them by
\begin{align}
\mathcal{E}^{(\pm)}_n(\lambda)=\frac{1}{\lambda \pi R}\left(\pm\sqrt{1-2\pi\lambda R E_n+\lambda^2\pi^2 R^2 P_n^2}-1\right).
\end{align}
The solution which allows a well-defined CFT limit is $\mathcal{E}_n^{(+)}$. Taking the sum of the two solutions, we obtain
\begin{align}
\mathcal{E}^{(+)}_n(\lambda)+\mathcal{E}^{(-)}_n(\lambda)=-\frac{2}{\pi R\lambda}.
\end{align}
We find that
\begin{align}
-2\pi R\tau_2\left[\mathcal{E}^{(+)}_n(\lambda)+\mathcal{E}^{(-)}_n(\lambda)\right]=\frac{4\tau_2}{\lambda},
\end{align}
which is the exponent of the non-perturbative solution. We can consider more general solutions of this type by making the ansatz
\begin{align}
X(\tau,\bar{\tau}|\lambda)\re^{\frac{4\tau_2}{\lambda}}.
\end{align}
Plugging this ansatz into the flow equation, we find that $X(\tau,\bar{\tau}|\lambda)$ satisfies exactly the same flow equation, but with $\lambda\to-\lambda$. Namely,
\begin{align}
\label{eq:flowX}
\partial_{\lambda}X(\tau,\bar{\tau}|\lambda)=-\left[\tau_2\partial_{\tau}\partial_{\bar{\tau}}+\frac{1}{2}\left(\partial_{\tau_2}-\frac{1}{\tau_2} \right)
\lambda\partial_{\lambda}\right]X(\tau,\bar{\tau}|\lambda).
\end{align}

\paragraph{Different signs} Now let us consider the solution of the following kind
\begin{align}
\label{eq:ansatzNP}
\mathcal{Z}(\tau,\bar{\tau}|\lambda)=\sum_{k=0}^\infty Z_k(\tau,\bar{\tau})\lambda^k+\re^{\frac{4\tau_2}{\lambda}}X(\tau,\bar{\tau}|\lambda)
\end{align}
where the first term is our solution from the recursion relation (\ref{eq:recursionZppp}). Now we consider the second term for two different signs.\par

For the case $\lambda>0$, the second term in (\ref{eq:ansatzNP}) is divergent in the limit $\lambda\to 0^+$. Therefore the second term is not allowed by the initial condition $\mathcal{Z}(\tau,\bar{\tau}|0)=Z_0(\tau,\bar{\tau})$.\par

For the case $\lambda<0$, the factor $\lim_{\lambda\to 0^-}f_{\text{NP}}=0$ and is thus compatible with the initial condition. There is no restriction on the function $X(\tau,\bar{\tau}|\lambda)$ apart from the flow equation (\ref{eq:flowX}). Therefore we can choose any function $X(\tau,\bar{\tau}|\lambda)$ satisfying (\ref{eq:flowX}). This is the non-perturbative ambiguity.\par

This is another manifestation of the possible break down of unitarity, which is consistent with what we have seen in the spectrum. One interesting point is that similar phenomena can be seen in the string theory side \cite{Giveon:2017nie}. For the two signs of the coupling constant, the deformed geometry is different. For the bad sign, the geometry has certain pathologies, such as naked singularities and closed time-like trajectory. By considering also the $\rJ\overline{\rT}$ deformation, we believe that the ambiguity is related to closed time-like trajectory.

\subsubsection*{Asymptotic density of states and Hagedorn singularity}
In this section, we derive the asymptotic density of states for $\rT\overline{\rT}$ deformed CFTs. We will see the density of states interpolates between the Cardy behavior $\rho(E)\sim \re^{-\sqrt{{E}}}$ and the Hagedorn behavior $\rho(E)\sim \re^{-{E}}$. Our derivation is similar to the derivation of Cardy's formula in CFT.\par

To derive the asymptotic density of states, we apply the S-modular transformation
\begin{align}
\mathcal{Z}(\tau,\bar{\tau}|\lambda)=\mathcal{Z}\left(\left.-\frac{1}{\tau},-\frac{1}{\bar{\tau}}\right|\frac{\lambda}{|\tau|^2}\right).
\end{align}
In addition, we choose the modular parameter $\tau$ to be $\tau=\ri\beta/R$ which implies that
\begin{align}
\tau_1=0,\qquad \tau_2=\frac{\beta}{R}.
\end{align}
In this case, the partition sum simplifies to
\begin{align}
\mathcal{Z}(\lambda)=\sum_{n} \re^{-2\pi\beta\mathcal{E}_n(\lambda)}.
\end{align}
Now let us consider the low temperature limit $\beta\to\infty$. In this limit, the partition sum is dominated by the ground state energy. The ground state energy of the deformed theory is given by $E_0=-c/12$ and $P_0=0$. From the expression of the deformed energy, we obtain
\begin{align}
\mathcal{E}_0(\lambda)=\frac{1}{\pi R\lambda}\left(\sqrt{1-\frac{\pi c\lambda}{6}}-1\right).
\end{align}
Notice that for this formula to make sense, we need to restrict ourselves to the regime $\pi c\lambda<6$. We will make this assumption implicitly in what follows. The partition function in the low-temperature limit is thus given by
\begin{align}
\label{eq:lowTZ}
\lim_{\beta\to\infty}\mathcal{Z}(\lambda)\approx \exp\left[-\frac{2\beta}{R\lambda}\left(\sqrt{1-\frac{\pi c\lambda}{6}}-1\right) \right]
=\exp\left[-\frac{2\beta R}{t}\left(\sqrt{1-\frac{\pi c t}{6R^2}}-1\right) \right]
\end{align}
where in the second equality we have changed from the dimensionless parameter $\lambda$ to the dimensionful parameter $t$. This makes the modular transformation slightly simpler since $t$ does not change under modular transformation. Now we perform the S-modular transformation on (\ref{eq:lowTZ}), which is equivalent to exchange $R$ and $\beta$. Physically, the S-modular transformation maps the low temperature limit to the high temperature limit. In the high energy limit, $\beta\ll 1$ and almost all the state contributes. We can thus define the following spectral density
\begin{align}
\rho(\mathcal{E})=\sum_n\delta(\mathcal{E}-\mathcal{E}_n(\lambda)),
\end{align}
and write the partition function as
\begin{align}
\mathcal{Z}(\lambda)\approx \int_{\mathcal{E}_0(\lambda)}^{\infty}\rho(\mathcal{E})\re^{-2\pi\beta\mathcal{E}}\rd\mathcal{E}.
\end{align}
Identifying this expression with the S-modular transformation of (\ref{eq:lowTZ}), we obtain
\begin{align}
\exp\left[-\frac{2\beta R}{t}\left(\sqrt{1-\frac{\pi c t}{6\beta^2}}-1\right) \right]=\int_{\mathcal{E}_0(\lambda)}^{\infty}\rho(\mathcal{E})\re^{-2\pi\beta\mathcal{E}}\rd\mathcal{E}.
\end{align}
We see that the asymptotic density $\rho(\mathcal{E})$ can be extracted by performing an inverse Laplace transformation
\begin{align}
\rho(\mathcal{E})=\oint \rd\beta\exp\left[\beta\mathcal{E}-\frac{2\beta R}{t}\left(\sqrt{1-\frac{\pi c t}{6\beta^2}}-1\right) \right].
\end{align}
This integral can be computed by saddle-point, which is at
\begin{align}
\beta=\beta_s=\sqrt{\frac{c\pi R}{6\mathcal{E}}}\frac{1+\mathcal{E}t/(2R)}{(1+\mathcal{E}t/(4R))^{1/2}}.
\end{align}
Taking into account the Gaussian fluctuations, we find
\begin{align}
\rho(\mathcal{E})=\frac{2\pi R(2c/3\pi)^{1/4}}{\left[\mathcal{E}(\mathcal{E}t+4R)\right]^{3/4}}\times
\exp\left[\sqrt{\frac{2\pi c R\mathcal{E}}{3}\left(1+\frac{\mathcal{E}t}{4R}\right)}\right]
\end{align}
We can consider two limits,
\begin{itemize}
\item The IR limit $\mathcal{E}t\ll R$, we recover Cardy's formula for CFT
\begin{align}
\rho(\mathcal{E})\approx\mathcal{N}_{\rC}\,\mathcal{E}^{-3/4}\exp\left( \sqrt{\frac{2c\pi R\mathcal{E}}{3}}\right).
\end{align}
\item The UV limit $\mathcal{E}t\gg R$, we find the Hagedorn behavior
\begin{align}
\rho(\mathcal{E})\approx\mathcal{N}_{\rH}\,\mathcal{E}^{-3/2}\exp\left(\sqrt{\frac{\pi c t}{6}}\mathcal{E} \right).
\end{align}
We can identify the Hagedorn temperature to be
\begin{align}
\beta_{\rH}=\sqrt{\frac{\pi c t}{6}}.
\end{align}
\end{itemize}
The Hagedorn behavior in the UV implies that the partition function has a singularity near $\beta_{\rH}$. The scaling of the partition function near that point can be found by
\begin{align}
\mathcal{Z}(\beta,\lambda)\approx\int_{\mathcal{E}_0(\lambda)}^{\infty}\rd\mathcal{E}\mathcal{E}^{-3/2}\exp\left[(\beta_{\rH}-\beta)\mathcal{E}\right]
\approx (\beta-\beta_{\rH})^{-5/2}\,\Gamma\left(\frac{5}{2},(\beta-\beta_{\rH})\mathcal{E}_0(\lambda) \right),
\end{align}
where $\Gamma(a,b)$ is the incomplete Gamma function. It is regular at $b=0$. Therefore we see the nature of the singularity is a branch point. The Hagedorn growth of density of states is not the usual behavior of a local QFT, but rather the behavior of non-local theories such as the little string theory.

\subsection{Uniqueness of $\rJ\overline{\rT}$ deformation}
As we mentioned in the previous section, for theories with an $U(1)$ holomorphic current, one can define an irrelevant solvable deformation called the $\rJ\overline{\rT}$ deformation \cite{Guica:2017lia,Chakraborty:2018vja}. This deformation is triggered by the irrelevant $\rJ\overline{\rT}$ operator defined by
\begin{align}
\rJ\overline{\rT}=J\bar{T}-\bar{J}\Theta,
\end{align}
where $\bar{T}$ and $\Theta$ are the components of stress energy tensor which appeared before in the definition of $\rT\overline{\rT}$ before. $J$ and $\bar{J}$ are components of the $U(1)$ current $J_{\mu}$
\begin{align}
J=J_z,\qquad\bar{J}=J_{\bar{z}}.
\end{align}
For purely holomorphic current, we have $\bar{J}=0$. A factorization formula for the $\rJ\overline{\rT}$ can be proved in the same way as the $\rT\overline{\rT}$ operator, which we discussed in detail in the previous sections.
Putting the theory on an infinite cylinder, both the energy and the $U(1)$ charge are deformed and the deformed quantities can be determined explicitly, similar to the $\rT\overline{\rT}$ case. In this section, we show that the deformed spectrum can also be determined from modular properties of the torus partition sum. The idea is similar to the discussion in the previous subsection for the $\rT\overline{\rT}$ deformation, but it is technically more involved.

\subsubsection*{The set-up}
Let us first give the set-up. For a CFT with $U(1)$ current, we define the \emph{charged} torus partition sum
\begin{align}
Z(\tau,\bar{\tau},\nu)=\sum_n \re^{2\pi \ri\tau_1 R P_n-2\pi\tau_2 R E_n+2\pi \ri\nu Q_n},
\end{align}
where $\nu$ is the chemical potential. We denote the common eigenstate of operators ${P},{H},{Q}$ as $|n\rangle$ and
\begin{align}
{P}|n\rangle=P_n|n\rangle,\qquad {H}|n\rangle=E_n|n\rangle,\qquad {Q}|n\rangle=Q_n|n\rangle,
\end{align}
where for CFT we have
\begin{align}
{P}=\frac{1}{R}\left(L_0-\bar{L}_0\right),\qquad {H}=\frac{1}{R}\left(L_0+\bar{L}_0-\frac{c}{12}\right),\qquad {Q}=J_0.
\end{align}
Here $J_0$ is the time component of the $U(1)$ current.
The charged partition function transforms as a \emph{Jacobi form} of weight $(0,0)$ and \emph{index} $k$
\begin{align}
\label{eq:Ztrans}
Z\left(\frac{a\tau+b}{c\tau+d},\frac{a\bar{\tau}+b}{c\bar{\tau}+d},\frac{\nu}{c\tau+d}\right)=\exp\left(\frac{\pi \ri k c\nu^2}{c\tau+d}\right)Z(\tau,\bar{\tau},\nu),
\end{align}
with $a,b,c,d\in\mathbb{Z}$ and $ad-bc=1$. Here $k$ is the level of the $U(1)$ affine Lie algebra
\begin{align}
\label{eq:normJ}
[J_m,J_n] = k \, m \, \delta_{m+n,0}\, .
\end{align}
Note that (\ref{eq:Ztrans}) implies that the chemical potential $\nu$ transforms as a modular form of weight $(-1,0)$. This is due to the fact that it couples to a holomorphic current of dimension $(1,0)$ (see \cite[\textsection \,3.1]{Dijkgraaf:1996iy} for a discussion).

\subsubsection*{Deformations and perturbative expansion}
Now we consider a solvable deformation of the following form
\begin{align}
\label{eq:JTdeform}
{P}\mapsto {P},\qquad  {H}\mapsto \mathcal{H}({P},{H},{Q},\hat{\mu}),\qquad
{Q}\mapsto\mathcal{Q}({P},{H},{Q},\hat{\mu})
\end{align}
where $\mathcal{H}(x,y,z,\hat{\mu})$ and $\mathcal{Q}(x,y,z,\hat{\mu})$ are arbitrary smooth functions for the moment. The spectrum is deformed in a universal way, we denote
\begin{align}
\mathcal{E}_n(\hat{\mu})=\mathcal{H}(P_n,E_n,Q_n,\hat{\mu}),\qquad \mathcal{Q}_n(\hat{\mu})=\mathcal{Q}(P_n,E_n,Q_n,\hat{\mu}).
\end{align}
The deformed partition function is defined as
\begin{align}
\label{eq:JtbarZ}
\mathcal{Z}(\tau,\bar{\tau},\nu|\hat{\mu})=\sum_n \re^{2\pi \ri\tau_1 R P_n-2\pi\tau_2 R \mathcal{E}_n(\hat{\mu})+2\pi \ri\nu \mathcal{Q}_n(\hat{\mu})}.
\end{align}
We assume that $\mathcal{E}_n(\hat{\mu})$ and $\mathcal{Q}_n(\hat{\mu})$ have regular Taylor expansion
\begin{align}
\label{eq:expEQ}
\mathcal{E}_n(\hat{\mu})=\sum_{k=0}^{\infty}\rE_n^{(k)}(P_n,E_n,Q_n)\hat{\mu}^k,\qquad
\mathcal{Q}_n(\hat{\mu})=\sum_{k=0}^{\infty}\rQ_n^{(k)}(P_n,E_n,Q_n)\hat{\mu}^k.
\end{align}
Similar to the $\rT\overline{\rT}$ deformation, we require the deformed partition sum to be modular covariant
\begin{align}
	\label{eq:modularZJ}
	\mathcal{Z}\left(\left.\frac{a\tau+b}{c\tau+d},\frac{a\bar{\tau}+b}{c\bar{\tau}+d},\frac{\nu}{c\tau+d}\right|\frac{\hat{\mu}}{c\bar{\tau}+d}\right)
	=\exp\left(\frac{\ri\pi k  c\nu^2}{c\tau+d}\right)\mathcal{Z}(\tau,\bar{\tau},\nu|\hat{\mu}).
\end{align}
Plugging (\ref{eq:expEQ}) into (\ref{eq:JtbarZ}), we can perform the expansion of the deformed partition function
\begin{align}
\mathcal{Z}(\tau,\bar{\tau},\nu|\hat{\mu})=\sum_{p=0}^{\infty}Z_p(\tau,\bar{\tau},\nu)\hat{\mu}^p.
\end{align}
Modular covariance of the deformed partition sum, (\ref{eq:modularZJ}), implies that $Z_p$ transforms as a non-holomorphic Jacobi form of weight $(0,p)$ and index $k$,
\begin{align}\label{eq:formppp}
	Z_p\left(\frac{a\tau+b}{c\tau+d},\frac{a\bar{\tau}+b}{c\bar{\tau}+d},\frac{\nu}{c\tau+b}\right)=(c\bar{\tau}+d)^p\,\exp\left(\frac{\ri \pi  k c\nu^2}{c\tau+d}\right)\,Z_p(\tau,\bar{\tau},\nu).
\end{align}
The first few orders in the $\hat{\mu}$ expansion are given by
\begin{align}
	\label{eq:expZp}
	Z_p=\sum_{n}f_n^{(p)}\re^{2\pi \ri\tau_1 R P_n-2\pi \tau_2 R E_n+2\pi \ri\nu Q_n},
\end{align}
where
\begin{align}
	\label{eq:expfn}
	f_n^{(1)}=&\,(-2\pi R\rE_n^{(1)})\tau_2+2\ri\pi\nu\rQ_n^{(1)},\\\nonumber
	f_n^{(2)}=&\,\frac{1}{2!}\left(-2\pi R\rE_n^{(1)}\right)^2\tau_2^2-2\pi R\left[\rE_n^{(2)}+2\ri\pi\nu\rE_n^{(1)}\rQ_n^{(1)}\right]\tau_2
	-2\left[\pi^2\nu^2(\rQ_n^{(1)})^2-\ri\pi\nu\rQ_n^{(2)}\right],\\\nonumber
	f_n^{(3)}=&\,\frac{1}{3!}\left(-2\pi R\rE_n^{(1)}\right)^3\tau_2^3+4\pi^2R^2\left[\rE_n^{(1)}\rE_n^{(2)}+i\pi\nu(\rE_n^{(1)})^2\rQ_n^{(1)}\right]\tau_2^2\\\nonumber
	&\,+\left[2R\pi^2\nu^2\rE_n^{(1)}(\rQ_n^{(1)})^2-2\pi \ri R\nu(\rE_n^{(1)}\rQ_n^{(2)}+\rE_n^{(2)}\rQ_n^{(1)})-2\pi R\rE_n^{(3)}\right]\tau_2\\\nonumber
	&\,-\frac{4}{3}\ri\pi^3\nu^3(\rQ_n^{(1)})^3-4\pi^2\nu^2\rQ_n^{(1)}\rQ_n^{(2)}+2\pi \ri\nu\rQ_n^{(3)}.
\end{align}
We can write $Z_p$ as a differential operator made of $\partial_{\tau},\partial_{\bar{\tau}},\partial_{\nu}$ acting on $Z_0$. This is done by replacing  $\rE_n^{(k)}(E_n,P_n,Q_n)$, $\rQ_n^{(k)}(E_n,P_n,Q_n)$ in (\ref{eq:expfn})  by differential operators with the replacement rules
\begin{align}
	E_n\mapsto -\frac{1}{2\pi R}\partial_{\tau_2},\qquad P_n\mapsto \frac{1}{2\pi \ri R}\partial_{\tau_1},\qquad Q_n\mapsto\frac{1}{2\pi \ri}\partial_{\nu}.
\end{align}
The above procedure leads to a double expansion of $Z_p$ in powers of $\tau_2$ and $\nu$,
\begin{align}
	\label{eq:expandzp}
	Z_p=\sum_{l,m}\tau_2^l\nu^m {\mathcal{O}}_{lm}^{(p)}(\partial_{\tau},\partial_{\bar{\tau}},\partial_\nu)Z_0,
\end{align}
where the sum runs over the range $l,m=0,1,\cdots, p$; $0<l+m\le p$. As before, our strategy is to first prove uniqueness using induction and then give a way to construct the partition function which proves existence.

\subsubsection*{Uniqueness}
It is clear that the differential operator $ {\mathcal{O}}_{lm}^{(p)}(\partial_{\tau},\partial_{\bar{\tau}},\partial_\nu)$ for fixed $p$ is only sensitive to the energy and charge shifts $\rE_n^{(k)}$,  $\rQ_n^{(k)}$ with $k=1,2,\cdots, p$. Conversely, if we know all $ {\mathcal{O}}_{lm}^{(p)}$ with given $p$, we can determine all the energy and charge shifts with $k\le p$ by using (\ref{eq:expZp}) -- (\ref{eq:expandzp}).

We can use the expansion  (\ref{eq:expandzp}) to prove that if $Z_1,\cdots, Z_p$ have been determined, $Z_{p+1}$ can be determined uniquely as well. As in the previous section, we start by considering the first step in this process.  Equation (\ref{eq:expandzp}) (with $p=1$) takes the form
\begin{align}\label{eq:zone}
	Z_1=\left(\tau_2\widehat{O}^{(1)}_{1,0}(\partial_{\tau},\partial_{\bar{\tau}},\partial_\nu)
	+\nu\widehat{O}^{(1)}_{0,1}(\partial_{\tau},\partial_{\bar{\tau}},\partial_\nu)\right)Z_0.
\end{align}
We are looking for differential operators $\widehat{O}^{(1)}_{1,0}$, $\widehat{O}^{(1)}_{0,1}$, for which $Z_1$ transforms as a Jacobi form of weight $(0,1)$ and index $k$, for any $Z_0$ of weight $(0,0)$ and index $k$. To find them, we need a new covariant derivative with respect to the chemical potential $\nu$ in addition to the Maass-Shimura derivative. The new covariant derivative is given by
\begin{align}
	\label{eq:Dnu}
	\textsf{D}_\nu\equiv\partial_{\nu}+\frac{\pi k \nu}{\tau_2}.
\end{align}
Acting with $\mathsf{D}_\nu$ on a Jacobi form of weight $(r,\bar r)$ and index $k$ gives a Jacobi form of weight $(r+1,\bar r)$ and index $k$. The proof is basically the same as we did for the Maass-Shimura derivatives.

Using these covariant derivatives, it is straightforward to find a combination of the form (\ref{eq:zone}) that has the correct modular transformation properties,
\begin{align}\label{eq:first-ord}
	Z_1=\alpha\tau_2\textsf{D}_\nu\partial_{\bar{\tau}}Z_0.
\end{align}
Here $\alpha$ is a constant that can be absorbed in the definition of $\hat{\mu}$. We will set it to one below. It is not hard to check that (\ref{eq:first-ord})  is the unique object of the form (\ref{eq:zone}) with the correct modular transformation properties.

We are now ready to move on to the general induction step. We assume that $Z_1,\cdots, Z_p$ (with $p\ge 1$) have been determined, and want to show that $Z_{p+1}$ can be determined as well.

We saw before that from the form of $Z_1,\cdots, Z_p$ we can read off the energy and charge shifts $\rE_n^{(k)}$,  $\rQ_n^{(k)}$ with $k=1,2,\cdots, p$. Consider now the expansion (\ref{eq:expandzp}) of $Z_{p+1}$. Most of the terms in that expansion involve the energy and charge shifts with $k\le p$, which are assumed to be already known. There are only two terms in the sum, corresponding to $(l,m)=(1,0)$ and $(0,1)$, that involve the unknowns $\rE_n^{(p+1)}$,  $\rQ_n^{(p+1)}$.

To show that there is no more than one solution for the expansion (\ref{eq:expandzp}), suppose there were two different ones. Subtracting them, and using the fact that most terms in the expansion (\ref{eq:expandzp}) cancel between the two, we find that there must exist differential operators $\delta\widehat{O}^{(p+1)}_{1,0}(\partial_{\tau},\partial_{\bar{\tau}},\partial_\nu)$, $\delta\widehat{ O}^{(p+1)}_{0,1}(\partial_{\tau},\partial_{\bar{\tau}},\partial_\nu)$, such that
\begin{align}
	\left(\tau_2\,\delta\widehat{O}^{(p+1)}_{1,0}(\partial_\tau,\partial_{\bar{\tau}},\partial_\nu)
	+\nu\,\delta\widehat{ O}^{(p+1)}_{0,1}(\partial_\tau,\partial_{\bar{\tau}},\partial_\nu)\right)Z_0
\end{align}
is a Jacobi form of weight $(0,p+1)$ and index $k$, for any $Z_0$ which is a Jacobi form of weight (0,0) and index $k$. The fact that such differential operators do not exist (for $p>0$) can be proven by using the properties of the covariant derivatives in a similar way to the proof for the $\rT\overline{\rT}$ case. Since we have discussed this point in detail in the previous section we will not repeat it here.

\subsubsection*{Existence}
We have shown that if a modular covariant deformation of the type (\ref{eq:JTdeform}) exist, then it must be unique. Now we want to give a way to construct the deformed partition function and find the deformed spectrum. The idea is again very similar to the $\rT\overline{\rT}$ case. We first use the covariant ansatz approach to compute the first few orders and then give a recursion relation between different orders in $\hat{\mu}$ expansion. From the recursion relation, we can derive a flow equation for the partition function. From the latter, we can write down the flow equation for the deformed charges similar to the Burgers' equation. Finally we solve the equations to find the deformed spectrum and find a perfect match with the results obtained from other approaches.

\paragraph{Covariant ansatz} In order to find $Z_p$, we write down an ansatz with the desired modular properties, (\ref{eq:formppp}), and require it to be consistent with the general structure of the perturbative expansion, (\ref{eq:expandzp}). The leading term in $\tau_2$  is fixed by (\ref{eq:expandzp}), (\ref{eq:first-ord}), to be
\begin{align}
	Z_p=\frac{\tau_2^p}{p!}\mathscr{D}_{\nu}^{(p)}\mathscr{D}_{\bar{\tau}}^{(p)}Z_0+\cdots,
\end{align}
where
\begin{align}\label{eq:curly-D-def}
	\mathscr{D}_{\nu}^{(j)}\equiv\textsf{D}_{\nu}^j,\qquad \mathscr{D}_{\bar{\tau}}^{(j)}\equiv\prod_{m=0}^{j-1}\textsf{D}_{\bar{\tau}}^{(2m)}.
\end{align}
The other terms have lower powers of $\tau_2$ and can be written in terms of $\mathscr{D}_{\nu}^{(i)}$, $\mathscr{D}_{\bar{\tau}}^{(j)}$ with $0\le i,j\le p$. A term of the form $\mathscr{D}_{\nu}^{(i)}\mathscr{D}_{\bar{\tau}}^{(j)}Z_0$ with particular $i,j$ is multiplied by $\tau_2^a\nu^b$, such that its contribution to $Z_p$ transforms as a Jacobi form of weight $(0,p)$ and index $k$, (\ref{eq:formppp}). Since $\tau_2^a\nu^b\mathscr{D}_{\nu}^{(i)}\mathscr{D}_{\bar{\tau}}^{(j)}$ has weight
\begin{align}
	(i-a-b,2j-a),
\end{align}
we have the constraint
\begin{align}
	i-a-b=0,\qquad 2j-a=p.
\end{align}
The indices $i,j$ satisfy the constraits $0\le i,j\le p$ and $0\le b\le p$. This leads to
\begin{align}
	0\le p+i-2j\le p.
\end{align}
In addition, there are no terms with $a=b=0$. Taking into account these constraints, we can write down the ansatz for any $p$. The first few $Z_p$ take the form
\begin{align}
	Z_1=&\,a_1\,\tau_2\mathscr{D}_{\nu}^{(1)}\mathscr{D}_{\bar{\tau}}^{(1)}Z_0,\\\nonumber
	Z_2=&\,\left(b_4\,\tau_2^2\mathscr{D}_{\nu}^{(2)}\mathscr{D}_{\bar{\tau}}^{(2)}+b_3\,\nu^2\mathscr{D}_{\nu}^{(2)}\mathscr{D}_{\bar{\tau}}^{(1)}  +b_2\,\nu\mathscr{D}_{\nu}^{(1)}\mathscr{D}_{\bar{\tau}}^{(1)}+b_1\,\mathscr{D}_{\bar{\tau}}^{(1)}\right)Z_0,\\\nonumber
	Z_3=&\,\left(c_7\,\tau_2^3\mathscr{D}_{\nu}^{(3)}\mathscr{D}_{\bar{\tau}}^{(3)}+c_6\,\tau_2\nu^2\mathscr{D}_\nu^{(3)}\mathscr{D}_{\bar{\tau}}^{(2)}  +c_5\,\tau_2\nu\mathscr{D}_\nu^{(2)}\mathscr{D}_{\bar{\tau}}^{(2)}+c_4\,\tau_2\mathscr{D}_{\nu}^{(1)}\mathscr{D}_{\bar{\tau}}^{(2)}\right)Z_0\\\nonumber
	&\,+\frac{1}{\tau_2}\left(c_3\,\nu^3\mathscr{D}_{\nu}^{(2)}\mathscr{D}_{\bar{\tau}}^{(1)}+c_2\,\nu^2\mathscr{D}_{\nu}^{(1)}\mathscr{D}_{\bar{\tau}}^{(1)}  +c_1\,\nu\mathscr{D}_{\bar{\tau}}^{(1)}\right)Z_0
\end{align}
To fix the constants $a_k, b_k, c_k$ we impose the conditions which stem from the structure of the perturbative expansion. To be more explicit, we first expand the covariant derivatives $\mathscr{D}_{\nu}^{(j)}$ and $\mathscr{D}_{\bar{\tau}}^{(j)}$ in terms of $\partial_{\nu}$ and $\partial_{\bar{\tau}}$ in the ansatz. Comparing with the structure of the perturbative expansion, we impose the following conditions
\begin{itemize}
	\item The coefficient of $\tau_2^p\partial_{\nu}^{p}\partial_{\bar{\tau}}^{p}$ is fixed to be $1/p!$;
	\item The coefficients of the terms without $\tau_2$ and $\nu$, namely $\partial_{\nu}^n\partial_{\bar{\tau}}^m$ are zero;
	\item The coefficients of terms with negative powers of $\tau_2$, i.e. terms of the form $\tau_2^{-n}\nu^m\partial_{\nu}^{i}\partial_{\bar{\tau}}^{j}$ with $n>0$, vanish.
\end{itemize}
We find that these conditions are powerful enough to fix $Z_p$ completely at any given order. 
The solutions for the first few orders are given by
\begin{align}
	Z_1=&\,\left(\tau_2\mathscr{D}_{\nu}^{(1)}\mathscr{D}_{\bar{\tau}}^{(1)}\right)Z_0,\\\nonumber
	Z_2=&\,\left(\frac{1}{2}\tau_2^2\mathscr{D}_{\nu}^{(2)}\mathscr{D}_{\bar{\tau}}^{(2)} -\frac{\ri\pi}{2}\nu\mathscr{D}_{\nu}^{(1)}\mathscr{D}_{\bar{\tau}}^{(1)}-\frac{\ri\pi}{2}\mathscr{D}_{\bar{\tau}}^{(1)}\right)Z_0,\\\nonumber
	Z_3=&\,\left(\frac{1}{6}\tau_2^3\mathscr{D}_{\nu}^{(3)}\mathscr{D}_{\bar{\tau}}^{(3)}  -\frac{\ri\pi}{2}\tau_2\nu\mathscr{D}_\nu^{(2)}\mathscr{D}_{\bar{\tau}}^{(2)}-\frac{3\ri\pi}{4}\tau_2\mathscr{D}_{\nu}^{(1)}\mathscr{D}_{\bar{\tau}}^{(2)}
	-\frac{\pi^2}{4\tau_2}\nu^2\mathscr{D}_{\nu}^{(1)}\mathscr{D}_{\bar{\tau}}^{(1)}  -\frac{\pi^2}{2\tau_2}\nu\mathscr{D}_{\bar{\tau}}^{(1)}\right)Z_0.
\end{align}
This approach enables us to find explicitly $Z_p$ to any order in principle.
\paragraph{Recursion relation} In practice, it is more convenient to write down a recursion relation. It turns out that such a relation exists, but it is more complicated than the $\rT\overline{\rT}$ case. In particular, it relates $Z_p$ to all $Z_j$ with $0\le j<p$. It takes the form
\begin{align}
	\label{eq:vvv}
	Z_{p}=\frac{\tau_2}{p}\left[\textsf{D}_\nu\textsf{D}_{\bar{\tau}}^{(p-1)}-\frac{\ri\pi k \nu(p-1)}{2\tau_2^2}\right]Z_{p-1}
	-\frac{\ri\pi k}{2p}\sum_{j=0}^{p-2}\left(\frac{\pi\nu k }{2\ri\tau_2}\right)^j\textsf{D}_{\bar{\tau}}^{(p-j-2)}Z_{p-j-2}~.
\end{align}
We see that $Z_p$ depends on all of the lower order quantities $Z_1,\cdots,Z_{p-1}$ while in the $\rT\overline{\rT}$ case, $Z_p$ only depends on $Z_{p-1}$. Nevertheless, the recursion relation (\ref{eq:vvv}) can also be phrased as a differential equation for the partition sum.
Namely, if the partition sum $\mathcal{Z}(\tau,\bar{\tau},\nu|\hat{\mu})$ satisfies
\begin{align}
	\label{eq:david}
	\left(1+\frac{\ri\pi k \hat{\mu}\nu}{2\tau_2}\right)\partial_{\hat{\mu}}\mathcal{Z}=\tau_2\textsf{D}_\nu\mathcal{D}_{\bar{\tau}}\mathcal{Z}
	-\frac{\ri\pi k \hat{\mu}}{2}\frac{1}{1+\frac{\ri\pi k \hat{\mu}\nu}{2\tau_2}}\mathcal{D}_{\bar{\tau}}\mathcal{Z},
\end{align}
where
\begin{align}
	\label{eq:amit}
	\mathcal{D}_{\bar{\tau}}\equiv\partial_{\bar{\tau}}+{\ri\over 2\tau_2}\hat{\mu}\partial_{\hat{\mu}},
\end{align}
then expanding this equation in a power series in $\hat{\mu}$ reproduces (\ref{eq:vvv}). For the $\rT\overline{\rT}$ case, the flow equation for the torus partition sum can also be derived from a description with a dynamical metric \cite{Cardy:2018sdv,Dubovsky:2018bmo}. It would be interesting to derive (\ref{eq:david}) from a similar point of view, by including a dynamical gauge field as well.

\paragraph{Deformed spectrum} From the equation (\ref{eq:david}) we can read off a system of differential equations that describes the evolution of the energies and charges of states with the coupling $\hat{\mu}$. To do that, we plug the the partition sum (\ref{eq:JtbarZ}) into (\ref{eq:david}), and compare the coefficients of a given exponential on the left and right hand sides.
We also multiply by the factor $(1+\ri\pi k \hat{\mu}\nu/(2\tau_2))$ on both sides. The resulting equation then takes the form
\begin{align}
\mathcal{Y}_0 + \mathcal{Y}_1 \nu + \mathcal{Y}_2 \nu^2 =0.
\end{align}
Here, $\mathcal{Y}_i$ are functions containing $\mathcal{E}_n(\hat{\mu}),\,\mathcal{Q}_n(\hat{\mu}),\,   {P}_n$ and the derivatives $\mathcal{E}'_n(\hat{\mu}),\,\mathcal{Q}'_n(\hat{\mu})$. Since this should hold for all values of $\nu$, we have $\mathcal{Y}_{0,1,2}=0$. The equations $\mathcal{Y}_1- \mathcal{Y}_2=0$ and  $\mathcal{Y}_2=0$ respectively yield
\begin{align}\label{eq:burr}
&\mathbb{E}'_n(\hat{\mu})\left[1+\pi\hat{\mu}\mathcal{Q}_n(\hat{\mu})\right]=\pi\left[\mathbb{P}_n-\mathbb{E}_n(\hat{\mu})\right]\mathcal{Q}_n(\hat{\mu}),\\\nonumber
&\mathcal{Q}'_n(\hat{\mu})\left[1+\pi\hat{\mu}\mathcal{Q}_n(\hat{\mu})\right]=
\frac{\pi {k}}{2}\left[\mathbb{P}_n-\mathbb{E}_n(\hat{\mu})\right],
\end{align}
where $\mathbb{E}_n(\hat{\mu})=R\mathcal{E}_n(\hat{\mu})$, and $\mathbb{P}_n=RP_n$ is the quantized momentum. The equation $\mathcal{Y}_0=0$ gives rise to a equation which is consistent with the above two.

Dividing the two equations in (\ref{eq:burr}), one finds that
\begin{align}\label{eq:coset}
k\mathbb{E}_n(\hat{\mu})-\mathcal{Q}_n(\hat{\mu})^2={\rm independent\;of\;\hat{\mu},}
\end{align}
which reproduces one of the results of \cite{Chakraborty:2018vja}.

Equations (\ref{eq:burr}) can be expressed in a form that is closer to Burgers' equation by writing them in terms of the dimensionful coupling $\mu=\hat{\mu} R$, and using the fact that the dimensionless energies $\mathbb{E}_n$ depend only on the dimensionless coupling $\hat{\mu}$.

The resulting system of equations can be written as\footnote{To reproduce the equations given in \cite{Chakraborty:2018vja}, we need to make the replacement $\hat{\mu}=\mu/(2\pi R)$.}
\begin{align}\label{eq:ttburg}
	&\frac{\partial}{\partial \mu} (\mathcal{E}_n-P_n)=\,\pi\mathcal{Q}_n\frac{\partial}{\partial R} (\mathcal{E}_n-P_n),\\\nonumber
	&\frac{\partial \mathcal{Q}_n}{\partial \mu}=\,\pi\mathcal{Q}_n\frac{\partial \mathcal{Q}_n}{\partial R}-\frac{\pi {k}}{2}(\mathcal{E}_n-P_n).
\end{align}
The differential equation on the second line of (\ref{eq:ttburg}) looks like the inviscid Burgers' equation with a time-dependent source, where the coupling $\mu$ plays the role of time. The dynamics of this source is described by the first line of (\ref{eq:ttburg}).

The solution of (\ref{eq:ttburg}) with the boundary conditions $\mathcal{E}_n(0)=E_n$ and $\mathcal{Q}_n(0)=Q_n$ is given by
\begin{align}\label{eq:enqn}
	\mathcal{E}_n^{(+)}(\hat{\mu})=&\,-\frac{2}{\pi^2\hat{\mu}^2 {k}R}\sqrt{(1+\pi Q_n\hat{\mu})^2+\pi^2\hat{\mu}^2 {k}R(P_n-E_n)}\\\nonumber
	&\,+\frac{1}{\pi^2\hat{\mu}^2{k} R}\left(2+2\pi Q_n\hat{\mu}+\pi^2\hat{\mu}^2 {k} P_n R\right),\\\nonumber
	\mathcal{Q}_n^{(+)}(\hat{\mu})=&\,\frac{1}{\pi\hat{\mu}}\sqrt{(1+\pi Q_n\hat{\mu})^2+\pi^2\hat{\mu}^2{k}R(P_n-E_n)}-\frac{1}{\pi\hat{\mu}},
\end{align}
where we took the positive branch of the square root, so that
\begin{align}\label{eq:limeq}
\lim_{\hat{\mu}\to 0} \mathcal{E}_n^{(+)}(\hat{\mu})=E_n,\qquad \lim_{\hat{\mu}\to 0} \mathcal{Q}_n^{(+)}(\hat{\mu})=Q_n.
\end{align}
This spectrum indeed matches with the one of the $\rJ\overline{\rT}$ deformed CFT.\par

Let us notice that the $\rJ\overline{\rT}$ deformed spectrum does not depend on the sign of the deformation parameter $\hat{\mu}$. This is different from the case of the $\rT\overline{\rT}$ deformed spectrum. The spectrum for both signs of $\hat{\mu}$ behaves like the bad sign of the $\rT\overline{\rT}$ deformed spectrum. Namely, the spectrum become complex for high enough energies of the original theory.

\subsubsection*{Non-perturbative aspects}
Now we consider some non-perturbative features of the partition function using the flow equation (\ref{eq:david}). As one might expect, since the $\rJ\overline{\rT}$ deformed spectrum behaves like the bad sign of the $\rT\overline{\rT}$ deformation, there are non-perturbative ambiguities in the solution of the flow equation.\par

A simple way to investigate them is to consider the contribution to the partition sum of states for which we take the negative branch of the square root in (\ref{eq:enqn}). The two branches are related by
\begin{align}
	\mathcal{E}_n^{(+)}+\mathcal{E}_n^{(-)}=&\,\frac{2}{\pi^2\hat{\mu}^2 {k} R}\left(2+2\pi Q_n\hat{\mu}+\pi^2\hat{\mu}^2 {k} P_n R\right),\\\nonumber
	\mathcal{Q}_n^{(+)}+\mathcal{Q}_n^{(-)}=&\,-\frac{2}{\pi\hat{\mu}}.
\end{align}
While $\mathcal{E}_n^{(+)}$, $\mathcal{Q}_n^{(+)}$ approach finite limits as $\hat{\mu}\to 0$, (\ref{eq:limeq}), $\mathcal{E}_n^{(-)}$, $\mathcal{Q}_n^{(-)}$ diverge in this limit,
\begin{align}
\label{eq:limeqneg}
\mathcal{E}_n^{(-)}(\hat{\mu})\simeq \frac{4}{\pi^2\hat{\mu}^2 {k} R},\qquad \mathcal{Q}_n^{(-)}(\hat{\mu})\simeq -\frac{2}{\pi\hat{\mu}}.
\end{align}
The fact that the energy $\mathcal{E}_n^{(-)}$ goes to $+\infty$ in the limit, implies that states with these energies give non-perturbative contributions to the partition sum, which satisfy the correct boundary conditions $\lim_{\hat{\mu}\to0} \mathcal{Z}(\tau,\bar{\tau},\nu|\hat{\mu})=Z_0(\tau,\bar{\tau},\nu)$, as in the $\rT\overline{\rT}$ case with $t<0$.

One way to find consistent non-perturbative contributions is then to assume that we have some extra states in our theory labeled by ${\tilde n}$, whose energies and charges are given by ${\mathcal{E}}_{\tilde n}^{(-)}$ and ${\mathcal{Q}}_{\tilde n}^{(-)}$. These states can be the negative branch energies and charges of some other $\rJ\overline{\rT}$ deformed CFT, that a priori need not have anything to do with the one that gives the perturbative contributions discussed above. We find
\begin{align}\label{eq:zznnpp}
	\mathcal{Z}_{\text{np}}=&\,\sum_{\tilde n} \re^{2\pi \ri\tau_1RP_{\tilde n}-2\pi\tau_2 R {\mathcal{E}}_{\tilde n}^{(-)}+2\pi \ri\nu {\mathcal{Q}}_{\tilde n}^{(-)}}\\\nonumber
	=&\,\re^{-\frac{8\tau_2}{\pi {k}\hat{\mu}^2}-\frac{4\ri\nu}{\hat{\mu}}}\sum_{\tilde n} \re^{2\pi \ri\tau_1RP_{\tilde n}+2\pi\tau_2 R {\mathcal{E}}_{\tilde n}^{(+)}-2\pi \ri\nu {\mathcal{Q}}_{\tilde n}^{(+)}-8\tau_2 Q_{\tilde n}/\hat{\mu}-4\pi R\tau_2 P_{\tilde n}}.
\end{align}
Using the relation
\begin{align}
	Q_n=\frac{\pi\hat{\mu} {k} R}{2}\left({\mathcal{E}}_n^{(\pm)}(\hat{\mu})-P_n\right)+{\mathcal{Q}}_n^{(\pm)}(\hat{\mu}),
\end{align}
satisfied by both branches of (\ref{eq:enqn}), we can rewrite (\ref{eq:zznnpp}) as
\begin{align}
	\label{eq:Znp2}
	\mathcal{Z}_{\text{np}}=\re^{\frac{\pi{k}\tilde{\nu}^2}{2\tau_2}-\frac{\pi{k}\nu^2}{2\tau_2}}\sum_{\tilde n} \re^{2\pi i\tau_1RP_{\tilde n}-2\pi\tau_2 R{\mathcal{E}}_{\tilde n}^{(+)}+2\pi \ri\tilde{\nu}{\mathcal{Q}}_{\tilde n}^{(+)}},
\end{align}
where the shifted chemical potential is given by
\begin{align}\label{eq:nnt}
	\tilde{\nu}=-\nu+\frac{4 \ri\tau_2}{\pi{k}\hat{\mu}}~.
\end{align}
By construction, the partition sum (\ref{eq:Znp2}) must be modular invariant (since the original expression (\ref{eq:zznnpp}) is). This can be shown directly as follows. The prefactors in (\ref{eq:Znp2}) transform as
\begin{align}
	\label{eq:two-phases}
	\re^{-\frac{\pi{k}\nu^2}{2\tau_2}}\mapsto \re^{-\frac{\pi {k}\nu^2}{2\tau_2}}\times \re^{\frac{\ri c{k}\pi\nu^2}{c\tau+d}},\qquad
	\re^{+\frac{\pi{k}\tilde{\nu}^2}{2\tau_2}}\mapsto \re^{+\frac{\pi{k}\tilde{\nu}^2}{2\tau_2}}\times \re^{-\frac{\ri c\pi{k}\tilde{\nu}^2}{c\tau+d}}.
\end{align}
The partition sum on the right-hand side of (\ref{eq:Znp2}) transforms as
\begin{align}
	\sum_{\tilde n} \re^{2\pi \ri\tau_1P_{\tilde n}-2\pi\tau_2 R{\mathcal{E}}_{\tilde n}^{(+)}+2\pi \ri\tilde{\nu}{\mathcal{Q}}_{\tilde n}^{(+)}}
	\mapsto \re^{\frac{\ri c{k}\pi\tilde{\nu}^2}{c\tau+d}}
	\sum_{\tilde n} \re^{2\pi \ri\tau_1P_{\tilde n}-2\pi\tau_2 R{\mathcal{E}}_{\tilde n}^{(+)}+2\pi \ri\tilde{\nu}{\mathcal{Q}}_{\tilde n}^{(+)}}.
\end{align}
Combining these transformations, we see that $\mathcal{Z}_{\text{np}}$ (\ref{eq:Znp2}) indeed transforms as a Jacobi form,
(\ref{eq:modularZJ}).\par

The fact that $\mathcal{Z}_{\text{np}}$ is a non-perturbative contribution to the partition sum is due to the behavior as $\hat{\mu}\to 0$ of the prefactor on the right-hand side of (\ref{eq:Znp2}). The leading behavior of the partition sum in this limit is $\mathcal{Z}_{\text{np}}\sim \re^{-\frac{8\tau_2}{\pi k\hat{\mu}^2}}\tilde Z_0$, which is exponentially small for both signs of $\hat{\mu}$, as expected. Thus, we see that the non-perturbative completion of the partition sum of $\rJ\overline{\rT}$ deformed CFT has a similar ambiguity to that for a $\rT\overline{\rT}$ deformed CFT.

\section{Gravity and holography}
\label{lecture3}
In this section, we present several different perspectives of $\rT\overline{\rT}$ deformation. In particular, its relations to 2d gravity theories and holography. We first present Cardy's random geometry point of view, this provides us an alternative explanation for the solvability of $\rT\overline{\rT}$ deformation. We also derive the flow equation for the torus partition function from this perspective. We then discuss a path integral formulation of $\rT\overline{\rT}$ deformation, which is related to the 2d topological gravity. We will derive the deformed $S$-matrix and partition function from this definition. Finally we consider the holographic dual of the $\rT\overline{\rT}$ deformed CFT. The proposal is quite simple and has passed several non-trivial checks. Since in this section we consider several different topics, we will only present the main results and key steps of the derivations. For more details, we refer to the original papers.\par

\subsection{Random geometry}
The $\rT\overline{\rT}$ deformation can be interpreted as coupling the original QFT to a random geometry \cite{Cardy:2018sdv}. Using this interpretation, one can derive a flow equation (or evolution equation if we view the coupling $t$ as time) for the torus partition function. At the same time, it provides an alternative point of view on the solvability of $\rT\overline{\rT}$ deformation.\par

The analysis can be performed in two steps. In the first step, using the fact that the $\rT\overline{\rT}$ operator is \emph{quadratic} in components of the stress energy tensor, one performs a Hubbard-Stratonovich (HS) transformation of the variation of the partition function. The HS transformation introduces auxiliary variables $h_{ij}$ which couple to $T_{ij}$ and have the natural interpretation as the fluctuation of the metric, or infinitesimal fluctuation of the spacetime geometry. The auxiliary `gravity sector' can be solved by saddle-point approximation and the on-shell action turns out to be a total derivative. This is the reason for the solvability of the $\rT\overline{\rT}$ deformation. In the second step, if the variations of spacetime geometry is uniform, one can rewrite the fluctuation of the metric $h_{ij}$ in terms of the variation of \emph{parameters} that specify the geometry. This leads to a diffusion type equation which can be used to study the deformation of the partition function. In what follows, we mainly discuss the deformed partition function on a torus. For other interesting geometries, we refer to the original paper \cite{Cardy:2018sdv}.

\subsubsection*{From $\rT\overline{\rT}$ deformation to random geometry}
We explain how to interpret the $\rT\overline{\rT}$ deformation as coupling the theory to random geometry. From the definition of $\rT\overline{\rT}$ deformation, the infinitesimally deformed action near a point $t$ is given by\footnote{Note that our definition of the parameter $t$ is slightly different from the paper of Cardy. We have $t_{\text{Cardy}}=4t_{\text{here}}$}
\begin{align}
S^{(t+\delta t)}=S^{(t)}+\delta t\,\int_{\mathcal{M}}\det T_{ij} \rd^2 x
=S^{(t)}+\frac{\delta t}{2}\,\int_{\mathcal{M}}\epsilon_{ik}\epsilon_{jl}T^{ij}T^{kl} \rd^2x,
\end{align}
where $\mathcal{M}$ is the domain on which we define the theory. We work in the Cartesian coordinate system and use the relation
\begin{align}
\det T_{ij}=\frac{1}{2}\epsilon_{ik}\epsilon_{jl}T^{ij}T^{kl}
\end{align}
in the second equality. Since $\det T_{ij}$ is quadratic in components of the stress energy tensor, we can rewrite the infinitesimal deformation of the partition function by performing a Hubbard-Stratonovich transformation, which is nothing but a fancy name for Gaussian integral
\begin{align}
\label{eq:HStrans}
\re^{\frac{\delta t}{2}\int_{\mathcal{M}}\epsilon_{ik}\epsilon_{jl}T^{ij}T^{kl}\,\rd^2x}
\propto\int\mathcal{D}h\, \re^{-\frac{1}{2\delta t}\int_{\mathcal{M}}\epsilon^{ik}\epsilon^{jl}h_{ij}h_{kl}\,\rd^2x+\int_{\mathcal{M}}h_{ij}T^{ij}\,\rd^2x}.
\end{align}
By the definition of stress energy tensor $T_{ij}$, the second term is equivalent to an infinitesimal change in the metric $g_{ij}=\delta_{ij}+h_{ij}$. Several comments are in order.
\begin{enumerate}
\item We notice that the `gravity action'
\begin{align}
\label{eq:gravity-sector}
S[h]=-\frac{1}{2\delta t}\int_{\mathcal{M}}\epsilon^{ik}\epsilon^{jl}h_{ij}h_{kl}\,\rd^2x
\end{align}
is not the linearized Einstein-Hilbert action, so it is not really a standard gravity theory.
\item The infinitesimal parameter $\delta t$ appears in the action (\ref{eq:gravity-sector}) as $1/\delta t$. Since $\delta t\to 0$, the gravity sector is dominated by the saddle-point. So we can restrict our considerations to the saddle-point.\par
\item In principle the functional integral should be performed over all possible $h_{ij}$, including the ones with non-zero curvature. However, as we will show, it is enough to restrict ourselves to the flat metrics.
\end{enumerate}

\paragraph{Restriction to flat metrics} In two dimensions, any infinitesimal deformation of the metric can be written as
\begin{align}
\label{eq:h-decompose}
h_{ij}=\partial_i\alpha_j+\partial_j\alpha_i+g_{ij}\Phi.
\end{align}
This is because any 2d metric is locally conformal flat.
The first two terms in (\ref{eq:h-decompose}) correspond to an infinitesimal diffeomorphism $x_i\to x'_i=x_i+\alpha_i(x)$ of the Euclidean metric. The last term in (\ref{eq:h-decompose}) corresponds to an infinitesimal change of scale. We will show that we can take $\Phi=0$. The derivation is based on the saddle-point equation and the conservation of stress energy tensor. Due to the saddle-point equation,
\begin{align}
h_{ij}=\delta t\,\epsilon_{ik}\epsilon_{jl}T^{kl}=\delta t(\delta_{ij}T_k^k-T_{ij}).
\end{align}
Or equivalently,
\begin{align}
T_{ij}=\frac{1}{\delta t}\epsilon_{ik}\epsilon_{jl} {h}^{kl}=\frac{1}{\delta t}(\delta_{ij}{h}^k_k-h_{ij}).
\end{align}
Notice that in the above equations $h_{ij}$ are at the saddle point. The conservation of stress energy tensor $\partial^iT_{ij}=0$ can be translated to an equation about the metric
\begin{align}
\label{eq:eqPhi}
\partial^i\partial_i\alpha_j+\partial^i\partial_j\alpha_i-2\partial_j\partial^k\alpha_k-\partial_j\Phi=0,
\end{align}
where repeated indices are summed over. This is equivalent to
\begin{align}
\partial_j\Phi=\partial^i\partial_i\alpha_j-\partial_j\partial^k\alpha_k.
\end{align}
It is then easy to see that
\begin{align}
\label{eq:zerocurv}
\partial^j\partial_j\Phi=\partial^i\partial_i\partial^j\alpha_j-\partial^j\partial_j\partial^i\alpha_i=0.
\end{align}
This implies that $\Phi$ can be absorbed into a redefinition of $\alpha_i$. It is easier to see this in the complex coordinate $(\rz,\bar{\rz})$. The equation (\ref{eq:eqPhi}) can be written as
\begin{align}
\label{eq:eqPhi2}
\partial_{\rz}\Phi=&\,2\partial_{\rz}\left(\partial_{\bar{\rz}}\alpha_{\rz}-\partial_{\rz}\alpha_{\bar{\rz}}\right),\\\nonumber
\partial_{\bar{\rz}}\Phi=&\,2\partial_{\bar{\rz}}\left(\partial_{\rz}\alpha_{\bar{\rz}}-\partial_{\bar{\rz}}\alpha_{\rz} \right).
\end{align}
Since we know from (\ref{eq:zerocurv}) that $\partial_{\rz}\partial_{\bar{\rz}}\Phi=0$, we can write
\begin{align}
\Phi=2f(\rz)+2\bar{f}(\bar{\rz})
\end{align}
for some functions $f(\rz)$ and $\bar{f}(\bar{\rz})$. Combining with (\ref{eq:eqPhi2}), we find
\begin{align}
&\partial_{\rz}\left(\partial_{\bar{\rz}}\alpha_{\rz}-\partial_{\rz}\alpha_{\bar{\rz}}\right)=\partial_{\rz}f(\rz),\\\nonumber
&\partial_{\bar{\rz}}\left(\partial_{\bar{\rz}}\alpha_{\rz}-\partial_{\rz}\alpha_{\bar{\rz}}\right)=-\partial_{\bar{\rz}}\bar{f}(\bar{\rz}).
\end{align}
So we can take
\begin{align}
\partial_{\bar{\rz}}\alpha_{\rz}-\partial_{\rz}\alpha_{\bar{\rz}}=f(\rz)-\bar{f}(\bar{\rz}).
\end{align}
If we define
\begin{align}
\tilde{\alpha}_\rz=\alpha_\rz+\int_{\bar{\rz}_0}^{\bar{\rz}}\bar{f}(\bar{\rw})\rd\bar{\rw},\qquad
\tilde{\alpha}_{\bar{\rz}}=\alpha_{\bar{\rz}}+\int_{\rz_0}^{\rz}f(\rw)\rd\rw.
\end{align}
Then we have
\begin{align}
h_{\rz\bar{\rz}}=\partial_{\rz}\tilde{\alpha}_{\bar{\rz}}+\partial_{\bar{\rz}}\tilde{\alpha}_{\rz},\qquad
\partial_{\rz}\tilde{\alpha}_{\bar{\rz}}-\partial_{\bar{\rz}}\tilde{\alpha}_{\rz}=0.
\end{align}
To conclude the discussions so far, to define the path integral in (\ref{eq:HStrans}), we can restrict to the following class of metrics
\begin{align}
\label{eq:hijcons}
\boxed{h_{ij}=\partial_i\alpha_j+\partial_j\alpha_i,\qquad \partial_i\alpha_j=\partial_j\alpha_i.}
\end{align}
Using the second equation, we can write $h_{ij}=2\partial_i\alpha_j$.
Therefore we can write the gravity action as
\begin{align}
\epsilon_{ik}\epsilon_{jl}h_{ij}h_{kl}=4\epsilon_{ik}\epsilon_{jl}(\partial_i\alpha_j)(\partial_k\alpha_l)
=4\partial_i(\epsilon_{ik}\epsilon_{jl}\alpha_j\partial_k\alpha_l),
\end{align}
which is \emph{a total derivative}. We can thus write down the full action as
\begin{align}
S=&\,\frac{1}{2\delta t}\int_{\mathcal{M}}\epsilon^{ik}\epsilon^{jl}h_{ij}h_{kl}\, \rd^2x-\int_{\mathcal{M}} h_{ij}T^{ij}\,\rd^2 x\\\nonumber
=&\,\frac{2}{\delta t}\int_{\partial\mathcal{M}}(\epsilon^{ik}\epsilon^{jl}\alpha_j\partial_k\alpha_l)\,\rd n_i-2\int_{\partial\mathcal{M}}\alpha_j T^{ij}\,\rd n_i
\end{align}
where $\partial M$ is the boundary of the region and $n_i$ is the outward pointing normal vector. Alternatively, the boundary action can also be written as
\begin{align}
S=\frac{2}{\delta t}\int_{\partial\mathcal{M}}(\epsilon^{jl}\alpha_j\partial_k\alpha_l)\rd s_k-2\int_{\partial\mathcal{M}}\epsilon_{ik}\alpha_j T^{ij}\rd s_k,
\end{align}
where $s_k$ is now the tangent vector. We can perform the path integral over the new variables $\alpha_i$ since it is a rewriting of $h_{ij}$. However, we should not forget the constraint (\ref{eq:hijcons}) $\partial_i\alpha_j=\partial_j\alpha_i$. This can be rewritten as
\begin{align}
\oint\alpha_k \rd s_k=0.
\end{align}
From the discussion so far, we find that after rewriting in terms of $h_{ij}$, the gravity sector is in a sense topological and reduces to a boundary action. Topological theories are typically solvable, and this is the main reason of the solvability of the $\rT\overline{\rT}$ deformation. This already gives us an alternative explanation for the simplicity of $\rT\overline{\rT}$ deformation. In addition, it gives us a way to write down a flow equation for the deformed partition function.

\subsubsection*{Flow equation for partition function}
\begin{figure}[h!]
\begin{center}
\includegraphics[scale=0.5]{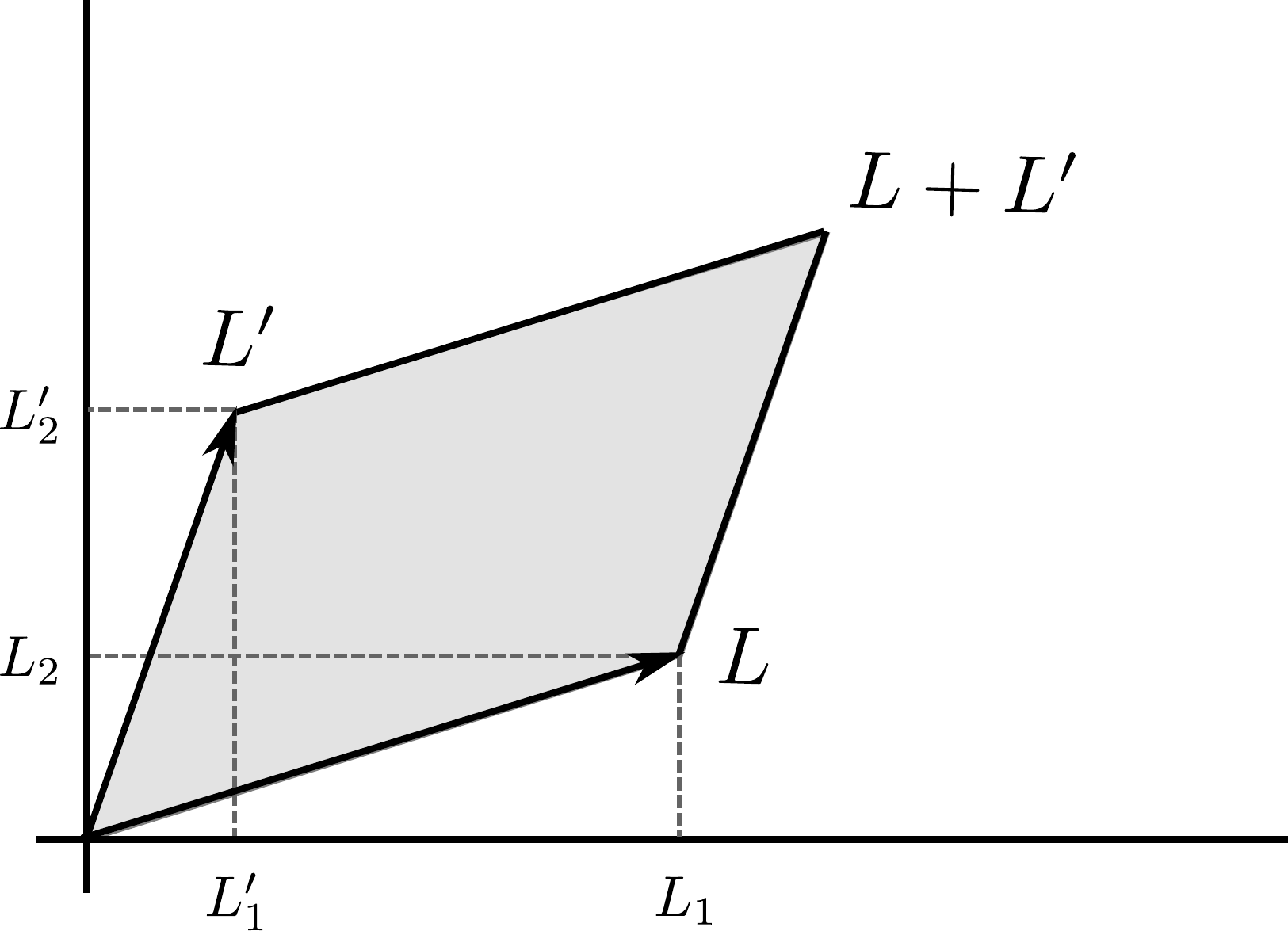}
\caption{The parallelogram which is identified with the torus.}
\label{fig:torus}
\end{center}
\end{figure}

For definiteness, we consider the torus partition function. Following Cardy \cite{Cardy:2018sdv}, we take the four end points of the parallelogram to be $0,L,L',L+L'$ with $L=L_1+\ri L_2$ and $L'=L'_1+\ri L'_2$, as is shown in figure~\ref{fig:torus}.
Firstly we want to argue that although we are performing a functional integral over the fluctuations of the metric $h_{ij}(x)$ in (\ref{eq:HStrans}), in fact only \emph{constant} flat metric $h_{ij}$ contributes. This is actually very natural. The random geometry is basically a rewriting of the effect of $\rT\overline{\rT}$ deformation on partition function at the linear order, namely $\delta t\,\rT\overline{\rT}$. In conformal perturbation theory, the effect is given by $\delta t\langle\rT\overline{\rT}(x)\rangle$. We have seen that in fact $\langle\rT\overline{\rT}(x)\rangle$ is a constant due to translational invariance. If we relate this deformation to the fluctuation of the spacetime geometry, the effect should be the same everywhere. In other words, this is a \emph{uniform} deformation. This implies that $h_{ij}$ should be position independent and the functional integral actually reduces to the usual integral. We can therefore write
\begin{align}
\re^{\delta t\int\det(T_{ij})\rd^2x}\propto\int\prod_{i,j=1}^2 \rd h_{ij} \re^{-\frac{A}{\delta t}(h_{11}h_{22}-h_{12}^2)+A h_{ij}T^{ij}},
\end{align}
where $A=L_1L'_2-L_2L'_1$ is the area of the torus. One can also find a more elaborated argument for this in the original paper. Let us then define the free energy as
\begin{align}
F^{(t)}(\{g_{ij}\})=-\log Z^{(t)}(\{g_{ij}\}).
\end{align}
We can write the second term in the exponent as
\begin{align}
A h_{ij}T_{ij}=-F^{(t)}(\{\delta_{ij}+h_{ij}\})+F^{(t)}(\{\delta_{ij}\}),
\end{align}
and we have the following equation for the deformed free energy
\begin{align}
\re^{-F^{(t+\delta t)}(\{\delta_{ij}\})}=\frac{A}{\delta t}\int\prod_{i,j=1}^2 \rd h_{ij} \re^{-\frac{A}{\delta t}(h_{11}h_{22}-h_{12}^2)-F^{(t)}(\{\delta_{ij}+h_{ij}\})},
\end{align}
where we have put in some factors to the measure which are not essential.

\paragraph{A small lemma} To proceed, we need a small lemma by Cardy. The proof is given in the original paper. The statement is as follow. Suppose we have an integral over $N$ variables $\{X_i\}$. If the quantity $Z^{(t)}(\{X\})$ satisfies the relation
\begin{align}
\label{eq:lemmaZ}
Z^{(t+\delta t)}(\{X\})=(4\pi\delta t)^{-N/2}(\det M)^{1/2}\int \prod_{j=1}^N \rd x_j \re^{-\frac{1}{4\delta t}\sum_{ij}x_i M_{ij} x_j}Z^{(t)}(\{X+x\}).
\end{align}
Then the quantity $Z^{(t)}(\{X\})$ obeys the following equation
\begin{align}
\partial_t Z^{(t)}=\sum_{ij} M_{ij}^{-1}\partial_{X_i}\partial_{X_j}Z^{(t)}.
\end{align}
In terms of the quantity $F^{(t)}=-\log Z^{(t)}$, the differential equation is given by
\begin{align}
\partial_t F^{(t)}=\sum_{i,j}M_{ij}^{-1}\partial_{X_i}\partial_{X_j}F^{(t)}-\sum_{i,j}M_{ij}^{-1}(\partial_{X_i}F^{(t)})(\partial_{X_j}F^{(t)}).
\end{align}

\paragraph{Variations in parameters} For a geometry without boundary such as a plane, the $\rT\overline{\rT}$ deformation does not change the spectrum, or equivalently it does not have any effects on the deformed partition function. On the other hand, when we put the theory on the torus, the constant fluctuation will deformed the shape of the torus in a uniform but non-trivial way. This affects the partition function. The shape of the torus is specified by $L$ and $L'$, which are shifted to $L\to L+\delta L$ and $L'\to L'+\delta L'$. These shifts are related to the uniform variation of the metric by
\begin{align}
\label{eq:reDeltaL}
\delta L_i=\frac{1}{2}\delta g_{ij} L_j=\frac{1}{2}h_{ij}L_j,\qquad \delta L'_i=\frac{1}{2}\delta g_{ij}L'_j=\frac{1}{2}h_{ij}L'_j.
\end{align}
Using these relations, we can rewrite the fluctuation of the metric in terms of fluctuation of the parameters $L_i$ and $L'_i$. From the lemma (\ref{eq:lemmaZ}), we can write down the following differential equation for the partition function
\begin{align}
\partial_t Z=\frac{1}{2A}\epsilon_{ik}\epsilon_{jl}\left(\frac{\partial^2Z}{\partial g_{ij}\partial g_{kl}} \right).
\end{align}
Or equivalently, in term of $F$
\begin{align}
\label{eq:Ft}
\partial_t F=\frac{1}{2A}\epsilon_{ik}\epsilon_{jl}\left(\frac{\partial^2 F}{\partial g_{ij}\partial g_{kl}}-\frac{\partial F}{\partial g_{ij}}\frac{\partial F}{\partial g_{kl}} \right).
\end{align}
The derivative of $F$ with respect to $g_{ij}$ is given by the one-point function of stress energy tensor
\begin{align}
\label{eq:partialFg}
\frac{\partial F}{\partial g_{kl}}=\int\langle T_{kl}(x)\rangle \rd^2x=\frac{1}{2}\left(L_k\partial_{L_l}+L'_k\partial_{L'_l} \right)F.
\end{align}
In the second equality, we have used (\ref{eq:reDeltaL}) and
\begin{align}
\frac{\partial F}{\partial g_{kl}}=\frac{\partial F}{\partial L_a}\frac{\partial L_a}{\partial g_{kl}}+\frac{\partial F}{\partial L'_a}\frac{\partial L'_a}{\partial g_{kl}}=\frac{1}{2}L_k\frac{\partial F}{\partial L_l}+\frac{1}{2}L'_k\frac{\partial F}{\partial L'_l}.
\end{align}
By translational invariance, (\ref{eq:partialFg}) can be written as
\begin{align}
\langle T_{kl} \rangle=\frac{1}{2A}\left(L_k\partial_{L_l}+L'_k\partial_{L'_l} \right)F.
\end{align}
In general, for any local operator $\mathcal{O}(0)$,
\begin{align}
\frac{\partial}{\partial g_{kl}}\langle\mathcal{O}(0)\rangle=\int\langle T_{kl}(x)\mathcal{O}(0)\rangle_c\,\rd^2x
=\frac{1}{2}\left(L_k\partial_{L_l}+L'_k\partial_{L'_l} \right)\langle\mathcal{O}(0)\rangle.
\end{align}
Therefore we have
\begin{align}
\frac{\partial^2F}{\partial{g_{ij}}\partial{g_{kl}}}=&\,\iint\langle T_{ij}(x)T_{kl}(x')\rangle_c\,\rd^2x \rd^2x'=A\int\langle T_{ij}(x)T_{kl}(0)\rangle_c\,\rd^2x,\\\nonumber
=&\,\frac{A}{4}(L_i\partial_{L_j}+L'_i\partial_{L'_j})\frac{1}{A} (L_k\partial_{L_l}+L'_k\partial_{L'_l})F.
\end{align}
Then (\ref{eq:Ft}) can be written as
\begin{align}
\partial_tF=&\,\frac{1}{2}\epsilon_{ik}\epsilon_{jl}\left[(L_i\partial_{L_j}+L'_i\partial_{L'_j})(1/A)(L_k\partial_{L_l}+L'_k\partial_{L'_l})F\right.\\\nonumber
&\,\quad\left.-((L_i\partial_{L_j}+L'_i\partial_{L'_j})F)(1/A(L_k\partial_{L_l}+L'_k\partial_{L'_l})F)\right].
\end{align}
Equivalently for the partition function
\begin{align}
\label{eq:flowZ}
\partial_t Z=\frac{1}{2}\epsilon_{ik}\epsilon_{jl}(L_i\partial_{L_j}+L'_i\partial_{L'_j})(1/A)(L_k\partial_{L_l}+L'_k\partial_{L'_l})Z.
\end{align}
To write this equation in a nicer form, we can introduce $\mathcal{Z}=Z/A$. The equation for $\mathcal{Z}$ is given by
\begin{align}
\partial_t\mathcal{Z}=(\partial_{L_1}\partial_{L'_2}-\partial_{L_2}\partial_{L'_1})\mathcal{Z}.
\end{align}
The flow equation (\ref{eq:flowZ}) is equivalent to (\ref{eq:flowEq}) that we derived in section~\ref{lecture2} by the simple identification
\begin{align}
\tau=\frac{L'}{L}=\frac{L'_1+\ri L'_2}{L_1+\ri L_2}.
\end{align}

\subsection{2d topological gravity}
The random geometry point of view rewrites the \emph{infinitesimal deformation} of the partition function in terms of fluctuations of spacetime geometry. In this way, we can only derive the flow equation for the partition function. Can one somehow `integrate' the flow equation and give a path integral definition of the $\rT\overline{\rT}$ deformation ? Such a definition was proposed in \cite{Dubovsky:2017cnj,Dubovsky:2018bmo}. The proposal is that $\rT\overline{\rT}$ deforming a 2d QFT is equivalent to coupling the theory to a 2d topological gravity which is similar to flat spacetime Jackiw-Teitelbolm (JT) gravity. In this section, we discuss this interesting proposal. Before going into the details, it is worth emphasising that the gravity theory which is proposed in \cite{Dubovsky:2017cnj,Dubovsky:2018bmo} is not exactly the JT gravity. On the plane, the gravity theory is almost the JT gravity. This is enough for deriving the deformed $S$-matrix. For computing torus partition function, however, one has to resort to the first order formalism and strictly speaking the theory is no longer the JT gravity.\par

There are several checks for this proposal. Firstly, one can show that under the $\rT\overline{\rT}$ deformation, the $S$-matrix is deformed in a simple way by multiplying a CDD factor. Secondly, by performing the path integral carefully, one can derive the flow equation which we derived in the previous section. The derivations are technical, so we will present the main ideas and some of the key steps for the derivation. More details are referred to the original papers and references therein.

\subsubsection*{The proposal}
The proposal of \cite{Dubovsky:2017cnj,Dubovsky:2018bmo} is as follows. The $\rT\overline{\rT}$ deformed action is given by coupling the theory to the flat spacetime Jackiw-Teitelboim gravity
\begin{align}
\label{eq:JT-gravity}
S_{\rT\overline{\rT}}=S_0(g_{\mu\nu},\psi)+\int\sqrt{-g}(\varphi R-\Lambda)\rd^2x,
\end{align}
where $S_0(g_{\mu\nu},\psi)$ is the original field theory which minimally couples to the dynamic metric $g_{\mu\nu}$, but it does not couple directly to the dilaton $\varphi$. The second term is the gravity sector. On the plane, this action can be written equivalently in the first order formalism which we will write down in (\ref{eq:finalformJT}) in the next section. The vacuum energy $\Lambda$ is related to the $\rT\overline{\rT}$ coupling constant $t$ by
\begin{align}
t=-\frac{2}{\Lambda}.
\end{align}
Unlike Cardy's random geometry interpretation, which is an alternative interpretation about the \emph{infinitesimal} deformation of the partition function around a given point $t$. The definition (\ref{eq:JT-gravity}) is a complete and non-perturbative definition of the $\rT\overline{\rT}$ deformed theories along the whole flow. In this formulation, the relation between $\rT\overline{\rT}$ deformation and gravity is manifest.

The gravity sector is purely topological and does not have propagating degrees of freedom. The dilaton field plays the role of Lagrangian multiplier and forces the spacetime to be flat. Since the gravity sector is topological, the action again can be written as a total derivative and interesting physics happens at the boundary.

\subsubsection*{Gravitational dressing}
In the previous sections, we have studied the deformed spectrum and partition function of the $\rT\overline{\rT}$ deformed theories. We now turn to another observable which is the $S$-matrix. The $S$-matrix relates incoming and outgoing asymptotic states in a scattering process and is one of the most important physical observables in quantum field theories. An $S$-matrix element can be parameterized by the four-momenta and other quantum numbers of the scattering particles. For our purpose, we consider the $S$-matrix in 1+1 dimensions. The four-momentum is given by $p^{\mu}=(E,p)$ where $E$ is the energy and $p$ is the momentum of the particle. We denote the collection of all four-momenta by $\{p_i\}=\{p_1^{\mu},p_2^{\mu},\cdots\}$. To simplify our notation, we omit other quantum numbers of the particles and write the $S$-matrix element as $S(\{p_i\})$.
As we will show below, under $\rT\overline{\rT}$ deformation, the $S$-matrix is deformed in the following simple way
\begin{align}
\label{eq:gravityS}
S(\{p_i\})\mapsto \left(\prod_{i<j}\re^{-\ri\delta^{(t)}_{ij}}\right) S(\{p_i\}),\qquad \delta^{(t)}_{ij}=t\,\epsilon_{\mu\nu}p^{\mu}_ip^{\nu}_j,
\end{align}
where the phase factor $\re^{-\ri\delta^{(t)}_{ij}}$ is the so-called CDD factor. We see that turning on $\rT\overline{\rT}$ deformation at the $S$-matrix level is simply multiplying a CDD factor. This $S$-matrix was first conjectured in \cite{Dubovsky:2012wk} and later proved by using  gravity proposal \cite{Dubovsky:2017cnj}. In what follows, we first give the derivation of the gravitational dressing formula (\ref{eq:gravityS}) and then comment on CDD factors and their physical effects.

\paragraph{Derivation of the deformed $S$-matrix}
There are several ways to arrive at the deformed $S$-matrix. The simplest way is via a geometrical interpretation of the $\rT\overline{\rT}$ deformation. From the random geometry point of view, we already see that turning on $\rT\overline{\rT}$ deformation has the same effect as deforming the spacetime geometry. Therefore we can take two different point of views for $\rT\overline{\rT}$ deformation.
\begin{enumerate}
\item The spacetime geometry is fixed, we perform an irrelevant deformation of the QFT on the fixed background. This is the original definition of the $\rT\overline{\rT}$ deformation.
\item The spacetime is deformed in a dynamical way via a dynamical change of coordinates. The new coordinates depend on the stress energy tensor of the QFT. \emph{On the new coordinates, the theory `looks' undeformed}.
\end{enumerate}
Of course, to make the second point of view more precise, we need to first specify the change of coordinate and then make clear the meaning of `looks undeformed' in the new coordinate. As we will see, coupling the theory to 2d gravity provides a convenient way to find the dynamical coordinates. We will make the meaning `looks undeformed' more precise when deriving the gravitationally dressed $S$-matrix.

We want to emphasis that the second point of view for the $\rT\overline{\rT}$ deformation is potentially very powerful and its range of application is beyond the deformed $S$-matrix. At the classical level, this change of coordinate is constructed recently in \cite{Conti:2018tca} from a different consideration. A similar interpretation for the $\rJ\overline{\rT}$ deformation as dynamical change of coordinate (or field depend change of coordinates) is also proposed recently in \cite{Guica:2019vnb,Bzowski:2018pcy}.

For our purpose, it is convenient to work in the light-cone coordinate
\begin{align}
x^{\pm}=\frac{1}{\sqrt{2}}(x^0\pm x^1).
\end{align}
The vacuum solution of the JT gravity is given by
\begin{align}
g_{\alpha\beta}=\eta_{\alpha\beta},\qquad \phi=-\frac{\Lambda}{4}\eta_{\alpha\beta}x^{\alpha}x^{\beta}+c=\frac{\Lambda}{2}\sigma^+\sigma^-+c,
\end{align}
where $\alpha,\beta=\pm$, $c$ is some constant. We have $\eta_{+-}=\eta_{-+}=-1$ and $\eta_{--}=\eta_{++}=0$. We want to study scattering around this vacuum. We see that although the metric is invariant under Poincar\'e group $x^{\pm}\mapsto x^{\pm}+a^{\pm}$, the vacuum solution of the dilaton is not. This implies that the deformed vacuum is not invariant under the Poincar\'e transformation in the original coordinate system $x^\pm$. The vacuum solution of the dilaton is invariant under the combined transformation
\begin{align}
\label{eq:defCombine}
x^{\pm}\mapsto x^{\pm}+a^{\pm},\qquad \phi\mapsto\phi-\frac{\Lambda}{2}(a^+ x^- + a^-x^+).
\end{align}
If there exist a new coordinate system $X^{\pm}(x^+,x^-)$ on which the theory looks like undeformed, then it must has the Poincar\'e invariance $X^{\pm}\to X^{\pm}+a^{\pm}$ and it should involve $\phi$ in some way. This is a requirement for the new coordinates. To proceed, we work in the conformal gauge
\begin{align}
g_{\alpha\beta}=\re^{2\Omega(x^+,x^-)}\eta_{\alpha\beta}.
\end{align}
Then the gravity action becomes
\begin{align}
S_{\text{grav}}=\int \rd x^+ \rd x^-(4\phi\partial_+\partial_-\Omega-\Lambda e^{2\Omega}).
\end{align}
The equations of motion for $\phi$ and $\Omega$ are given by
\begin{align}
\label{eq:EOMphi1}
\partial_+^2\phi=-\frac{1}{2}T_{++},\qquad \partial_-^2\phi=-\frac{1}{2}T_{--},\qquad \partial_+\partial_-\phi=\frac{1}{2}(\Lambda \re^{2\Omega}+T_{+-})
\end{align}
and
\begin{align}
\partial_+\partial_-\Omega=0.
\end{align}
The general solution for $\Omega$ is that $\Omega(x^+,x^-)=f(x^+)+\bar{f}(x^-)$ for two arbitrary function $f,\bar{f}$. We impose the boundary condition that $g_{\alpha\beta}\to\eta_{\alpha\beta}$ at infinity. This implies that $\Omega=0$ everywhere. So (\ref{eq:EOMphi1}) simplifies to
\begin{align}
\label{eq:EOMphi2}
\partial_+^2\phi=-\frac{1}{2}T_{++},\qquad \partial_-^2\phi=-\frac{1}{2}T_{--},\qquad \partial_+\partial_-\phi=\frac{1}{2}(\Lambda+T_{+-}).
\end{align}
The main proposal of \cite{Dubovsky:2017cnj} is that we should take dynamical coordinate $X^\pm(x^+,x^-)$ to be
\begin{align}
X^+(x^+,x^-)=\frac{2}{\Lambda}\partial_-\phi,\qquad X^-(x^+,x^-)=\frac{2}{\Lambda}\partial_+\phi.
\end{align}
Let us separate the original coordinates and write
\begin{align}
X^{\pm}(x^+,x^-)=x^{\pm}+Y^{\pm}(x^+,x^-).
\end{align}
Then the equations of motion for the dilaton imply
\begin{align}
\label{eq:diffY}
\partial_+Y^-=-\frac{T_{++}}{\Lambda},\qquad \partial_-Y^+=-\frac{T_{--}}{\Lambda},\qquad \partial_+Y^+=\partial_-Y^-=\frac{T_{+-}}{\Lambda}.
\end{align}
This is equivalent to the following change of coordinates
\begin{align}
\label{eq:XX}
\frac{\partial X^+}{\partial x^+}=1-t\,T_{+-},\qquad \frac{\partial X^+}{\partial x^-}=t\,T_{--},\\\nonumber
\frac{\partial X^-}{\partial x^-}=1-t\,T_{-+},\qquad \frac{\partial X^-}{\partial x^+}=t\,T_{++},
\end{align}
which coincides with the proposal in \cite{Conti:2018tca}. We see that the transformation $X^{\pm}+\text{const}$, which is equivalent to the combined transformation (\ref{eq:defCombine}) indeed leaves the vacuum invariant.\par

Despite this nice feature, we see that the change of coordinates (\ref{eq:XX}) is rather unusual. Normally if we specify a change of coordinate, we write $X^{\pm}=f^{\pm}(x^{+},x^-)$ with some given function. Here, however, we specify the change of coordinates in a differential form $\rd X^{\alpha}=\hat{f}^{\alpha}_{\phantom{a}\beta}(T_{\mu\nu})\rd x^{\beta}$ with $\hat{f}$ being an operator that depends on the stress energy tensor. Strictly speaking, this `change of coordinates' is an operator equation. This makes is a bit subtle to interpret at the quantum level. At classical level, we can treat ${f}^{\alpha}_{\phantom{a}\beta}(T_{\mu\nu})$ as usual functions. The differential form of (\ref{eq:XX}) is related to non-locality of $\rT\overline{\rT}$ deformation. In order to find the dynamical coordinate $X^{\alpha}$ explicitly from (\ref{eq:XX}), we need to perform integrals, which is a non-local operation.
\par

Now we consider the scattering problem in 1+1 dimensions and see how we can make (\ref{eq:XX}) more precise in this simple context. This can be done in two steps. Firstly we work out the dynamical coordinates explicitly using some special features of scattering in 1+1 dimensions; Then we use the dynamical coordinate to perform the mode expansion of the free field, which leads to the dressing of the asymptotic states.

\begin{figure}[h!]
\begin{center}
\includegraphics[scale=0.55]{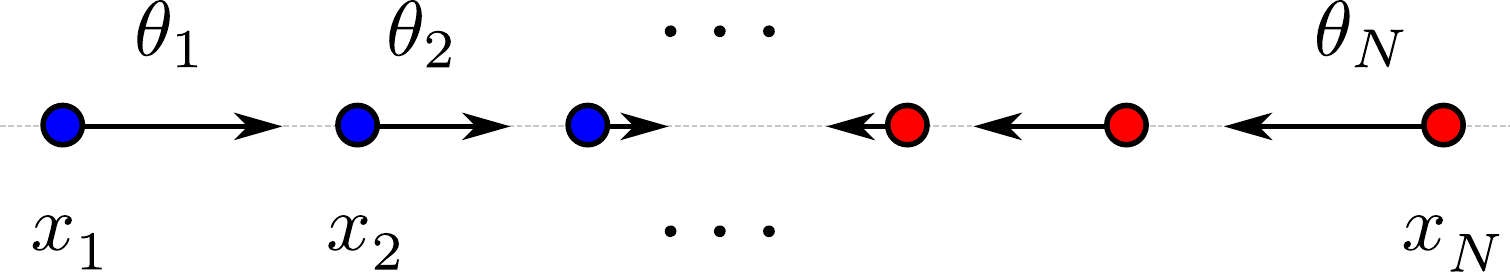}
\caption{The ordering of particles. The ordering of the rapitities are the same as the orderings of the spacetime positions.}
\label{fig:scattering}
\end{center}
\end{figure}

\paragraph{Dynamical coordinate} Let us denote the set of on-shell momenta of the incoming and outgoing particles as $\{p_i\}$ and $\{q_i\}$ respectively. Then the stress energy tensor is that of a collection of particles. We can find the quantities $Y^{\pm}$ by integrating (\ref{eq:diffY}). For example,
\begin{align}
Y^+(x^\mu)=-\frac{1}{\Lambda}\int_{-\infty}^{x^+} T_{++}(y^+,y^-)\rd y^+ +C^+
\end{align}
where $C^+$ is an integration constant. We have similar expression for $Y^-(x^{\mu})$. In order to fix $C^+$, we need to impose boundary conditions. In what follows, we denote the spacetime coordinate by $x^{\mu}=(t,x)$. The total momentum of the light-cone components are given by
\begin{align}
P^{+}=\int_{-\infty}^{\infty}T_{--}(x^+,x^-)\rd x^{-},\qquad P^{-}=\int_{-\infty}^{\infty}T_{++}(x^+,x^-)\rd x^{+}.
\end{align}
We can impose the condition that at infinite left, the coordinates $Y^{\pm}$ are given by
\begin{align}
\lim_{x\to-\infty}Y^+(x^{\mu})=-\frac{1}{2\Lambda}P^+,\qquad \lim_{x\to-\infty}Y^-(x^\mu)=\frac{1}{2\Lambda}P^-.
\end{align}
These conditions fix the constants $C^{\pm}$ completely. From this, we can deduce that at infinite right, we have
\begin{align}
\lim_{x\to+\infty}Y^+(x^{\mu})=\frac{1}{2\Lambda}P^+,\qquad \lim_{x\to+\infty}Y^-(x^{\mu})=-\frac{1}{2\Lambda}P^-.
\end{align}
We focus on the incoming particles first. In 1+1 dimensions, we can parameterize the momenta of the particle by rapidities. Let us denote the rapidities by $\theta_1,\cdots,\theta_N$. One special feature of scatterings in 1+1 dimension is that the rapidities are \emph{ordered}. We can take the following ordering
\begin{align}
\theta_1\ge\theta_2\ge\cdots\ge\theta_N,
\end{align}
where the positive rapidity corresponds moving towards right and the negative rapidity correspond to moving towards left. For incoming particles, this ordering is equivalent to the ordering of particles in space. The reason is simple. In 1+1 dimensions, all the particles are placed along a line. In order for the scattering to take place, for the incoming particles, the leftmost particle (moving towards right) must move faster than the other right moving particles, otherwise it cannot catch up with them and will miss the scattering. This is shown in figure~\ref{fig:scattering}.
The positions of the particles are denoted by $x_1,\cdots,x_N$, we have
\begin{align}
x_1\le x_2\le\cdots\le x_N.
\end{align}
In the far past where $t\to-\infty$, the particles are well separated, we have
\begin{align}
\label{eq:pastP}
\lim_{t\to-\infty}Y^+(x^{\mu})=&\,\frac{1}{2\Lambda}\left(-P^+ + 2P_L^+(x^{\mu})\right),\\\nonumber
\lim_{t\to-\infty}Y^-(x^{\mu})=&\,\frac{1}{2\Lambda}\left(+P^- - 2P_L^-(x^{\mu})\right).
\end{align}
Here $P_L^{\pm}(x^{\mu})$ denote the total momentum of all particles to the left of position $x$ at the given time $t$. This definition is ambiguous when $x=x_i$ because it is not clear whether the particle is to the left or to the right of $x$. So at these points, we adopt a democratic prescription and define $P_L^{\pm}(x^{\mu})$ as
\begin{align}
\label{eq:pres}
P_L^{\pm}(x_i^{\mu})=\frac{1}{2}p_i^{\pm}+\sum_{j<i}p_j^{\pm}.
\end{align}
Plugging the prescription (\ref{eq:pres}) in (\ref{eq:pastP}) and recalling that
\begin{align}
P^{\pm}=\sum_{i=1}^N p_i,
\end{align}
we find that $Y^{\pm}(x_i^{\mu})$ is independent of $x_i^{\mu}$. Therefore, at the position of the $i$-th particle, the coordinate $Y^{\pm}$ can be written as
\begin{align}
Y^-(x_i^{\mu})\sim &\,Y^-(p_i^{\mu})=\frac{1}{2\Lambda}\left[\mathcal{P}^-_{<}(p_i^{\mu})-\mathcal{P}^-_{>}(p_i^{\mu})\right],\\\nonumber
Y^+(x_i^{\mu})\sim &\,Y^+(p_i^{\mu})=\frac{1}{2\Lambda}\left[\mathcal{P}^+_{>}(p_i^{\mu})-\mathcal{P}^+_{<}(p_i^{\mu})\right],
\end{align}
where $\mathcal{P}^{\alpha}_{<}(p_i)$ and $\mathcal{P}^{\alpha}_{>}(p_i)$ are the total momentum of particles with smaller and larger rapidities as compared to $\theta_i$ respectively. Notice that we are allowed write $Y^{\pm}(x_i^{\mu})\sim Y^{\pm}(p_i^{\mu})$ because the ordering of momenta is equivalent to the ordering of particle positions in the far past. This gives us an explicit form of the new coordinate $X^{\pm}$ in terms of the momenta of the incoming particles $\{p_i\}$.

\paragraph{Dressing asymptotic states} To compute the $S$-matrix, one needs to prepare asymptotic states in the far past. In this asymptotic regime, before introducing gravity, the matter field can be decomposed as
\begin{align}
\label{eq:modexp1}
\psi=\int_{-\infty}^{\infty}\frac{\rd p}{\sqrt{2\pi}}\frac{1}{\sqrt{2E}}\left(a^{\dagger}_{\text{in}}\re^{-\ri p_{\mu}x^{\mu}}+\text{h.c.}\right).
\end{align}
As we discussed before, after introducing gravity, the effect can be encoded in a dynamical change of coordinates. Therefore if we use the new coordinates $X^{\mu}$, the theory should `look the same' in the asymptotic regime. More precisely, all we need to do is to perform the same decomposition, but using new coordinate system $X^{\mu}$. Accordingly, we need to define a set of new creation and annihilation operators $A_{\text{in}}(p)$ and $A^{\dagger}_{\text{in}}(p)$ such that
\begin{align}
\label{eq:modexp2}
\psi=\int_{-\infty}^{\infty}\frac{\rd p}{\sqrt{2\pi}}\frac{1}{\sqrt{2E}}\left(A^{\dagger}_{\text{in}}(p)\re^{-\ri p_{\mu}X^{\mu}}+\text{h.c.}\right).
\end{align}
The precise meaning of `looks undeformed' in our context is to identify the $\psi$ in (\ref{eq:modexp1}) and (\ref{eq:modexp2}). We then find
\begin{align}
A^{\dagger}_{\text{in}}(p)=a_{\text{in}}^{\dagger}\re^{\ri p_{\alpha}Y^{\alpha}(p)}=a_{\text{in}}^\dagger(p)\re^{-\ri(p^+Y^-(p)+p^-Y^+(p))}.
\end{align}
The dressed incoming asymptotic state is generated by the action of $A_{\text{in}}^{\dagger}(p)$ as
\begin{align}
|\{p_i\}\rangle\!\rangle_{\text{in}}\equiv\prod_{i=1}^N A_{\text{in}}^{\dagger}(p_i)|0\rangle=\exp\left(-\frac{\ri}{2\Lambda}\sum_{i<j}\epsilon_{\alpha\beta}p_i^{\alpha}p_j^{\beta}\right)|\{p_i\}\rangle_{\text{in}}.
\end{align}
A similar analysis can be done for the outgoing states. The only difference is that the ordering is reversed which leads to a sign flip
\begin{align}
{_{\text{out}}\langle}\!\langle\{q_i\}|=\exp\left(\frac{\ri}{2\Lambda}\sum_{i<j}\epsilon_{\alpha\beta}q_i^{\alpha}q_j^{\beta}\right){_{\text{out}}\langle}\{q_i\}|
\end{align}
From the definition of $S$-matrix,
\begin{align}
{S}(\{p_i\},\{q_i\})=&\,{_{\text{out}}\langle}\!\langle\{q_i\}|\{p_i\}\rangle\!\rangle_{\text{in}}\\\nonumber
=&\,\re^{-\frac{\ri}{2\Lambda}\sum_{i<j}\epsilon_{\alpha\beta}p_i^{\alpha}p_j^{\beta}
-\frac{\ri}{2\Lambda}\sum_{i<j}\epsilon_{\alpha\beta}q_i^{\alpha}q_j^{\beta}}
{_{\text{out}}\langle}\{q_i\}|\{p_i\}\rangle_{\text{in}}\\\nonumber
=&\,\re^{-\frac{\ri}{2\Lambda}\sum_{i<j}\epsilon_{\alpha\beta}p_i^{\alpha}p_j^{\beta}
-\frac{\ri}{2\Lambda}\sum_{i<j}\epsilon_{\alpha\beta}q_i^{\alpha}q_j^{\beta}}S_0(\{p_i\},\{q_i\}).
\end{align}
This is the gravitational dressing formula for the $S$-matrix in (\ref{eq:gravityS}).

\paragraph{CDD factor and S-matrix}
The dressing factor in (\ref{eq:gravityS}) is a CDD factor. Here we discuss in more detail about these factors and their physical effects. The CDD factor (sometimes called CDD ambiguity), which is a short name for the Castillejo-Dalitz-Dyson factor \cite{Castillejo:1955ed} is a special kind of function in the context of $S$-matrix bootstrap. For simplicity, we focus on the scattering of massive particles in integrable field theories where we only need to consider 2 to 2 scatterings. We parameterize energy and momentum in terms of rapidity
\begin{align}
E=m\cosh\theta,\qquad p=m\sinh\theta.
\end{align}
Due to Lorentz invariance, the 2 to 2 $S$-matrix can be written as $S_{ij}^{kl}(\theta_i-\theta_j)=S_{ij}^{kl}(\theta)$ where `$i,j$' and `$k,l$' denote the quantum numbers of incoming and outgoing particles, respectively and $\theta\equiv\theta_i-\theta_j$. The idea of $S$-matrix bootstrap is to fix the $S$-matrix as much as possible from general physical principles and assumptions about analyticity. One requires that $S_{ij}^{kl}(\theta)$ satisfy a set of bootstrap axioms such as unitarity and crossing symmetry. These axioms lead to a set of functional equations for the $S$-matrix. However, in general they are not sufficient to fix the $S$-matrix completely. If we multiply $S_{ij}^{kl}(\theta)$ by a scalar factor $\Phi(\theta)$
\begin{align}
\tilde{S}_{ij}^{kl}(\theta)=S_{ij}^{kl}(\theta)\Phi(\theta),
\end{align}
such that $\Phi(\theta)$ is a meromorphic function which is analytic and bounded in the ``physical strip'' $0\le\text{Im}\theta\le \pi$ and satisfies the equations
\begin{align}
\label{eq:Phi}
\Phi(\theta)\Phi(-\theta)=1,\qquad \Phi(\ri\pi+\theta)\Phi(\ri\pi-\theta)=1.
\end{align}
Then the deformed $S$-matrix $\tilde{S}_{ij}^{kl}(\theta)$ still satisfies the same set of axioms. Bound state structures and other considerations might impose further constraints, but in general this ambiguity cannot be completely fixed. The families of scalar functions $\Phi(\theta)$ which satisfy (\ref{eq:Phi}) are the CDD factors. A generic CDD factor admits the representation in the exponential form
\begin{align}
\Phi(\theta)=\exp\left(\ri\sum_{s=1}^{\infty}\alpha_s \sinh(s\theta)\right),
\end{align}
where $\alpha_s$ are parameters. Alternatively, it has the representation in the rational form
\begin{align}
\Phi(\theta)=\prod_p^N\frac{\beta_p-\ri\sinh\theta}{\beta_p+\ri\sinh\theta}.
\end{align}
It is easy to verify that the factor from gravitational dressing can be written as
\begin{align}
\re^{-\ri\delta^{(t)}_{ij}}=\re^{\ri tm_im_j\sinh(\theta_i-\theta_j)}=\re^{\ri tm_im_j\sinh(\theta)},
\end{align}
which is indeed a CDD factor in the exponential form. Although our discussion is for integrable field theories, the proposal of gravitational dressing is applicable to any $S$-matrix.

\paragraph{Effects of CDD factor} The CDD factor from gravitational dressing is quite simple. Multiplying such factors to the $S$-matrix has some interesting consequences which is most easily analyzed in integrable quantum field theories.
\begin{enumerate}
\item The deformed $S$-matrix still satisfies unitarity, crossing symmetry and Yang-Baxter equations, so it is still integrable. This implies that the $\rT\overline{\rT}$ deformation should preserve integrability. This is confirmed by the fact that $\rT\overline{\rT}$ deformation preserves all the higher spin charges \cite{Smirnov:2016lqw}.
\item Having the deformed $S$-matrix, we can compute the deformed spectrum by thermodynamic Bethe ansatz. This has been done and the result is the same as the deformed spectrum we found in section~\ref{sec:def-spectrum} using factorization formula.
\item The connection between CDD factors and irrelevant deformations of CFTs was already noticed by Sasha Zamolodchikov in \cite{Zamolodchikov:1991vx} when studying the RG flow between Ising and tri-cricial Ising. There he also found that at the leading order of perturbation theory, the corresponding operator is the $\rT\overline{\rT}$ operator. In the early 90's, Alyosha Zamolodchikov studied extensively the spectrum of IQFTs that are deformed by CDD factors using TBA. He observed that for generic CDD factors, the spectrum always has a square root branch point at finite $R$. This singularity was also observed in \cite{Mussardo:1999aj}. In general, given a CDD factor, it is not clear which irrelevant deformation it corresponds to in the Lagrangian formulation. The $\rT\overline{\rT}$ deformation and the CDD factor we found in this section provides a very neat example of such correspondences.
\item As we mentioned before, multiplying CDD factors to the $S$-matrix corresponds to irrelevant deformations of QFTs. Under irrelevant deformation, a QFT becomes an effective field theory and one expects that there are infinitely many terms in the Lagrangian, which is the case for $\rT\overline{\rT}$ deformation. For relevant perturbation of CFTs, it is known that in general different signs of the coupling constant lead to very different IR theories. Therefore we should not be surprised that similar things happen for irrelevant deformations, which is indeed the case for $\rT\overline{\rT}$ deformation.
\end{enumerate}

\subsubsection*{Torus partition function}
In this section, we explain how to compute the torus partition function directly from the path integral of 2d gravity. This calculation is rather technical and have many subtleties. We will give the idea and sketch the main steps. More details are referred to the original paper \cite{Dubovsky:2018bmo}. Before going into any details, it is worth pointing out that the computations below are \emph{not} the usual path integral that one does for 2d gravity. First of all, one restricts the path integral to torus topology. Secondly, the final result of the computation, which is the $\rT\overline{\rT}$ deformed torus partition function still depends on the parameters of a fixed torus. Therefore, what the path integral really does is summing over all possible mappings from a \emph{worldsheet torus} to a fixed \emph{target space torus}. The mappings are restricted to have winding number 1 and the weight of the mapping is specified by the gravity action.

\paragraph{The first order formalism} Due to several technical reasons, in particular, to consider the system in finite volume, it is necessary to use the so-called first order formalism. In this formalism, instead of integrating over the metric $g_{\alpha\beta}$, one integrates over the veilbein $e_{a\alpha}$, the spin connection $\omega_{\alpha}$ and a pair of Lagrange multipliers $\lambda^a$. Here the indices $\alpha,\beta=0,1$ and $a,b=1,2$. We write the action as
\begin{align}
S_{\rT\overline{\rT}}=S_{\text{grav}}+S_{\text{m}},
\end{align}
where
\begin{align}
S_{\text{grav}}=\int\sqrt{-g}\phi R\,\rd^2x,\qquad S_{\text{m}}=S_0(g_{\alpha\beta},\psi)-\Lambda\int\sqrt{-g}\,\rd^2x.
\end{align}
Notice that we put the vacuum energy term to the matter part of the action. The veilbein is defined as the `square root' of the metric
\begin{align}
g_{\alpha\beta}=e_{a\alpha}e_{b\beta}\eta^{ab},
\end{align}
where $\eta^{ab}$ is the standard metric for Minkowski spacetime. Using the veilbein, the action of the vacuum energy becomes
\begin{align}
-\Lambda\int\sqrt{-g}\,\rd^2x=-\frac{\Lambda}{2}\int\epsilon^{\alpha\beta}\epsilon^{ab}e_{a\alpha}e_{b\beta}\,\rd^2x.
\end{align}
The action for the 2d gravity becomes
\begin{align}
\label{eq:JTfirstorder}
S_{\text{grav}}=\int\epsilon^{\alpha\beta}\left(\lambda^a(\partial_{\alpha}e_{a\beta}-\epsilon_a^b\omega_{\alpha}e_{b\beta})
+\phi\partial_{\alpha}\omega_{\beta}  \right)\rd^2x.
\end{align}
The variation with respect to $\lambda^a$ enforces the relation between the veilbein $e_{a\alpha}$ and the spin connection $\omega_{\alpha}$
\begin{align}
\partial_{\alpha}e_{a\beta}-\epsilon_a^b\omega_{\alpha}e_{b\beta}=0.
\end{align}
As before, a variation with respect to the dilaton $\phi$ forces the spacetime to be flat
\begin{align}
\epsilon^{\alpha\beta}\partial_{\alpha}\omega_{\beta}=0.
\end{align}
This equation implies that we can take $\omega_{\alpha}=\partial_{\alpha}\omega$. Plugging this into (\ref{eq:JTfirstorder}) and after some manipulations, one arrives at the following simple action
\begin{align}
S_{\text{grav}}=-\int\epsilon^{\alpha\beta}\partial_{\alpha}\lambda^a e_{a\beta}\,\rd^2x.
\end{align}
Furthermore, one introduces an auxiliary variable $X^a$ defined as
\begin{align}
X^a=\Lambda^{-1}\epsilon_a^b\lambda^b.
\end{align}
By adding a total derivative term, one can rewrite the total action in the final form
\begin{align}
\label{eq:finalformJT}
S_{\rT\overline{\rT}}=\frac{\Lambda}{2}\int\epsilon^{\alpha\beta}\epsilon_{ab}(\partial_{\alpha}X^a-e^a_{\alpha})
(\partial_{\beta}X^b-e^b_\beta)\rd^2x.
+S_0(\psi,g_{\alpha\beta})
\end{align}
This is the action that the authors of \cite{Dubovsky:2018bmo} propose to take in the path integral.

\paragraph{Partition function and integration} The partition function is now defined as
\begin{align}
Z_{\rT\overline{\rT}}=\int\frac{\mathcal{D}e\mathcal{D}X\mathcal{D}\psi}{V_{\text{diff}}}\re^{-S_{\rT\overline{\rT}}}
=\int\frac{\mathcal{D}e\mathcal{D}X}{V_{\text{diff}}}
\re^{-\frac{\Lambda}{2}\int\epsilon^{\alpha\beta}\epsilon_{ab}(\partial_{\alpha}X^a-e^a_{\alpha})(\partial_{\beta}X^b-e^b_\beta)\rd^2x}
Z_0(g_{\alpha\beta})
\end{align}
where we need to integrate over the vielbein $e_{a\alpha}$ and the Lagrange multiplier $X^a$ and $V_{\text{diff}}$ is the volume of reparameterization group. The idea of performing the path integral is similar to computing the one-loop vacuum energy in string theory \cite{Polchinski:1985zf}. For the different variables, we decompose them to various pieces that we need to integrate over. For the vielbein, it can be shown that a generic vielbein on a torus can be decomposed in the following way
\begin{align}
e^a_{\alpha}(x)=\re^{\Omega(x)}\left(e^{\epsilon\phi(x)}\right)^a_b\hat{e}^b_{\alpha}(\tau),
\end{align}
where $\epsilon$ in the exponent is the Levi-Civita symbol and $\hat{e}^b_{\alpha}(\tau)$ is a constant vielbein depending on the modular parameter $\tau=\tau_1+\ri\tau_2$ given by
\begin{align}
\hat{e}^b_{\alpha}(\tau)=\left(
                           \begin{array}{cc}
                             1 & \tau_1 \\
                             0 & \tau_2 \\
                           \end{array}
                         \right)
\end{align}
From this decomposition, it is clear that integrating over $e^a_{\alpha}(x)$ is equivalent to performing functional integration over $\Omega(x)$, $\phi(x)$ together with usual integral over modular parameter $\tau$.\par

The quantity $X^a(x)$ can be decomposed as
\begin{align}
X^a(x)=|\Lambda|^{1/2}L^a_{\mu}x^\mu+Y^a(x)
\end{align}
where
\begin{align}
L_{\mu}^a=(L_{\mu},L'_{\mu})
\end{align}
are the parameters that parameterize the parallelogram (or the torus) and $Y^a(x)$ are scalar fields that are periodic in $x^{\mu}$. Notice that on a torus the coordinates $x^\mu$ are periodic and we can choose
\begin{align}
0\le x^{\mu}<|\Lambda|^{-1/2}.
\end{align}
So the integration over $X^a(x)$ can be traded by the integration over $x^\mu$ and functional integration over $Y^a(x)$.\par

To proceed, one further decomposes the integrations over $\Omega(x)$ and $\phi(x)$ into constant part (slow) and its orthogonal part (fast) as
\begin{align}
\Omega(x)=\bar{\Omega}+\Omega'(x),\qquad \phi(x)=\bar{\phi}+\phi'(x)
\end{align}
The upshot of the calculation is that, as it turns out, the functional integrals over $Y^a(x)$, $\Omega'(x)$, $\phi'(x)$ can be performed and we are left with only the usual integrals over the constant parts $\tau,\bar{\Omega},\bar{\phi}$.\par

Notice that the functional $Z_0(g_{\alpha\beta})$ also depends on $\Omega'(x)$ and $\phi'(x)$, so it might seem a bit surprising that such integrals can be performed. The reason is the following. The functional $Z_0(g_{\alpha\beta})$ does not depend on $Y^a(x)$, so we can first integrate out $Y^a(x)$. This can be done straightforwardly. The integrand is Gaussian and the exponent turns out to be linear in $Y^a(x)$ after neglecting total derivatives. Therefore the integration over $Y^a(x)$ leads to certain $\delta$-functional. Written out explicitly, the $\delta$-functional contains the fast moving factors $\delta(\Omega'(x))\delta(\phi'(x))$, but not the constant pieces $\bar{\Omega}$ and $\bar{\phi}$. That's the reason why we can get rid of the functional integrals over $\Omega'(x)$ and $\phi'(x)$.\par

After performing the functional integrals, we are left with the following usual integral
\begin{align}
\label{eq:resultInt}
Z_{\rT\overline{\rT}}=\frac{|\Lambda|Ae^{-|\Lambda|A}}{(2\pi)^2}\int_{-\infty}^{\infty}\rd\bar{\Omega}\int_0^{2\pi}\rd\bar{\phi}
\int_{\rP}\frac{\rd^2\bar{\tau}}{\bar{\tau}_2}
\re^{\frac{\Lambda}{|\Lambda|}\epsilon^{\alpha\beta}\epsilon_{ab}\left(\sqrt{|\Lambda|}L^a_{\alpha}\bar{e}^b_{\beta}-\frac{1}{2}\bar{e}^a_{\alpha}\bar{e}^b_{\beta} \right)}Z_0(\bar{g}_{\alpha\beta}),
\end{align}
where $A$ is the area of the torus. We put a bar on the quantities to indicate that these are usual variables instead of functions of $x^{\mu}$. The above equation looks like an integral transformation of the original partition function. The fact that we can get rid of the functional integrations and are left with only constant deformations is also consistent with random geometry point of view, where we argued that the deformation of the geometry is uniform and does not depend on the position. This can be traced back to the fact that the expectation value of the $\rT\overline{\rT}$ operator is constant.

\paragraph{Equivalence to $\rT\overline{\rT}$ deformed partition sum} Now we need to show that the integral (\ref{eq:resultInt}) is indeed equivalent to the $\rT\overline{\rT}$ deformed partition sum. We recall that a constant vielbein can be parameterized by $\bar{\Omega},\bar{\phi}$ and $\bar{\tau}_1,\bar{\tau}_2$ which are four parameters. We can make a change of variables to bring the integral (\ref{eq:resultInt}) to different forms. For example, one interesting parameterization of the vielbein is
\begin{align}
\bar{e}^a_{\alpha}=\sqrt{\Lambda}(\bar{L}_{\alpha},\bar{L}'_{\alpha}).
\end{align}
Intuitively, these can be seen as the parameters that specify the shape of deformed torus which we want to integrate over. Then the integral (\ref{eq:resultInt}) can be interpreted as summing over all possible shapes of the torus with a specific weight. More explicitly, we have
\begin{align}
\label{eq:defTorus}
Z_{\rT\overline{\rT}}=\Lambda^2 A \re^{-\Lambda A}\int_{\bar{A}>0}\frac{\rd^4\bar{L}}{(2\pi)^2\bar{A}}
\re^{\Lambda\epsilon^{\alpha\beta}\epsilon_{ab}(L^a_{\alpha}\bar{L}^b_{\beta}-\frac{1}{2}\bar{L}^a_{\alpha}\bar{L}^b_{\beta})}
Z_0(\bar{g}_{\alpha\beta}),
\end{align}
where $\bar{A}$ is the surface area of the deformed torus. This representation makes the random geometry point of view quite manifest.\par

To bring $Z_{\rT\overline{\rT}}$ to a form that is closer to our discussions in section~\ref{lecture2}, we first assume that the original partition function can be written in the form of sum over Boltzmann weights
\begin{align}
Z_{0}=\sum_n \re^{2\pi \ri\tau_1 R P_n-2\pi R\tau_2 R\mathcal{E}_n(R)},
\end{align}
where
\begin{align}
R=\sqrt{|t|}e^{\Omega},\qquad P_n=\frac{2\pi \ri k_n}{R}.
\end{align}
We can perform another change of variables and parameterize the torus by $L_1,L_2,\tau_1,\tau_2$ by
\begin{align}
L_1=\sqrt{|t|}\re^{\Omega}\cos\phi,\qquad L_2=\sqrt{|t|}\re^{\Omega}\sin\phi.
\end{align}
where we have used $t=-2/\Lambda$. Similarly, for the deformed torus, we define $\bar{L}_1,\bar{L}_2$. Then the deformed partition function can be written as
\begin{align}
Z_{\rT\overline{\rT}}=\sum_n\frac{A}{(2\pi)^2t}\re^{A/t}\int_{-\infty}^{\infty}\rd^2\bar{L}\int_{\rP}\frac{\rd^2\bar{\tau}}{\bar{\tau}_2}
\re^{\frac{1}{t}\left(\bar{R}^2\bar{\tau}_2-R(\bar{L}_1(\bar{\tau}_2+\tau_2)+\bar{L}_2(\bar{\tau}_1-\tau_1))\right)}
\re^{2\pi \ri k_n\bar{\tau}_1-2\pi\bar{\tau}_2\bar{R}E_n(\bar{R})}.
\end{align}
Notice that we have
\begin{align}
\bar{R}^2=\bar{L}_1^2+\bar{L}_2^2.
\end{align}
Let us consider the case where the original theory is a CFT. Then the dependence on $\bar{R}$ takes a simple form
\begin{align}
E_n(R)=\frac{1}{R}\left(n+\bar{n}-\frac{c}{12}\right)=\frac{1}{R}\mathbb{E}_n,
\end{align}
where $\mathbb{E}_n$ is a pure number and does not depend on $\bar{R}$. Then the integral over $\bar{L}_1$ and $\bar{L}_2$ can be performed since it is Gaussian and we obtain
\begin{align}
Z_{\rT\overline{\rT}}^{\text{CFT}}=\sum_n\frac{A}{\pi t}\re^{A/t}\int_{\rP}\frac{\rd^2\bar{\tau}}{\bar{\tau}_2^2}
\re^{-\frac{R^2}{4t\tau_2}\left((\bar{\tau}_1-\tau)^2+(\tau_2+\bar{\tau}_2)^2 \right)}
\re^{-2\pi\bar{\tau}_2\mathbb{E}_n+2\pi \ri\bar{\tau}_1k_n}.
\end{align}
We can then perform the integration over $\bar{\tau}_1$ and $\bar{\tau}_2$ and obtain the expected result
\begin{align}
Z_{\rT\overline{\rT}}^{\text{CFT}}=\sum_n \re^{2\pi \ri\tau_1 k_n-2\pi\tau_2\mathcal{E}_n(R,t)},
\end{align}
where $\mathcal{E}_n(R,t)$ is the $\rT\overline{\rT}$ deformed spectrum which we found in section~\ref{sec:def-spectrum}.

\subsection{Holography}
\begin{figure}[h!]
\begin{center}
\includegraphics[scale=0.5]{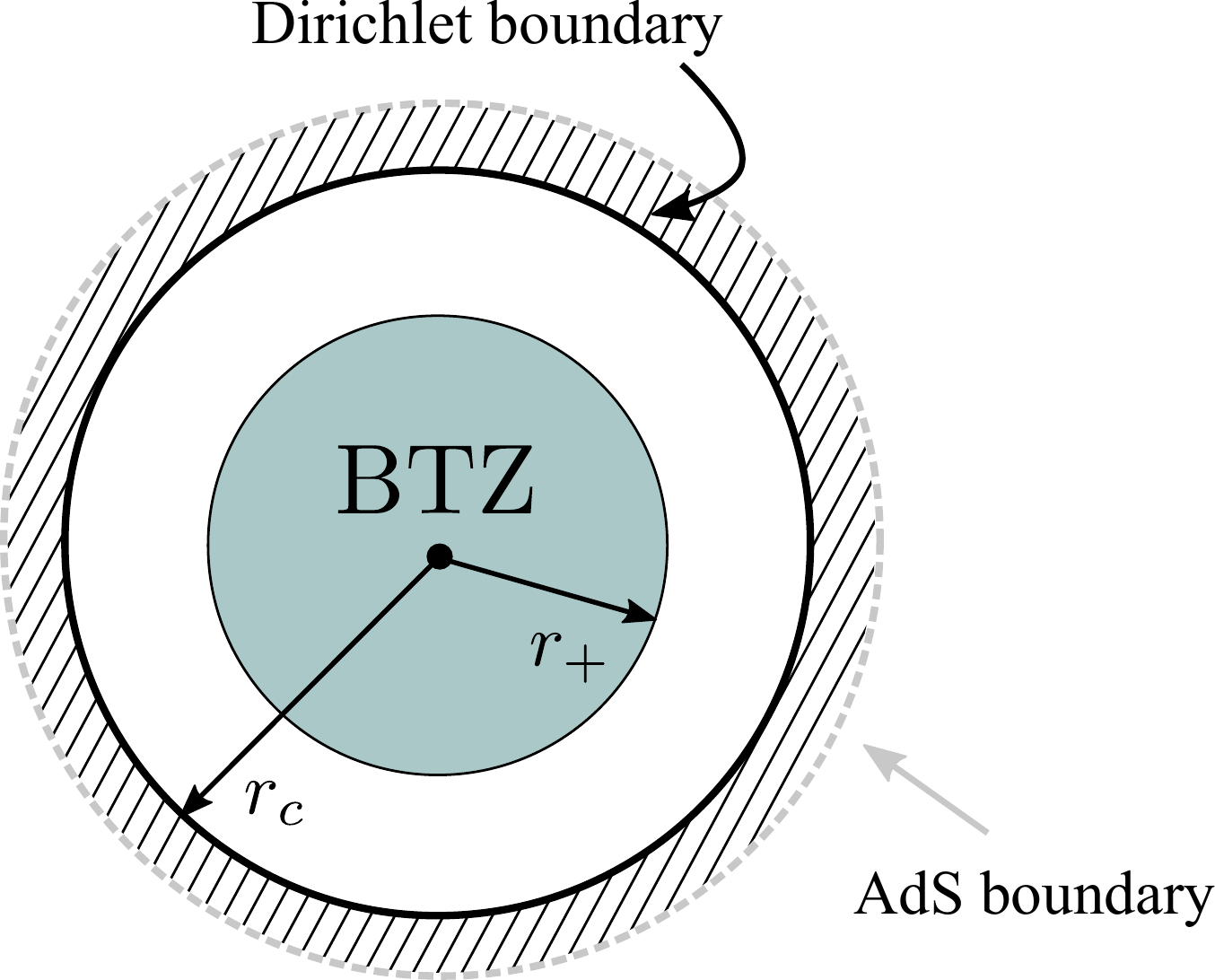}
\caption{The cut-off geometry. Turning on $\rT\overline{\rT}$ deformation for the bad sign is equivalent to putting a Dirichlet boundary condition at finite radius.}
\label{fig:cutoff}
\end{center}
\end{figure}

The $\rT\overline{\rT}$ deformation can be defined for any 2d QFT and in particular CFTs. From AdS/CFT point of view, it is a very natural question to ask what is the holographic interpretation of the $\rT\overline{\rT}$ deformation. This question was partly answered by a proposal by McGough, Mezei and Verline \cite{McGough:2016lol} (see also \cite{Kraus:2018xrn}). The proposal is rather simple, which is that the $\rT\overline{\rT}$ deformation correspond to putting Dirichlet boundary condition in the bulk at \emph{finite} radius, as is shown in figure~\ref{fig:cutoff}.
In what follows we first introduce the main proposal in more detail and then discuss some of the evidences that support this proposal. We mainly follow the paper \cite{Kraus:2018xrn} in the derivations below.

\subsubsection*{The cut-off picture}
We give more details about the MMV proposal \cite{McGough:2016lol}. Let us consider the action for pure gravity in AdS$_3$ space
\begin{align}
S=-\frac{1}{16\pi G}\int_{M}\rd^3x\sqrt{g}(R+2\ell^{-2})-\frac{1}{8\pi G}\int_{\partial M}\rd^2 x\sqrt{h}(K-\ell^{-1}),
\end{align}
where the second term is the Gibbons-Hawking-York boundary term. We need this term in a geometry with boundary. Different from the usual case, where the boundary is put at infinity, we put the boundary or Dirichlet wall at finite radius direction. The tensor $h_{ij}$ is the induced metric on the boundary and $K_{ij}$ is the extrinsic curvature. We write the metric in the following coordinate system
\begin{align}
\rd s^2=\rd r^2+g_{ij}(x,r)\rd x^i\rd x^j.
\end{align}
In this coordinate system, the extrinsic curvature is given by
\begin{align}
K_{ij}=\frac{1}{2}\partial_{r}g_{ij}.
\end{align}
We take the AdS radius to be $\ell=1$, the Einstein equation in this coordinate system
\begin{align}
R_{\mu\nu}-\frac{1}{2}R g_{\mu\nu}-g_{\mu\nu}=0
\end{align}
is given by
\begin{align}
\label{eq:EinsteinEq}
E_j^i=&\,-\partial_{r}(K^i_j-\delta^i_j K)-K K^i_j+\frac{1}{2}\delta^i_j\left[K^{mn}K_{mn}+K^2\right]-\delta^i_j=0,\\\nonumber
E_j^{r}=&\,\nabla^i\left(K_{ij}-K g_{ij}\right)=0,\\\nonumber
E_{r}^{r}=&-\frac{1}{2}R^{(2)}+\frac{1}{2}\left[K^2-K^{ij}K_{ij}\right]-1=0.
\end{align}
In terms of $g_{ij}$, the action after proper integration by parts takes the following form
\begin{align}
S=-\frac{1}{16\pi G}\int \rd^3x\sqrt{g}\left(R^{(2)}+K^2-K^{ij}K_{ij}+2\right)+\frac{1}{8\pi G}\int_{\partial M}\rd^2x\sqrt{h}.
\end{align}
The quasi-local stress energy tensor on the boundary is defined via the on-shell variation of the action
\begin{align}
\delta S=\frac{1}{4\pi}\int \rd^2x\sqrt{h}T^{ij}\delta h_{ij}.
\end{align}
This yields the stress energy tensor of the form
\begin{align}
\label{eq:TKK}
T_{ij}=\frac{1}{4G}(K_{ij}-Kg_{ij}+g_{ij}).
\end{align}
In AdS/CFT dictionary, we identify this quasi-local stress energy tensor with the $\rT\overline{\rT}$ deformed stress energy tensor of the boundary theory. The crucial point is that via (\ref{eq:TKK}) the Einstein equation (\ref{eq:EinsteinEq}) can be rewritten in terms of the tensor $T_{ij}$. This leads to equations about $T_{ij}$. For example, the Einstein equation $E_i^{r}=0$ leads to the conservation of stress energy tensor
\begin{align}
\nabla^i T_{ij}=0.
\end{align}

\subsubsection*{Trace flow equation}
Now let us see what does $E^{r}_{r}=0$ tell us about the quasi-local stress energy tensor. The trace of the stress energy tensor is
\begin{align}
\label{eq:traceT}
T_{ij}g^{ij}=\frac{1}{4G}(K-2K+2)=\frac{1}{4G}(2-K).
\end{align}
The $\rT\overline{\rT}$ operator can be written as\footnote{Here we follow the convention of \cite{Kraus:2018xrn} for the definition of $\rT\overline{\rT}$ operator, which is slightly different from the previous sections.}
\begin{align}
\rT\overline{\rT}=\frac{1}{8}\left(T^{ij}T_{ij}-(T^i_i)^2\right)=\frac{1}{128G^2}\left(K_{ij}K^{ij}-K^2+2K-2\right)
\end{align}
Now using the Einstein equation $E^{r}_{r}=0$, we have
\begin{align}
K^2-K^{ij}K_{ij}=R^{(2)}-2.
\end{align}
Therefore we can write
\begin{align}
\label{eq:TTbvalue}
\rT\overline{\rT}=-\frac{1}{64G^2}(2-K)-\frac{R^{(2)}}{128G^2}.
\end{align}
When the boundary metric is flat, we have $R^{(2)}=0$. From (\ref{eq:traceT}) and (\ref{eq:TTbvalue}) we see that
\begin{align}
\label{eq:trceflowBulk}
\Theta=-16G\,\rT\overline{\rT}.
\end{align}
This is the trace flow equation for the $\rT\overline{\rT}$ deformed CFT which can also be derived from the boundary QFT, as we will show shortly. Notice that the discussion above is general and works for any solution of Einstein equation with flat boundary metric.\par

Now we derive the trace flow equation from the boundary QFT. Let us denote the action of the theory as $S(t)$. From the definition of $\rT\overline{\rT}$ deformation,
\begin{align}
\label{eq:dSt}
\frac{\rd S(t)}{\rd t}=\int \rd^2x\,\sqrt{g}\,\rT\overline{\rT}
\end{align}
If the theory has a single mass scale $\mu$ (which is the case for the $\rT\overline{\rT}$ deformed CFT), the infinitesimal scale transformation is given by the trace of the stress energy tensor as
\begin{align}
\label{eq:dmumu}
\mu\frac{\rd}{\rd\mu}S(\lambda)=\int \rd^2x\sqrt{g}\,\Theta.
\end{align}
The mass scale is set by the dimensionful parameter $t=1/\mu^2$. Therefore by comparing (\ref{eq:dSt}) and (\ref{eq:dmumu}), we find that
\begin{align}
\label{eq:traceflowBd}
\Theta=-2t\,\rT\overline{\rT}.
\end{align}
Notice that this is an operator statement and is true inside another correlation functions. We find that (\ref{eq:trceflowBulk}) and (\ref{eq:traceflowBd}) matches exactly if we identify $t=8G$. We see that the bulk and boundary derivations of the trace flow equation has very different starting points, yet they lead to the same equation. This is an evidence for the cut-off geometry proposal.

Another comment is that using the trace flow equation (\ref{eq:traceflowBd}) and the factorization formula for the expectation value of the $\rT\overline{\rT}$ operator, we can also derive the spectrum of the $\rT\overline{\rT}$ deformed CFTs. Therefore, we can view the trace flow equation (\ref{eq:traceflowBd}) as an alternative definition of the $\rT\overline{\rT}$ deformation for CFTs.

The trace flow equation can be generalized to curved spacetime. For maximally symmetric spacetime with constant curvature, it is simply given by
\begin{align}
\Theta=-2t\rT\overline{\rT}-\frac{c}{24\pi}R^{(2)},
\end{align}
where $R^{(2)}$ is the Ricci scalar of the 2d boundary spacetime.\par

\subsubsection*{Deformed spectrum from the bulk}
In this subsection, we show how to obtain the deformed finite volume spectrum from the bulk. To this end, we consider the Euclidean black hole solution. The metric of the Euclidean BTZ is given by
\begin{align}
\rd s^2=\frac{\rd r^2}{f(r)^2}+f(r)^2\rd\tau^2+r^2(\rd\phi-i\omega(r)\rd\tau)^2
\end{align}
where
\begin{align}
f(r)^2=r^2-8GM+\frac{16G^2J^2}{r^2},\qquad \omega(r)=\frac{4GJ}{r^2}.
\end{align}
The quasi-local energy is given by
\begin{align}
\label{eq:int}
\mathcal{E}=\int\frac{\rd\phi}{2\pi}\sqrt{g_{\phi\phi}}u^iu^jT_{ij}
\end{align}
where $T_{ij}$ is the quasi-local stress energy tensor (\ref{eq:TKK}) and $u^i=(u^\tau,u^\phi)$ is the unit vector normal to the constant $\tau$ slice. We have
\begin{align}
u^\tau=1/f,\qquad u^\phi=\ri\omega/f.
\end{align}
The size of the spacial circle is given by
\begin{align}
L=\int \rd\phi\sqrt{g_{\phi\phi}}=2\pi r.
\end{align}
Using the prescription for the definition of the quasi-local stress energy tensor and the explicit form of the metric, we can compute the integral (\ref{eq:int}). The result is given by
\begin{align}
\label{eq:defEbulk}
\mathcal{E}=&\,\frac{L}{\pi t}(1-t^{-1}f(r)),\\\nonumber
=&\,\frac{L}{\pi t}\left(1-\sqrt{1-\frac{2\pi t}{L}M+\left(\frac{\pi t}{L}J\right)^2}\right),
\end{align}
where we have used the relation $t=8G$. Identifying the parameters properly, this indeed takes the form of the deformed finite volume spectrum, which constitute another strong evidence for the proposal.

Before we end this section. We make two comments about the cut-off geometry proposal. First of all, the cut-off geometry only works for the bad sign of the deformation parameter. This can be seen with the identification $t=8G$ and the formula for the deformed energy (\ref{eq:defEbulk}).\par

Secondly, there are other checks which provides further evidence for the cut-off geometry. These include signal propagation speed \cite{McGough:2016lol}, sphere partition function \cite{Caputa:2019pam} and entanglement entropy \cite{Donnelly:2018bef,Banerjee:2019ewu,Murdia:2019fax}. However, the cut-off geometry proposal is only valid for pure gravity sector. If there are other fields, say a bulk scalar field, then the proposal is no longer true \cite{Hartman:2018tkw,Kraus:2018xrn}.

\section{New developments}
\label{newDevp}
\subsection{Introduction}
\label{sec:gen}
The contents in the previous sections constitute the lectures which the author gave in December 2018. The lecture notes were completed in March 2019. As we mentioned at the beginning, these lectures were intended for non-experts who wanted to study the subject for the first time. Therefore more emphasis is put on pedagogy instead of completeness. Due to this reason, the topics chosen in these lectures are rather limited and are mainly related to the authors own works. However, $\rT\overline{\rT}$ deformation has been a highly active research area in the past few years and many exciting new developments have been made since March 2019. Any attempt for reviewing a quickly developing field runs the potential danger of being out-of-dated quickly. To compensate the very limited choice of the subjects discussed so far in the previous sections, and also to keep the readers more up to date, we give a concise summary for more recent developments. By recent, we mean the developments up to January 2021. Our main goal here is to provide a convenient literature guide for the reader with some short comments.\par

To make the discussions more organized, we classify the developments into the following loose categories
\begin{itemize}
\item \textbf{Generalizations}. The $\rT\overline{\rT}$ deformation is a solvable irrelevant deformation for \emph{2d relativistic QFTs in flat spacetime}. It can be generalized in various directions. In fact, every word in the emphasized part have been generalized to some extent. For example, there have been attempts to generalize $\rT\overline{\rT}$ deformations to \emph{curved spacetime}, to \emph{non-relativistic} systems, to \emph{non-QFTs}, and to \emph{higher dimensions}. We will summarize these generalizations in section~\ref{sec:gen}.
\item \textbf{Other physical quantities}. In the previous sections, we have discussed a few physical quantities including the deformed classical Lagrangian, the deformed finite volume spectrum, quantum S-matrix and torus partition function. There are many other interesting physical quantities that one would like to compute in order to gain deeper understanding of $\rT\overline{\rT}$ deformation. For example, two classes of quantities of great interest are \emph{correlation functions} and \emph{quantum information measures}. These quantities are typically harder to compute than the ones we have discussed, but solid progress have been made in the past few years, which will be summarized in section~\ref{sec:quantity}.
\item \textbf{Holography}. Shortly after the original papers of $\rT\overline{\rT}$ deformation, the cut-off geometry was proposed as the holographic dual. Important developments have been made in several directions in the past few years. The first direction is \emph{refinements} and \emph{further discussions} of the cut-off geometry picture; the second is applying the holographic proposal to compute more physical quantities in the bulk. The last direction is going \emph{beyond the cut-off geometry} proposal. It is known that the cut-off geometry proposal has serious limits such as it is valid only for one sign of the deformation parameter and only for the pure gravity sector. Therefore the cut-off geometry cannot be the whole story and alternative holographic dictionaries are called for. Developments in these directions will be summarized in section~\ref{sec:holo}.
\item \textbf{Single-trace $\rT\overline{\rT}$ deformation}. The $\rT\overline{\rT}$ and other solvable irrelevant deformations we discussed so far belong to the double-trace deformations. There has been a parallel development for a closely related deformation called the \emph{single-trace} $\rT\overline{\rT}$ deformation shortly after the original papers of $\rT\overline{\rT}$ deformation. The single-trace $\rT\overline{\rT}$ deformation is defined for orbifold CFTs, which are holographic dual of string theories on $AdS_3\times\mathcal{N}$ background. This deformation share many properties of the double trace $\rT\overline{\rT}$ deformation. Moreover, it can be realized on the string worldsheet as a marginal deformation. This allows one to apply powerful worldsheet techniques to study this deformation. We will summarize some developments in this direction in section~\ref{sec:singletrace}.
\item \textbf{String theory and 2d gravity}. One of the early motivations for studying $\rT\overline{\rT}$ deformation comes from the study of effective string theory. Therefore the connections between $\rT\overline{\rT}$ deformation, string theory and 2d gravity have been expected to some extent. In particular, as we discussed in section~\ref{lecture3}, $\rT\overline{\rT}$ deformation of a QFT can be re-interpreted as coupling the theory to a 2d topological gravity. These important intuitions about $\rT\overline{\rT}$ deformation have been clarified further and generalized and will be summarized in section~\ref{sec:holo}.
\end{itemize}
We would like to emphasis that this classification is not strict because some works fall into several categories while some other works cannot be easily classified.

\subsection{Generalizations}
\label{sec:gen}
In this section, we briefly summarize the generalizations of $\rT\overline{\rT}$ deformation of 2d relativistic QFTs in various directions. All the generalizations share certain common features such as non-locality and solvability in some proper sense. At the same time, different generalizations come with their own motivations and new features.

\paragraph{More general bilinear deformations} $\rT\overline{\rT}$ deformation can be defined for any QFT with a local stress energy tensor. In this sense, it is the most general solvable irrelevant deformation. If the QFT under consideration enjoys \emph{more symmetries}, it is possible to construct other solvable deformations.\par

The first deformation of this kind proposed in the literature is the $\rJ\overline{\rT}$ deformation \cite{Guica:2017lia}. A slight variant is the very special $\rJ\overline{\rT}$ deformation \cite{Nakayama:2019mvq,Nakayama:2018ujt}. A generalization of the $\rJ\overline{\rT}$ deformation is the $\rJ\rT_a$ deformation \cite{Anous:2019osb}. All these deformations require an additional U(1) symmetry, which is preserved along the flow of the deformation. These deformations are also irrelevant solvable deformations and are somewhat simpler than the $\rT\overline{\rT}$ deformation. For example, the $\rJ\overline{\rT}$ deformed 2d CFT preserves one copy of the Virasoro algebra, which allows the non-perturbative computation of correlation functions \cite{Guica:2019vnb}.
In addition, we can combine $\rT\overline{\rT}$ and $\rJ\overline{\rT}$ deformations and define a more general joint $\rT\overline{\rT}+\rJ\overline{\rT}+\overline{\rJ}\rT$ deformation. This joint deformation is obviously more complicated, yet it is still solvable. For example, the deformed spectrum \cite{Frolov:2019xzi} and torus partition function \cite{Aguilera-Damia:2019tpe,Chakraborty:2019mdf} have been studied.

For integrable QFTs and 2d CFTs, there are even more possibilities because these theories possess an infinite tower of conserved currents.
Irrelevant deformations triggered by operators which are constructed from higher conserved currents have been discussed in the original paper by Smirnov and Zamolodchikov \cite{Smirnov:2016lqw}. Such deformations have been studied further in \cite{LeFloch:2019rut} for CFTs and in \cite{Hernandez-Chifflet:2019sua,Conti:2019dxg} for integrable quantum field theories.\par

Yet another interesting additional symmetry is supersymmetry. $\rT\overline{\rT}$ deformation of supersymmetric quantum field theories have also been studied in \cite{Baggio:2018rpv,Chang:2018dge,Jiang:2019hux,Chang:2019kiu,Coleman:2019dvf,Jiang:2019trm,Ferko:2019oyv,Ebert:2020tuy} for different numbers of supercharges.

\paragraph{Other spacetimes} The original $\rT\overline{\rT}$ deformation is defined for QFTs in 2d flat spacetime. Therefore it is natural to ask whether we can define $\rT\overline{\rT}$ deformation for QFTs on curved spacetime and in higher dimensions.\par

The generalization of $T\bar{T}$ deformation to curved spacetimes \emph{at the classical level} is straightforward. The deformed classical Lagrangian on curved spacetime was first derived in \cite{Bonelli:2018kik} and later found by different approaches in \cite{Caputa:2020lpa}. At the quantum level, it is easier to consider the large central charge limit where the large $c$ factorization ensures the factorization of the expectation value of the $T\bar{T}$ operator, which underlies the solvability of the quantum theory. One important application in this limit is the calculation of partition functions for maximally symmetric spaces by using the trace flow equation \cite{Donnelly:2018bef,Caputa:2019pam}. An interesting deformation which is related but different from the $T\bar{T}$ deformation can be defined in the context of dS/dS correspondence \cite{Gorbenko:2018oov}. It is more difficult to take into account finite $c$ corrections. The factorization property of the expectation value of $T\bar{T}$ operator is studied \cite{Jiang:2019tcq,Brennan:2020dkw}. A definition of $T\bar{T}$ deformation in curved spacetime inspired from 3D gravity is proposed in \cite{Mazenc:2019cfg}.

The generalization of $\rT\overline{\rT}$ deformation to higher dimension is basically an open question. One of the reason that, even before any kind of $\rT\overline{\rT}$-like solvable deformation, there are very few examples of higher dimensional QFTs that are solvable apart from free theories. Therefore, defining a `solvable' deformation for generic higher dimensional QFTs seems a bit pointless. One possibility is to focus on certain specific theory that is solvable, and try to define a solvable deformation similar to $\rT\overline{\rT}$ deformation. One of the most interesting theories of this type is the $\mathcal{N}=4$ SYM theory, which is integrable in the planar limit. An interesting irrelevant deformation of $\mathcal{N}=4$ SYM is proposed in \cite{Caetano:2020ofu}. The possible string theory dual are discussed in \cite{Sfondrini:2019smd,Chakraborty:2020nme}.

\paragraph{Specific theories} It is also interesting to study $\rT\overline{\rT}$ deformation for specific models, although strictly speaking, these studies are not `generalizations' of $\rT\overline{\rT}$ deformation. The specific models that have been studied under $\rT\overline{\rT}$ deformation include chiral bosons \cite{Ouyang:2020rpq,Chakrabarti:2020dhv}, compactified boson \cite{Beratto:2019bap}, WZW model \cite{He:2020hhm}, Sinh-Gordon models \cite{Conti:2018jho}, Heisenberg model \cite{Nastase:2020evb}, large $N$ vector model \cite{Haruna:2020wjw}, 2d VA model \cite{Cribiori:2019xzp}, Liouville related theories \cite{Leoni:2020rof,Okumura:2020dzb}, 2d dilaton gravity \cite{Ishii:2019uwk} and the 2d Yang-Mills theory \cite{Santilli:2018xux,Ireland:2019vvj,Santilli:2020qvd,Gorsky:2020qge}.

\paragraph{Beyond relativistic QFTs} $\rT\overline{\rT}$ and other deformations have been generalized beyond relativistic QFTs. The first generalization is to non-relativistic QFTs \cite{Cardy:2018jho,Cardy:2020olv,Blair:2020ops,Hansen:2020hrs,Ceschin:2020jto,Chen:2020jdi} where one does not assume Loretnz invariance. Similar deformations can be defined even for non-QFTs. Examples include quantum mechanics \cite{Gross:2019uxi,Gross:2019ach,Chakraborty:2020xwo}, quantum integrable spin chains \cite{Bargheer:2008jt,Pozsgay:2019ekd,Marchetto:2019yyt} and the 1d Bose gas \cite{Jiang:2020nnb}. Deformations which are analogous but different from $\rT\overline{\rT}$ deformation have also been considered. These include DBI action analogue \cite{Brennan:2019azg} and non-linear electrodynamics \cite{Babaei-Aghbolagh:2020kjg}.

\subsection{Other quantities}
\label{sec:quantity}
In the previous sections, we have discussed several physical quantities of $\rT\overline{\rT}$ deformed theories. In the past few years, more quantities have been studied. Most notably, correlation functions of local operators and quantum information measures such as entanglement entropies.

\paragraph{More on torus} Before moving to the discussion of new quantities, let us mention that there have been more progress on torus partition functions and thermodynamics of the deformed theories. The torus partition function has been computed perturbatively in \cite{He:2020cxp}. Deformation of KdV charges was first studied in \cite{LeFloch:2019wlf} and later generalized to the torus case in \cite{Asrat:2020jsh}. Some general comments on thermal instability is given in \cite{Barbon:2020amo}. An elegant method to obtain the torus partition function for the $\rT\overline{\rT}$ and more general $\rT\overline{\rT}+\rJ\overline{\rT}+\overline{\rJ}\rT$ deformed CFTs via integral transformation have been proposed. The integral transformation method for $\rT\overline{\rT}$ deformation first appeared in \cite{Dubovsky:2018bmo} from 2d gravity proposal. Later it was developed largely in the single trace deformed theories \cite{Chakraborty:2020xyz,Hashimoto:2019hqo,Hashimoto:2019wct} where the integral kernels were derived from worldsheet techniques. Due to universality, the results are also applicable to the double trace deformations. An alternative derivation for the integral kernel based on 3d gravity perspective similar to \cite{Dubovsky:2018bmo} is given in \cite{Aguilera-Damia:2019tpe}.

\paragraph{Quantum information measures} Quantum information has played an important role in recent developments of theoretical physics. It is therefore a natural and interesting question how $\rT\overline{\rT}$ and other solvable irrelevant deformations affect quantum information measures. The most widely studied measures are the R\'enyi and entanglement entropies. The $\rT\overline{\rT}$ deformed entanglement entropy was first studied in \cite{Donnelly:2018bef} and further developed in \cite{Banerjee:2019ewu,Grieninger:2019zts,Donnelly:2019pie,Lewkowycz:2019xse} on a spherical geometry in the large $c$ limit. Perturbative calculations of the R\'enyi and entanglement entropies have been performed in \cite{Chen:2018eqk,Sun:2019ijq,Jeong:2019ylz,He:2019vzf}. Holographic computations of $\rT\overline{\rT}$ deformed quantum information measures and their implications have been studied further in \cite{Lewkowycz:2019xse,Chen:2019mis,Ota:2019yfe,Asrat:2019end,Park:2018snf,Asrat:2020uib,Paul:2020gou,Caputa:2020fbc,Li:2020zjb}. Another important quantum information measure, the holographic complexity have been studied in \cite{Geng:2019yxo,Hashemi:2019xeq,Goto:2018iay,Akhavan:2018wla,Chakraborty:2020fpt}.

\paragraph{Correlation functions} Correlation functions are much more complicated than spectrum or partition functions, even in the undeformed theories. Correlation functions for $\rT\overline{\rT}$ deformed CFTs was first studied in the large $c$ limit in \cite{Aharony:2018vux}. The correlation functions for the $\rJ\overline{\rT}$ deformed CFTs was first considered in \cite{Guica:2019vnb}. An important non-perturbative result is the flow equation for the $\rT\overline{\rT}$ deformed correlation functions, which was first derived in \cite{Cardy:2019qao}. It was later derived from a different approach in \cite{Kruthoff:2020hsi}. Perturbative calculations based on conformal perturbation theory to the first order for the $\rT\overline{\rT}$ and $\rJ\overline{\rT}$ deformations have been performed in \cite{He:2019ahx,He:2020udl,Li:2020pwa,Hirano:2020ppu,He:2020qcs}. Perturbative calculations based on Feynmann diagrams have been studied in \cite{Rosenhaus:2019utc,Dey:2020gwm}.

\paragraph{Other aspects} Another interesting direction is to consider physics out of equilibrium. Quench dynamics has been considered in \cite{Medenjak:2020ppv,Medenjak:2020bpe} for $\rT\overline{\rT}$ deformed CFTs. Finally, it is important to understand how symmetry algebra deforms under $\rT\overline{\rT}$ and related deformations, especially for deformed 2d CFTs. This will shed new lights on solvability of the deformations. The deformed classical algebra for $\rT\overline{\rT}$, $\rJ\overline{\rT}$ are considered in \cite{Guica:2020uhm,Guica:2020eab}.

\subsection{Holography}
\label{sec:holo}
The cut-off geometry picture was first proposed in \cite{McGough:2016lol}. It was further refined and developed in \cite{Kraus:2018xrn,Hartman:2018tkw,Shyam:2017znq,Taylor:2018xcy,Cottrell:2018skz,Shyam:2018sro,Jafari:2019qns,Coleman:2020jte}. This proposal is very intuitive and attracted a lot of interest. Due to its simplicity, one can apply it readily to compute various quantities in the bulk and compare with the boundary results. Works in this direction include \cite{Wang:2018jva,He:2019glx,Khoeini-Moghaddam:2020ymm,Belin:2020oib,Caputa:2020fbc,Bzowski:2020umc}.

The cut-off geometry proposal has a number of shortcomings. It only works for pure gravity sector, and it only works for one sign of the deformation parameter. From general principles of AdS/CFT, one expect that $\rT\overline{\rT}$ and $\rJ\overline{\rT}$ deformations, being double-trace deformations, should correspond to a change of the boundary conditions. The proposal in this direction was first formulated for the $\rJ\overline{\rT}$ deformation \cite{Bzowski:2018pcy} and later generalized to the $\rT\overline{\rT}$ deformation \cite{Guica:2019nzm}. Another holographic proposal based on random geometry picture was given in \cite{Hirano:2020nwq}.

\subsection{Single-trace deformations}
\label{sec:singletrace}
Soon after the proposal of $\rT\overline{\rT}$ deformation. It was realized that in the context of holography of AdS$_3$, one can define a \emph{single trace} $\rT\overline{\rT}$ deformation \cite{Giveon:2017myj}. Such deformation is defined for orbifold CFTs where one needs to take $N$ copies of a CFT. From the stress tensor of the each copy, one can define two $\rT\overline{\rT}$ operators for the orbifold theory. One is the usual double-trace $\rT\overline{\rT}$ operator, and the other one is the single trace $\rT\overline{\rT}$ operator which triggers the single-trace $\rT\overline{\rT}$ deformation.\par

From general principles of AdS/CFT, double trace deformations modify the boundary conditions and do not change the bulk geometry. On the other hand, single trace deformations do change the bulk geometry. Therefore, we expect that the singe trace $\rT\overline{\rT}$ deformation modifies the bulk geometry, which is indeed the case. The undeformed orbifold CFT corresponds to strings on AdS$_3$ geometry. It is found that for one sign of the deformation parameter, the deformed geometry on which the strings propagate interpolates between AdS$_3$ in the IR and a linear dilaton space in the UV. This geometry is the two dimensional vacuum of the little string theory \cite{Giveon:2017myj}. For the other sign, the resulting bulk geometry has curvature singularity at finite radius. The single trace $\rT\overline{\rT}$ deformation is dual to a marginal deformation on the string worldsheet, which is pointed out in \cite{Giveon:2017nie}. This deformation has been studied further in parallel to the double trace $\rT\overline{\rT}$ deformation. Many physical quantities have been computed, such as correlation functions \cite{Asrat:2017tzd,Giribet:2017imm}, entanglement entropy \cite{Chakraborty:2018kpr}, holographic Wilson loops \cite{Chakraborty:2018aji} and complexity \cite{Chakraborty:2020fpt}. Generalizations to the case with conformal boundaries have been considered in \cite{Babaro:2018cmq,Giribet:2020kde}. String dynamics for such theories are studied in \cite{Chakraborty:2020swe,Chakraborty:2020cgo,Chakraborty:2020yka}. The relation between the single trace $\rT\overline{\rT}$ deformation and the TsT transformation was pointed out in \cite{Araujo:2018rho,Apolo:2019zai}.\par

The single trace $\rT\overline{\rT}$ deformation can also be generalized to more general solvable irrelevant deformations. The single trace $\rJ\overline{\rT}$ deformation is defined and studied in \cite{Apolo:2018qpq,Chakraborty:2018vja}. Thermodynamics of this deformation is studied in \cite{Apolo:2019yfj}. Penrose limit is considered in \cite{Roychowdhury:2020vzd}, and worldsheet integrability has been studied in \cite{Roychowdhury:2020zer}. One can also combine the single trace $\rT\overline{\rT}$ and $\rJ\overline{\rT}$ deformations to define joint deformation $\rT\overline{\rT}+\rJ\overline{\rT}+\overline{\rJ}\rT$, which is studied in \cite{Chakraborty:2019mdf,Chakraborty:2020xyz,Chakraborty:2020udr}.

\subsection{String theory and 2d gravity}
\label{sec:string}
$\rT\overline{\rT}$ deformation is intimately related to string theories and 2d gravity, which are also non-local theories. Apart from the single trace $\rT\overline{\rT}$ deformation which has a direct string worldsheet realization, there are other connections between $\rT\overline{\rT}$ deformation and string theories. The relation between effective string theory and $\rT\overline{\rT}$ deformation of free bosons have been noticed in  \cite{Dubovsky:2012sh,Dubovsky:2012wk,Caselle:2013dra}, even before the activities in $\rT\overline{\rT}$ flourished. See also \cite{Callebaut:2019omt} for some more recent developments. Later, it was noticed in \cite{Baggio:2018gct} that $\rT\overline{\rT}$ deformation is closely related to the uniform lightcone. This observation was further developed in \cite{Frolov:2019nrr,Frolov:2019xzi,Jorjadze:2020ili,Sfondrini:2019smd}.

As we discussed in section~\ref{lecture3}, $\rT\overline{\rT}$ deformation of a QFT is equivalent to coupling the theory to a 2d topological gravity \cite{Dubovsky:2018bmo,Dubovsky:2017cnj}. This intuition was further clarified in \cite{Tolley:2019nmm} where the 2d topological gravity is identified with massive ghost free dRGT theory in 2d. Generalization of the 2d gravity interpretation to curved spacetime has been discussed in \cite{Tolley:2019nmm,Caputa:2020lpa}. Such an interpretation is also valid for the $\rT\overline{\rT}$ deformation of non-relativistic QFTs \cite{Hansen:2020hrs}. The difference is that the gravity theory is the 2d Newton-Cartan geometry. It is also pointed out in \cite{Hansen:2020hrs} that for non-relativistic QFTs one can actually define three fundamental solvable deformations including $\rT\overline{\rT}$ deformation, all of which have gravity interpretations.\par

A related interpretation of the $\rT\overline{\rT}$ deformation is that it can be seen as a dynamical or field dependent change of coordinates. This was first pointed out in \cite{Dubovsky:2017cnj} from 2d gravity interpretation. Later is was derived from a different perspective in \cite{Conti:2018tca} which was mainly motivated from integrability. The dynamical change of coordinate point of view was further clarified in \cite{Caputa:2020lpa} and generalized to the non-relativistic set-ups in \cite{Cardy:2020olv,Hansen:2020hrs,Ceschin:2020jto}.

\section*{Acknowledgement}

It is a pleasure to thank Ofer Aharony, Shouvik Datta, Amit Giveon and David Kutasov for collaborations on the relevant projects that lead to this review. I thank Gang Yang for kind invitation and hospitality at ITP-CAS. I'm also indebted to Luis Apolo, Wei Li, Pujian Mao, Wei Song, Junbao Wu and Gang Yang for various helpful discussions.

Many thanks to Alex Belin, Shouvik Datta, Amit Giveon, Madalena Lemos, Kostas Siampos, Wei Song, Roberto Tateo, Junbao Wu and Gang Yang for valuable feedbacks.

\end{document}